\documentclass[11pt, oneside]{article}   	
\usepackage{jheppub}
\usepackage{amsmath}
\usepackage{amssymb}
\usepackage{amsfonts}
\usepackage{amsthm}
\usepackage{amsbsy}
\usepackage{array}
\usepackage{mathtools}
\usepackage[usenames,dvipsnames]{xcolor}
\usepackage{subcaption}
\usepackage{dsdshorthand}
\usepackage{graphicx}
\usepackage[vcentermath]{youngtab}

\def\eone{e_1}
\def\etwo{e_2}
\usepackage{tikzit}
\tikzstyle{small circle}=[shape=circle, fill=white, draw=black]
\tikzstyle{medium circle}=[fill=white, draw=black, shape=circle, minimum width=1cm, minimum height=1cm]
\tikzstyle{twoPoint}=[fill=white, draw=black, shape=circle, twopt]
\tikzstyle{threePoint}=[fill=white, draw=black, shape=circle, threept]
\tikzstyle{m vert rect}=[fill=white, draw=black, shape=rectangle, minimum height=2cm, minimum width=.5cm]
\tikzstyle{s box}=[fill=white, draw=black, shape=rectangle]
\tikzstyle{celestialThreePoint}=[fill=white, draw=red, shape=circle, threept]
\tikzstyle{large circle}=[fill=white, draw=black, shape=circle, minimum height=1.5cm, minimum width=1.5cm]
\tikzstyle{green circle of inversion}=[draw={black!20!green}, shape=circle, minimum size=2cm, dashed, tikzit draw={black!20!green}]
\tikzstyle{large green circle}=[draw={black!20!green}, shape=circle, dashed, minimum size=3cm, semithick]
\tikzstyle{small triangle}=[fill=white, draw=black, regular polygon, regular polygon sides=3, inner sep=1.5pt, rotate=90]
\tikzstyle{small blue triangle}=[fill=white, draw={black!20!blue}, regular polygon, regular polygon sides=3, inner sep=1.5pt, rotate=-90, tikzit draw=blue]
\tikzstyle{small green circle}=[draw={black!20!green}, shape=circle, tikzit draw={black!20!green}, dashed, minimum size=.5cm]
\tikzstyle{celestialTwoPoint}=[fill=red, draw=red, shape=circle, inner sep=1pt, minimum size=1pt]

\tikzstyle{arrow}=[->]
\tikzstyle{op}=[-, spinning]
\tikzstyle{dashedOp}=[-, scalar]
\tikzstyle{celestial}=[-, color=red, scalar, tikzit draw=red]
\tikzstyle{minkowski}=[-, spinning, color={black!20!blue}, tikzit draw=blue]

\input{harmonic_analysis.tikzdefs}
\tikzstyle{pole}=[fill=black, draw=black, shape=circle, inner sep=0pt, minimum size=2pt]

\tikzstyle{Arrow}=[->, draw=blue]
\tikzstyle{anti arrow}=[<-, draw=blue]
\tikzstyle{axis}=[->]
\tikzstyle{dashed_arrow}=[->, draw=red]
\tikzstyle{dashed_only}=[-, draw=red]
\tikzstyle{dashed_anti}=[<-, draw=red]

\tikzstyle{point}=[fill=black, draw=black, shape=circle, inner sep=0pt, minimum size=3pt]
\tikzstyle{lr_point}=[fill=red, draw=red, shape=circle, inner sep=0pt, minimum size=2pt]
\tikzstyle{sh_point}=[fill={rgb,255: red,128; green,128; blue,128}, draw={rgb,255: red,128; green,128; blue,128}, shape=circle, inner sep=0pt, minimum size=3pt]
\tikzstyle{sh_lr_point}=[fill={red!40}, draw={red!40}, shape=circle, inner sep=0pt, minimum size=2pt, tikzit fill={rgb,255: red,255; green,128; blue,0}, tikzit draw={rgb,255: red,255; green,128; blue,0}]

\tikzstyle{dashed_st}=[-, dashed]
\tikzstyle{arrow}=[->]
\tikzstyle{dashed_grey}=[-, draw=black, dotted]
\tikzstyle{blue_line}=[draw=blue, ->]
\tikzstyle{energy}=[draw={rgb,255: red,255; green,128; blue,0}, decoration={{snake,amplitude=1pt,segment length=6pt,post length=1pt}}, decorate, ->]
\tikzstyle{blue_line_0}=[-, draw=blue]
\tikzstyle{redline}=[-, draw=red]

\def\bpm{\begin{pmatrix}}
\def\epm{\end{pmatrix}}

\newcommand\scalemath[2]{\scalebox{#1}{\mbox{\ensuremath{\displaystyle #2}}}}

\newcommand\wL{\mathbf{L}}

\renewcommand\vol{\mathop{\mathrm{vol}}}

\definecolor{energycolor}{RGB}{230,50,10}

\tikzset{
  energy/.style={->,
  energycolor,
  decoration={
      snake,
      amplitude=1pt,
      segment length=6pt,
      post length=1pt
    },
  decorate
  }
}

\makeatletter
\def\@fpheader{\ }
\makeatother

\title{Spinning dispersive CFT sum rules and bulk scattering}
\author{Cyuan-Han Chang, Yakov Landau, and David Simmons-Duffin}
\affiliation{Walter Burke Institute for Theoretical Physics, Caltech, Pasadena, California 91125, USA}
\emailAdd{cchang7@caltech.edu}
\emailAdd{ylandau@caltech.edu}
\emailAdd{dsd@caltech.edu}

\date{}
\abstract{We use commutativity of null-integrated operators on the same null plane to construct dispersive CFT sum rules for spinning operators. The contribution of heavy blocks to these sum rules is dominated by a saddle configuration that we call the ``scattering crystal." Correlators in this configuration have a natural flat-space interpretation, which allows us to build a dictionary between dispersive CFT sum rules for stress-tensors and flat-space dispersion relations for gravitons.
This dictionary is a crucial step for establishing the HPPS conjecture for stress tensor correlators.}

\preprint{CALT-TH 2023-043}

\begin{document}

\maketitle
\pagenumbering{roman}
\setcounter{page}{2}
\newpage
\pagenumbering{arabic}
\setcounter{page}{1}

\section{Introduction} 

Causality and unitarity imply constraints on the space of low-energy effective field theories (EFTs) \cite{Aharonov:1969vu,Pham:1985cr}. In a $2\to 2$ scattering process, causality and unitarity manifest as analyticity, Regge boundedness, and crossing symmetry of the S-matrix. Using these properties, one can derive dispersion relations that express the Wilson coefficients of a low-energy EFT in terms of data in the ultraviolet (UV), which have positive spectral density thanks to unitarity \cite{Adams:2006sv,Camanho:2014apa}. By applying functionals (such as expanding around the forward limit) to dispersion relations, positivity of the UV spectral density can be used to obtain two-sided bounds on EFT Wilson coefficients, with the correct scaling in the EFT cutoff $M$ expected by dimensional analysis. A systematic exploration of this approach was initiated recently in \cite{Bellazzini:2020cot, Caron-Huot:2020cmc, Arkani-Hamed:2020blm, Tolley:2020gtv, Sinha:2020win}. 

Including gravitational interactions introduces another layer of complexity, since graviton exchange produces divergences in the forward limit. This issue was overcome in \cite{Caron-Huot:2021rmr} by considering functionals that measure the scattering amplitude at small impact parameter. This method has paved the way for deriving bounds on higher-derivative corrections to General Relativity in flat space \cite{Caron-Huot:2022ugt, Caron-Huot:2022jli,Haring:2022cyf,Henriksson:2022oeu,McPeak:2023wmq}.

Similar questions can be explored in Anti-de Sitter (AdS) space as well. Furthermore, via the AdS/CFT correspondence, these questions can be phrased  --- and potentially answered --- in CFT language. For example, HPPS conjectured in \cite{Heemskerk:2009pn} that a large-$N$ CFT should have a local bulk EFT dual if its single trace spectrum has a large gap $\De_{\mathrm{gap}}=M R_\mathrm{AdS}\gg 1$ for $J>2$, where $R_{\mathrm{AdS}}$ is the AdS radius and $M$ is the cutoff scale of the bulk EFT.  Significant progress towards establishing this conjecture using CFT techniques was made in e.g.~\cite{Camanho:2014apa,Hartman:2015lfa,Afkhami-Jeddi:2016ntf,Meltzer:2017rtf,Kulaxizi:2017ixa,Caron-Huot:2017vep,Costa:2017twz,Belin:2019mnx,Kologlu:2019bco}. Recently, part of the HPPS conjecture was established in \cite{Caron-Huot:2021enk} by building a dictionary between conformal bootstrap functionals and flat space dispersion relations. The authors of \cite{Caron-Huot:2021enk} considered a weakly-coupled EFT in AdS with a massless graviton and a scalar with mass $m_{\f} =O(\frac{1}{R_{\mathrm{AdS}}})$. Schematically, the low-energy effective action is
\be\label{eq:scalarAdS_action}
S=S_{\mathrm{gravity}} + \int d^Dx\sqrt{-g} \p{\frac{1}{2}\f(\ptl^2-m_{\f}^2)\f+ \sum_{n} g_n D^{2n} \f^4 + \ldots},
\ee
where $S_{\mathrm{gravity}}$ contains graviton interactions and will be given below. Using bootstrap methods, they derived bounds on the higher-derivative interactions $g_n$ in \eqref{eq:scalarAdS_action} with the expected suppression in $\De_{\mathrm{gap}}$.

The main tool in this analysis is dispersive CFT sum rules \cite{Mazac:2016qev,Mazac:2018mdx, Carmi:2019cub, Mazac:2019shk, Penedones:2019tng, Caron-Huot:2020adz, Carmi:2020ekr, Meltzer:2021bmb, Knop:2022viy}, which have double zeros at the locations of most double-twist operators. Such sum rules allow one to separate out light double-trace contributions in holographic CFTs, and express the light contribution ($\De<\De_{\mathrm{gap}}$) described by \eqref{eq:scalarAdS_action} in terms of a sum of heavy conformal blocks with positive coefficients. Remarkably, in a certain ``flat space limit" that we review in section \ref{sec:review_superconvergence}, dispersive CFT sum rules reduce to the flat space sum rules previously studied in \cite{Caron-Huot:2021rmr}. As a result, flat space functionals can be uplifted to AdS and lead to bounds on the EFT couplings in \eqref{eq:scalarAdS_action}.\footnote{The ``uplifting" procedure only guarantees positivity of the functional in the ``bulk point" regime discussed below. One must additionally check other regimes to derive a rigorous CFT bound.}

Given the bounds on graviton interactions in flat space obtained recently \cite{Caron-Huot:2022ugt, Caron-Huot:2022jli,Haring:2022cyf,Henriksson:2022oeu,McPeak:2023wmq}, we would like to derive similar bounds on graviton interactions in AdS, in particular establishing HPPS for purely gravitational theories. More precisely, let us consider an EFT with only a massless graviton with the effective action
\be
S_{\mathrm{gravity}} = \frac{1}{16\pi G}\int d^Dx \sqrt{-g}\p{-2\Lambda + R + \a_2 R^2 + \a_3 R^3 + \ldots},
\ee
where $\Lambda = -(D-1)(D-2)/(2R_{\mathrm{AdS}}^2)$. We would like to obtain bounds like
\be\label{eq:alpha2_bound}
|\a_2| \leq \frac{\#}{\De_{\mathrm{gap}}^{2}}. 
\ee
On the CFT side, this means we must construct dispersive sum rules for four-point functions of stress tensors whose flat space limit agrees with flat space sum rules for gravitons.

Conceptually, this task is similar to the one undertaken in \cite{Caron-Huot:2021enk}. However, stress tensor four-point functions are technically more complicated due to the profusion of tensor structures, and we need to organize the calculation carefully. Our approach is to begin with ``subtracted superconvergence" sum rules, which exploit the fact that the commutator of null-integrated operators on the same null plane should vanish \cite{Kologlu:2019bco}. 
In particular, we focus on the action of such sum rules on a heavy conformal block (with $\De > \De_{\mathrm{gap}}$). This action is a spacetime integral that, in the flat space limit, localizes to a certain saddle configuration that we call the ``scattering crystal." Thus, the computation of the heavy action turns into evaluating conformally-invariant structures at this saddle, which is straightforward for spinning operators. Our calculation gives a ``spacetime interpretation" of the results of \cite{Caron-Huot:2021enk} for scalar operators.

The result of our saddle analysis has a simple interpretation in flat space. The inserted conformal blocks become flat space partial waves, and the subtracted superconvergence sum rule becomes a sum rule for flat space ``shock" amplitudes \cite{Camanho:2014apa, Kologlu:2019bco}.
Thus, by comparing the action of CFT sum rules on heavy blocks to the action of flat-space sum rules on heavy states, we can deduce a concrete dictionary between dispersive CFT functionals and flat space sum rules.\footnote{Once we know the dictionary, it follows that the contributions of light states must match between AdS and flat space as well, though we leave exploration of light states to future work.}  Our sum rules for stress tensors become a subset of the known flat space sum rules for gravitons \cite{Caron-Huot:2022jli}. The full set of flat space sum rules for gravitons includes additional sum rules that cannot be expressed in terms of ``shock amplitudes" (because they have different choices of external polarizations). We leave the problem of obtaining these additional sum rules using CFT techniques to future work. Our dictionary will be needed to convert the flat space functionals with positive action on partial waves into CFT functionals with positive action on heavy blocks.

This paper is organized as follows. In section \ref{sec:review_superconvergence}, we  briefly review superconvergence sum rules and the derivation of the scalar bounds in \cite{Caron-Huot:2021enk}. In section \ref{sec:heavyaction_CFT}, we derive a simple formula for the flat space limit of superconvergence sum rules for scalars using a spacetime saddle-point analysis. We then generalize the formula to spinning operators. In section \ref{sec:flat_space}, we explain how this formula can be matched to flat space sum rules, and obtain the dictionary between spinning CFT sum rules and flat space sum rules for photons and gravitons. We conclude in section \ref{sec:discussion}. We summarize our conventions in appendix \ref{app:conventions}, and present an alternative derivation of the flat space limit formula in appendix \ref{app:derivation_largenu}. Technical details on matching the CFT sum rules to flat space are given in appendices \ref{app:partial_waves_more}, \ref{app:dual_large_nu}, and \ref{app:4pt_structures}.

\section{Review: superconvergence sum rules}\label{sec:review_superconvergence}

Let us start by reviewing the idea of superconvergence in CFT, and explain how it can be related to flat space dispersive sum rules. Along the way we will introduce notation that we use throughout this work.

\subsection{The light transform and superconvergence}\label{sec:light_transform}

Subtracted superconvergence sum rules come from studying the commutator of two null-integrated operators on the same null plane \cite{Kologlu:2019bco}. For simplicity we first consider a four-point function of scalar operators. Let us choose lightcone coordinates $x=(u,v,\vec y)$ with $x^2=-uv+\vec y^2$. A subtracted superconvergence sum rule can be written explicitly as
\be\label{eq:superconvergence_vdirection}
\int_{-\oo}^{\oo} dv_1 \int_{-\oo}^{\oo} dv_3 f(v_1,v_3) \<0|\f_4[\f_3(0,v_3,\vec y_3),\f_1(0,v_1,\vec y_1)]\f_2|0\> = 0,
\ee
where $\f_1,\f_3$ are placed on the same null plane $u=0$. This integral vanishes for a simple reason: since $x_{13}^2=\vec y_{13}^2 >0$, the two operators $\f_1,\f_3$ are spacelike separated in the entire integration range, and therefore their commutator must vanish.

To obtain a sum rule, we would like to separate the two orderings $\f_3 \f_1$ and $\f_1 \f_3$ and perform the integral separately for each ordering. However, in order for this to be valid, we must check that each integral converges --- in particular that there is no divergence from the endpoints of the integration contours. See \cite{Kologlu:2019bco} for a detailed analysis of this convergence condition. In the end, convergence can be achieved by a suitable choice of the function $f(v_1,v_3)$, which we call a ``subtraction" factor. Expressions for subtraction factors will be given in section \ref{sec:heavyaction_CFT}.

Our goal in this section is to rewrite \eqref{eq:superconvergence_vdirection} as a conformally-invariant spacetime integral, using tools from \cite{Kravchuk:2018htv}. First, let us introduce index-free notation. For an ordinary integer-spin operator $\cO$, we contract the indices with a null polarization vector $z$:
\be
\cO(x,z) = \cO^{\mu_1\cdots\mu_J}z_{\mu_1}\cdots z_{\mu_J},\quad z^2=0.
\ee
More generally, for a representation $\r$ represented by a Young diagram with more than one row, we introduce polarizations $z,w,\tl{w},\ldots$ for each row of the Young diagram. They should be null and mutually orthogonal. Due to antisymmetry between indices in different rows, we have gauge redundancies $w\sim w+\#z$, $\tl{w}\sim \tl{w}+\#w+\#z$, etc. The number of boxes in each row of the Young diagram encodes the homogeneity of the corresponding polarization vector. Generalizing to arbitrary homogeneity in the polarization vector $z$ allows us to describe representations with continuous spin, which will be crucial in later calculations.\footnote{For the Lorentzian conformal group $\SO(d,2)$, their unitary principal representations have two continuous parameters, $\De$ and $J$ \cite{Kravchuk:2018htv}.}

Index-free notation can be viewed as the embedding space formalism \cite{436861d5-35f6-36cf-9470-cb587eced490,Costa:2011mg} for the Lorentz group $\SO(d-1,1)$. In the embedding space of the $d$-dimensional conformal group $\SO(d,2)$, one promotes the position $x$ and polarization $z$ to $X,Z \in \R^{d,2}$ satisfying $X^2=X\.Z=Z^2=0$ and $Z\sim Z+\# X$. Then, conformal transformations act linearly on $X,Z$, and conformally-invariant structures are simply built from dot products of $X$ and $Z$. We can recover the Minkowski space operator by the dictionary
\be
\cO(x,z)=\cO(X=(1,x^2,x^{\mu}),Z=(0,2x\.z,z^{\mu})),
\ee
where we again use lightcone coordinates $X=(X^{+},X^{-},X^{\mu})$, and  $X^2=-X^{+}X^{-}+X^{\mu}X_{\mu}$. We will often go between embedding space and Minkowski space when writing expressions for conformally-invariant structures. In \eqref{eq:embeddingvec_conditions}, we review the embedding space description of operators with more complicated representations.

The embedding space vector $X$ lives on a projective null cone in $\R^{d,2}$, which is topologically $S^1\x S^{d-1}$. Correlation functions of Lorentzian CFTs live on the universal cover $\tl{\cM}_d=\R\x S^{d-1}$, also called the Lorentzian cylinder. More precisely, CFT correlation functions on $\R^{d-1,1}$ can be analytically continued to their Lorentzian cylinder counterparts \cite{Luscher:1974ez}. The Lorentzian cylinder is tiled by Poincar\'{e} patches, where each patch represents a Minkowski space $\R^{d-1,1}$. The conformal group acting on $\tl{\cM}_d$ is the universal covering group $\tl{\SO}(d,2)$. There exists a symmetry $\cT$ such that for each point $p\in \tl{\cM}_d$, all light-rays emanating from $p$ will converge at $\cT p$ in the next Poincar\'{e} patch. The action of $\cT$ on an operator is most easily described in the embedding space, where we have
\be
\cT\cO(X,Z_1,Z_2,\ldots)\cT^{-1} = \cO(-X,-Z_1,-Z_2,\ldots).
\ee
We will also use the notation $p^{+}\equiv \cT p,p^{-}\equiv \cT^{-1}p$.

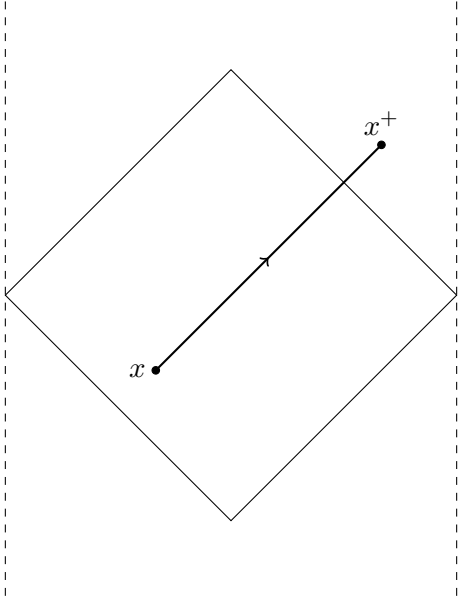
\begin{figure}[t!]
	\centering
	\begin{tikzpicture}	
	\draw[dashed] (-3,-4) -- (-3,4);
	\draw[dashed] (3,-4) -- (3,4);
	\draw[] (-3,0) -- (0,3) -- (3,0) -- (0,-3) -- cycle;

	\draw[fill=black] (-1,-1) circle (0.05);
	\node[left] at (-1,-1) {$x$};
	\draw[fill=black] (-1+3,-1+3) circle (0.05);
	\node[above] at (-1+3,-1+3) {$x^{+}$};
	
	\begin{scope}[thick,decoration={
    	markings,
   	 mark=at position 0.5 with {\arrow{>}}}
    	] 
    	\draw[postaction={decorate}]  (-1,-1) -- (-1+3,-1+3);
    	\end{scope}
	
	\end{tikzpicture}
	\caption{Contour of the light transform on the Lorentzian cylinder, where the two dashed lines should be identified. The contour starts at point $x$ and ends at $x^{+}$ on the next Poincar\'{e} patch.}
	\label{fig:light_transform}
\end{figure}

The null-integrated operators in \eqref{eq:superconvergence_vdirection} are an example of a conformally-invariant integral transform called the light transform \cite{Kravchuk:2018htv}. The light transform of a local operator $\cO$ with quantum numbers $(\De,J)$ is defined as
\be\label{eq:light_transform_definition}
\wL[\cO](x,z) = \int_{-\oo}^{\oo}d\a (-\a)^{-\De-J} \cO\p{x-\frac{z}{\a},z}.
\ee
Under conformal transformations, $\wL[\cO](x,z)$ transforms like a primary operator at $x$. The light transform contour is shown in figure \ref{fig:light_transform}. It starts at the point $x$, goes along the direction of the polarization $z$, and eventually ends at the point $\cT x=x^{+}$. The previous description of the null-integrated operator in \eqref{eq:superconvergence_vdirection}, where one puts the operator at $u=0$ and integrates along $v$, can be obtained by putting $x$ at past null infinity.

In the embedding space, the light transform \eqref{eq:light_transform_definition} becomes
\be
\wL[\cO](X,Z)= \int_{-\oo}^{\oo}d\a\ \cO(Z-\a X,-X).
\ee
From this expression, it is not hard to see that the quantum numbers of $\wL[\cO]$ are $(1-J,1-\De)$. The light transform of a local operator annihilates the vacuum \cite{Kravchuk:2018htv}:
\be\label{eq:light_transform_vacuum}
\wL[\cO](x,z)|\O\> = 0.
\ee

With all this in place, we can now write the subtracted superconvergence sum rule as a conformally-invariant spacetime integral. We define a functional $\Psi_{k,\nu}$ by\footnote{This functional is equivalent to the $\tl{C}_{k,\nu}$ functional defined in the appendix of \cite{Caron-Huot:2021enk}.}
\be\label{eq:scalarfunctional_nokernelexpression}
&\Psi_{k,\nu}[\cG] \equiv \int \frac{d^d x_2 d^d x_4 d^d x_0 D^{d-2}z_1 D^{d-2}z_3}{\vol(\tl \SO(d,2))}K_{k,\nu}(x_2,x_4,x_0,z_1,z_3) \nn \\
&\qquad\qquad \x\<\O|\f_4(x_4)[\wL[\f_3](x_0,z_3),\wL[\f_1](x_0,z_1)]\f_2(x_2)|\O\>,
\ee
where $K_{k,\nu}$ is a kernel chosen such that the integral is conformally-invariant. Its explicit expression is given below in \eqref{eq:scalarfunctional_definition}. The integral measure for the polarization $D^{d-2}z$ is defined as \cite{SimmonsDuffin:2012uy}
\be
D^{d-2}z \equiv \frac{2d^dz\de(z^2)\th(z^0)}{\vol \R_{+}},
\ee
where $\R_{+}$ acts by rescaling $z$. Then, the superconvergence condition \eqref{eq:superconvergence_vdirection} is equivalent to
\be
\Psi_{k,\nu}[(\mathrm{subtractions})\x \cG] = 0,
\ee
where $\cG$ is a four-point function and (subtraction) corresponds to the factor $f(v_1,v_3)$ in \eqref{eq:superconvergence_vdirection} (we discuss them in more detail below).

Using \eqref{eq:light_transform_vacuum}, we can rewrite \eqref{eq:scalarfunctional_nokernelexpression} as
\be
&\Psi_{k,\nu}[\cG] = \int \frac{d^d x_2 d^d x_4 d^d x_0 D^{d-2}z_1 D^{d-2}z_3}{\vol(\tl \SO(d,2))}K_{k,\nu}(x_2,x_4,x_0,z_1,z_3) \nn \\
&\qquad\qquad \x\<\O|[\f_4(x_4),\wL[\f_3](x_0,z_3)][\wL[\f_1](x_0,z_1),\f_2(x_2)]|\O\> - (1\leftrightarrow 3).
\ee
When we replace the four-point function $\cG$ with an s-channel conformal block $G^s_{\De,J}$, the double commutator $\<\O|[\f_4,\wL[\f_3]][\wL[\f_1],\f_2]|\O\>$ will give a $\sin^2(\pi\tfrac{\De-J-2\De_\f}{2})$ factor. For double-twist operators $\De-J=2\De_\f+2n$, the $\sin^2$ factor then becomes a double zero. Therefore, $\Psi_{k,\nu}$ is a dispersive functional, meaning that it gives double zeros for all but finitely many double-twist operators.\footnote{When there is a subtraction factor, the $\sin^2$ factor gets modified and can become nonzero at finitely many double twist locations \cite{Caron-Huot:2020adz}.}

\subsection{The flat space limit of superconvergence sum rules}\label{sec:flat_space_limit}

Let us now review the argument in \cite{Caron-Huot:2021enk} that leads to bounds on the scalar AdS EFT \eqref{eq:scalarAdS_action}. As explained in the introduction, the idea is that in a certain ``flat space limit,"  dispersive CFT sum rules reduce to flat space sum rules.

\subsubsection{Review: flat space sum rules}
Let us first briefly review flat space sum rules for identical real scalars. Flat space sum rules for photons and gravitons will be reviewed in section \ref{sec:flat_space}.

Let $\cM(s,u)$ be a $2\to 2$ scattering amplitude for identical scalars, which we assume satisfies analyticity, crossing symmetry, and Regge boundedness. This allows one to write down flat space dispersion relations for $\cM(s,u)$ (see e.g., \cite{Caron-Huot:2020cmc, Caron-Huot:2021rmr} for more details),
\be\label{eq:dispersion_relation_scalar}
-\oint_{\oo}\frac{ds}{2\pi i}\frac{1}{s}\frac{\cM(s,u)}{(s(s+u))^{\frac{k}{2}}} = 0,
\ee
where $s,u$ are the Mandelstam variables,
\be\label{eq:Mandelstams_definition}
s=-(p_1+p_2)^2,\quad t=-(p_2+p_3)^2,\quad u=-(p_1+p_3)^2.
\ee
The contour in (\ref{eq:dispersion_relation_scalar}) can be deformed and separated into two parts: the low energy part, which can be computed using an EFT, and the high energy part, which can be decomposed into partial waves. 

The partial wave decomposition of a scalar amplitude $\cM(s,u)$ is given by
\be\label{eq:amp_partialwave_expansion_scalar}
\cM(s,u) = s^{\frac{4-D}{2}} \sum_{J} n_{J}^{(D)} a_J(s)\cP_{J}\p{1+\frac{2u}{s}}.
\ee
Here, $n_{\r}^{(D)}=\frac{2^{d+1}(2\pi)^{d-1}\dim\r}{\vol S^{d-1}}$, where in our case $\rho$ is the spin-$J$ traceless symmetric tensor representation of $\SO(d)$, and $d=D-1$. Meanwhile $\cP_J$ is a Gegenbauer polynomial,
\be\label{eq:PJ_expr}
\cP_J(x)={}_2F_1(-J,J+d-2,\tfrac{d-1}{2},\tfrac{1-x}{2}).
\ee

After the partial wave decomposition, the high energy part of the the dispersion relation \eqref{eq:dispersion_relation_scalar} becomes
\be\label{eq:heavystate_expression_flatspace_scalar}
\left\< \frac{2m^2+u}{m^2+u}\frac{\cP_J(1+\tfrac{2u}{m^2})}{(m^2(m^2+u))^{\frac{k}{2}}}\right\>,
\ee
where $\<\cdots\>$ is a heavy average, defined as a sum with positive coefficients over heavy states with mass $m$ and spin $J$,
\be\label{eq:heavyaverage_flatspace_scalar}
\<\cdots\> = \frac{1}{\pi} \sum_{J} n_{J}^{(D)} \int_{M^2}^{\oo}  \frac{dm^2}{m^2}m^{4-D}  \Im a_{J}(s)\p{\cdots}.
\ee 

Unitarity implies that the spectral density $\Im a_{J}(s)$ should be positive (or more generally a positive semidefinite matrix). Therefore, by applying a functional that has positive action on all partial wave contributions \eqref{eq:heavystate_expression_flatspace_scalar}, one can derive inequalities for Wilson coefficients of the low energy EFT. Without gravity, the functional can simply include taking the forward limit $u\to 0$. For amplitudes with graviton exchange, one can integrate \eqref{eq:heavystate_expression_flatspace_scalar} over $u$ against some kernel (see \cite{Caron-Huot:2021rmr,Caron-Huot:2022ugt, Caron-Huot:2022jli}).

\subsubsection{The flat space limit}
Consider a large-$N$ CFT whose single-trace spectrum consists of a scalar operator $\f$, the stress tensor $T^{\mu\nu}$, and operators with twists $\tau=\De-J$ greater than $\De_{\mathrm{gap}} \gg 1$. By the HPPS conjecture, this theory is expected to be dual to an EFT in AdS with the effective action \eqref{eq:scalarAdS_action}. The s-channel conformal block decomposition of $\<\f\f\f\f\>$ is given by
\be
\<\f\f\f\f\> = G^s_{1} +f^2_{T} G^s_{T^{\mu\nu}} +\sum_{n,J} f^2_{[\f\f]_{n,J}} G^s_{[\f\f]_{n,J}} + \sum_{\tau >\De_{\mathrm{gap}}}f^2_{\De,J} G^s_{\De,J},
\ee
where $f_{\cO}$ is the OPE coefficient, and $[\f\f]$ are double-trace operators built from $\f$. Consider now a dispersive functional $\omega$ whose action on the four-point function vanishes. Applying it to the conformal block decomposition gives us
\be
-\left.\omega\right|_{\mathrm{light}}&= \sum_{\tau >\De_{\mathrm{gap}}}f^2_{\De,J} \omega[G^s_{\De,J}],
\ee
where
\be
\left.\omega\right|_{\mathrm{light}}&\equiv \omega[G^s_{1}] +f^2_{T} \omega[G^s_{T^{\mu\nu}}] +\sum_{n,J} f^2_{[\f\f]_{n,J}} \omega[G^s_{[\f\f]_{n,J}}].
\ee
One can argue that $\omega|_\textrm{light}$ is determined by the Wilson coefficients of the low-energy EFT in AdS. Therefore, if we can find a functional $\omega$ such that
\be
\omega[G^s_{\De,J}] \geq 0,\quad \tau >\De_{\mathrm{gap}},
\ee
then positivity of $f^2_{\De,J}$ implies that
\be\label{eq:omega_on_light}
-\left.\omega\right|_{\mathrm{light}} \geq 0,
\ee
which becomes an inequality on EFT Wilson coefficients.

To study the action of $\omega$ on heavy blocks, it is useful to consider two special limits.
\begin{itemize}
\item The ``bulk point"/``flat space" limit is the regime of large $\De$ with $J/\Delta\ll 1$. As explained in \cite{Caron-Huot:2021enk}, this corresponds to an AdS scattering process where the energy $m$ and impact parameter $\beta$ are given by 
\be
m^2 \approx \De^2, \quad \beta\approx \frac{2J}{\De} \ll 1.
\ee
The term ``flat space" comes from the fact that the impact parameter $\beta$ is much smaller than the AdS radius (which is 1 in our conventions).

\item The Regge limit is the regime of large $\De,J$ with fixed ratio $\De/J$. This corresponds to an AdS scattering process with impact parameter comparable to the AdS radius.
\end{itemize}

A necessary condition for a functional $\w$ to be positive is that it should be positive in the flat space regime. In this regime, one can relate the action of $\w$ on conformal blocks to the action of simple flat-space dispersion relations. For example,
\be\label{eq:superconvergence_bulkpoint_review}
\Psi_{k,\nu;f_k}[\De,J] = \frac{2m^2-\nu^2}{m^2-\nu^2}\frac{\cP_J(1-\tfrac{2\nu^2}{m^2})}{(m^2(m^2-\nu^2))^{\frac{k}{2}}}\p{1 + O\p{\frac{J^2}{m^2}}},
\ee
where $\cP_J$ is a Gegenbauer polynomial given by \eqref{eq:PJ_expr}. On the left-hand side $\w[\De,J]$ denotes the action of $\w$ on a conformal block, rescaled by some positive factors:
\be\label{eq:heavyaction_scalar_definition}
\omega[\De,J] \equiv \frac{1}{q_{\De,J}}\frac{\omega[G^s_{\De,J}]}{2\sin^2(\pi\tfrac{\De-J-2\De_\f}{2})},
\ee
where
\be
q_{\De,J} &= \frac{1}{\pi p_{\De,J}^{\mathrm{MFT}}}\frac{2n_J}{m^{2d-4}}\frac{\G(\De-1)(2\De-d)}{\G(\De-d+2)}, \nn \\
n_J&= \frac{2^{d+1}\pi^{\frac{d-1}{2}}}{\G(\tfrac{d-1}{2})}(J+1)_{d-3}(2J+d-2).
\ee 
$ p_{\De,J}^{\mathrm{MFT}}$ is the OPE coefficient of the Mean Field Theory, whose expression can be found in \cite{Fitzpatrick:2011dm}.

The functional $\Psi_{k,\nu;f_k}$ denotes a subtracted version of $\Psi_{k,\nu}$, where we insert an additional factor $f_k$ into the integrand. As discussed below \eqref{eq:superconvergence_vdirection} this factor is needed to modify the behavior near the endpoints of the integral. For example, for $k=2$, an appropriate subtraction factor is given by
\be
f_{k=2}(u',v') = \frac{v'-u'}{u'v'},
\ee
where $u',v'$ are conformally-invariant cross-ratios,
\be\label{eq:cross_ratios}
u'=\frac{x_{12}^2x_{34}^2}{x_{13}^2x_{24}^2},\quad v'=\frac{x_{14}^2x_{23}^2}{x_{13}^2x_{24}^2}.
\ee
We give subtraction factors for general $k$ in section \ref{sec:scalar_sum_rules}.

The right hand side of \eqref{eq:superconvergence_bulkpoint_review} is exactly the contribution of a state with mass $m$ and spin $J$ to the flat space scalar sum rule \eqref{eq:heavystate_expression_flatspace_scalar} after identifying $s=m^2,u=-\nu^2$. This guarantees that the flat space functionals with positive action on \eqref{eq:heavystate_expression_flatspace_scalar} can be uplifted to AdS and will give us a CFT functional with positive action on all blocks in the bulk-point limit ($\De \gg 1$).

As explained in \cite{Caron-Huot:2021enk}, to obtain bounds on the bulk EFT Wilson coefficients, one must also ensure that the functional $\w$ is positive outside the bulk point limit, in particular in the Regge regime (and other regimes if necessary). We leave an exploration of the Regge regime for four-point functions of spinning operators to future work. The goal in this work is to derive the bulk-point limit result \eqref{eq:superconvergence_bulkpoint_review} and its generalization to spinning operators from a saddle analysis of the spacetime integral.

Before we proceed, let us comment on a subtlety regarding \eqref{eq:superconvergence_bulkpoint_review}. In \cite{Caron-Huot:2021enk}, \eqref{eq:superconvergence_bulkpoint_review} was derived (for a slightly different functional $C_{k,\nu}$ which is equivalent to $\Psi_{k,\nu}$ at large $\nu$) for all $\nu\in [0,m)$. However, the derivation in this work will be in the limit where both $\nu$ and $m$ are large (with fixed ratio). (Therefore, sometimes we will also refer to the bulk-point limit as the $\nu,m \gg 1$ limit.) In \cite{Caron-Huot:2021enk} it was shown that in the scalar case, the regimes of finite $\nu$ and large-$\nu$ are continuously connected at large $m$. In other words, we can safely compute at large $\nu$, and then later consider all $\nu\in [0,m)$. We expect the same to be true in the spinning case as well.

\section{Heavy action from a spacetime saddle-point analysis}\label{sec:heavyaction_CFT}

In this section, we define our functional in position space, and study its action on a conformal block $G_{\De,J}$ with large scaling dimension $\De$. We will show that in the bulk point limit $\nu, \De \gg 1$, the integral completely localizes to a saddle point, and the heavy action can be obtained by evaluating conformally-invariant structures at the saddle. This allows us to derive a simple formula for the functional in the bulk point limit. We will first use this formula to reproduce the results summarized in section \ref{sec:review_superconvergence}, and then write down the generalized functional for spinning operators.

\subsection{Functionals as spacetime integrals}
As explained in section \ref{sec:review_superconvergence}, our starting point for the position space functional is the fact that the commutator of light-transformed operators vanishes,
\be\label{eq:L_commutator}
\<\O|\cO_4(x_4)[\wL[\cO_3](x_0,z_3),\wL[\cO_1](x_0,z_1)]\cO_2(x_2)|\O\> =0.
\ee
To make sure the light transform integrals $\wL[\cO_1]\wL[\cO_3]$ converge, we also need the condition $J_1+J_3-1>J_0$, where $J_0$ is the Regge intercept \cite{Kologlu:2019bco}. For the moment, let us assume this condition is satisfied, and we will come back to this issue later in section \ref{sec:scalar_sum_rules} and \ref{sec:spinning_sum_rules} by introducing subtraction factors.

\begin{figure}[t]
	\centering
		\begin{subfigure}[t]{0.4\textwidth}
		\begin{tikzpicture}
			\draw[blue,thick] (-3.3,-1) -- (-3.3+3,-1+3);
			\draw[red,thick] (-3.3,-1) -- (-3.3-0.9,-1+0.9);
			\draw[red,thick] (-3.3-0.9+6,-1+0.9) -- (-3.3+3,-1+3);
			\draw[dashed] (-4.2,-3.8) -- (-4.2,3.5);
			\draw[dashed] (1.8,-3.8) -- (1.8,3.5);	
			\draw[fill=black] (-2.6,-0.3) circle (0.05);
			\node[below] at (-2.6+0.1,-0.3) {$3$};
			\draw[fill=black] (-3.3,-1) circle (0.05);
			\node[below] at (-3.3,-1) {$0$};
			\draw[fill=black] (-3.3+3,-1+3) circle (0.05);
			\node[above] at (-3.3+3,-1+3) {$0^+$};
			\draw[fill=black] (-2.6+3,-1.7+3) circle (0.05);
			\node[above] at (-2.6+3+0.25,-1.7+3+0.05) {$1$};
			\draw[fill=black] (-2.6+3,-3+3) circle (0.05);
			\node[below] at (-2.6+3,-3+3) {$2$};
			\draw[fill=black] (-2.6+3,1-3) circle (0.05);
			\node[above] at (-2.6+3,1-3) {$4$};
		\end{tikzpicture}
		\caption{Original configuration}
		\label{fig:causality_config_1}
	\end{subfigure}
	\hspace{1.5cm}
	\begin{subfigure}[t]{0.4\textwidth}
		\begin{tikzpicture}
			\draw[blue,thick] (-3.3,-1) -- (-3.3+3,-1+3);
			\draw[red,thick] (-3.3,-1) -- (-3.3-0.9,-1+0.9);
			\draw[red,thick] (-3.3-0.9+6,-1+0.9) -- (-3.3+3,-1+3);
			\draw[dashed] (-4.2,-3.8) -- (-4.2,3.5);
			\draw[dashed] (1.8,-3.8) -- (1.8,3.5);	
			\draw[fill=black] (-2.6,-0.3) circle (0.05);
			\node[below] at (-2.6+0.1,-0.3) {$3$};
			\draw[fill=black] (-3.3,-1) circle (0.05);
			\node[below] at (-3.3,-1) {$0$};
			\draw[fill=black] (-2.6,-1.7) circle (0.05);
			\node[below] at (-2.6,-1.7) {$1$};
			\draw[fill=black] (-3.3+3,-1+3) circle (0.05);
			\node[above] at (-3.3+3,-1+3) {$0^+$};
			\draw[fill=black, opacity=0.4] (-2.6+3,-1.7+3) circle (0.05);
			\node[above, opacity=0.4] at (-2.6+3+0.25,-1.7+3+0.05) {$1^{+}$};
			\draw[fill=black] (-2.6,-3) circle (0.05);
			\node[below] at (-2.6,-3) {$2$};
			\draw[fill=black] (-2.6,1) circle (0.05);
			\node[above] at (-2.6,1) {$4$};
			\draw[fill=black, opacity=0.4] (-2.6+3,-3+3) circle (0.05);
			\node[below, opacity=0.4] at (-2.6+3,-3+3) {$2^{+}$};
			\draw[fill=black, opacity=0.4] (-2.6+3,1-3) circle (0.05);
			\node[above, opacity=0.4] at (-2.6+3,1-3) {$4^{-}$};
		\end{tikzpicture}
		\caption{Configuration after applying $\cT$'s}
		\label{fig:causality_config_2}
	\end{subfigure}
	\caption{Causality configuration of the functional. The left figure shows the configuration in the original definition \eqref{eq:scalarfunctional_definition}. Points $1,3$ both start from $0$ and are integrated along two different null directions. The right figures shows the configuration after applying $\cT_1^{-1},\cT_2^{-1},\cT_4$, which becomes $4>3>0>1>2$.}
	\label{fig:causality_config}
\end{figure}
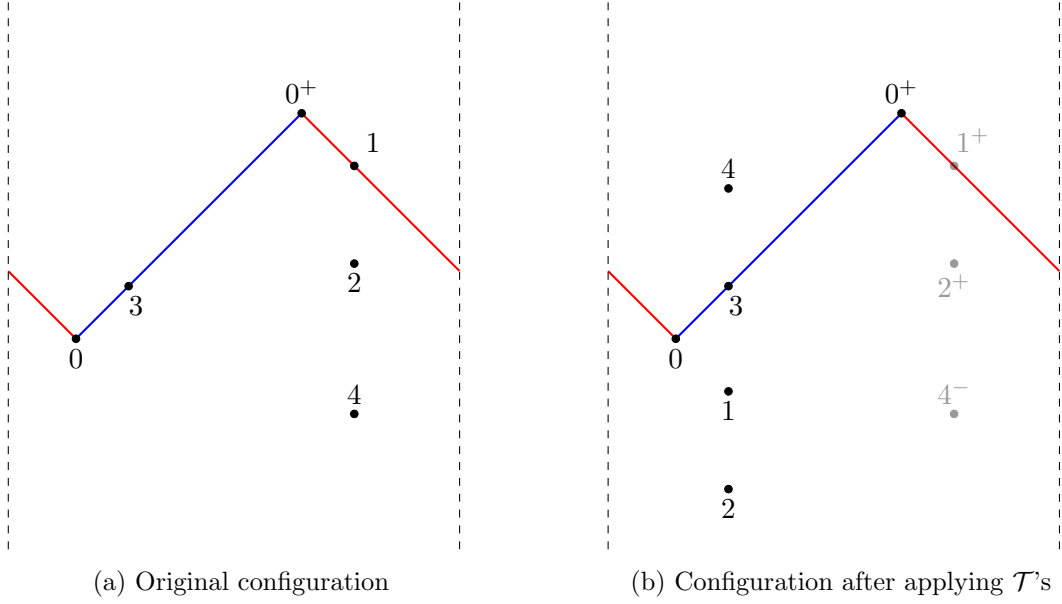

We can integrate \eqref{eq:L_commutator} against appropriate conformally-invariant structures to get different sum rules. We first consider the functional for scalar four-point functions. As we will see shortly, once we have derived this formula, the generalization to spinning operators will be obvious. For simplicity, we assume the external operators all have scaling dimension $\De_\f$, but we will not impose that they are identical until section \ref{sec:scalar_sum_rules}. Our functional is defined as
\be\label{eq:scalarfunctional_definition}
&\Psi_{k,\nu}[\cG] \equiv A_{k,\nu} \int_{\substack{2>4 \\ 0\approx 2,4}} \frac{d^d x_2 d^d x_4 d^d x_0 D^{d-2}z_0}{\vol(\tl \SO(d,2))} \<0|\f_4(x_4^{+})\wL[\cO](x_0,z_0)\f_2(x_2)|0\>^{-1} \nn \\
&\int D^{d-2}z_1 D^{d-2}z_3 \<\tl \cP_{\de_1}(z_1) \tl \cP_{\de_3}(z_3) \cP_{\de}(z_0)\> \<\O|\f_4(x_4^{+})[\wL[\f_3](x_0,z_3),\wL[\f_1](x_0,z_1)]\f_2(x_2)|\O\>,
\ee
where the coefficient $A_{k,\nu}$ is given by
\be\label{eq:Aknu_expression}
A_{k,\nu} = 2^{-8+4\De_\f}\pi^{-2-\frac{d}{2}}e^{\pi\nu}\nu^{2+d-2k-4\De_\f}\G\p{\tfrac{d-2}{2}}\G(\De_\f)^2\G\p{\De_\f-\tfrac{d-2}{2}}^2.
\ee
The coefficient is chosen such that the heavy action of $\Psi_{k,\nu}$ agrees with the flat space sum rule. We have also applied $\cT_4$ to the integrand for later convenience. The causality relations between the points are shown in figure \ref{fig:causality_config_1}. In \eqref{eq:scalarfunctional_definition}, $2>4$ indicates that $x_2$ is in the future lightcone of $4$, and $0\approx 2,4$ indicates that $x_0$ is spacelike separated from $x_2$ and $x_4$.

Let us be explicit about the quantum numbers and structures in the definition \eqref{eq:scalarfunctional_definition}. The operator $\cO$ in the first line has scaling dimension $\De=\frac{d}{2}+i\nu$, and its spin is fixed by symmetry to be $J=-1$ \cite{Hofman:2008ar, Kologlu:2019mfz}. The structure $\<0|\f_4(x_4^{+})\wL[\cO](x_0,z_0)\f_2(x_2)|0\>^{-1}$ is a dual structure of a light transformed three-point function with respect to a Lorentzian three-point pairing,
\be
&\p{\<\cO_1\cO_2\cO\>,\<\tl{\cO}^{\dag}_1\tl{\cO}^{\dag}_2\cO^{S\dag}\>}_L\nn \\
&\equiv \int_{\substack{2<1 \\ x\approx 1,2}} \frac{d^d x_1 d^d x_2 d^d x D^{d-2}z}{\vol(\tl \SO(d,2))} \<\cO_1(x_1)\cO_2(x_2)\cO(x,z)\>\<\tl{\cO}^{\dag}_1(x_1)\tl{\cO}^{\dag}_2(x_2)\cO^{S\dag}(x,z)\>.
\ee
The dual structure is defined by
\be
\p{\<0|\f_4(x_4^{+})\wL[\cO](x_0,z_0)\f_2(x_2)|0\>^{-1},\<0|\f_4(x_4^{+})\wL[\cO](x_0,z_0)\f_2(x_2)|0\>}_L = 1.
\ee
The explicit expression is \cite{Kravchuk:2018htv}\footnote{Here, we choose the structures to be an analytic continuation from odd $J$, so the $(-1)^J$ factor in \cite{Kravchuk:2018htv} becomes $-1$.}
\be\label{eq:dual_struct_scalar}
&\<0|\f_4(x_4^{+})\wL[\cO_{\De,J}](x_0,z_0)\f_2(x_2)|0\>^{-1} \nn \\
&= \frac{2^{2d-2}\G(\tfrac{\De+J}{2})^2 \vol(\SO(d-2))}{2\pi i \G(\De+J-2)}\frac{(2z_0\.x_{40}x_{20}^2-2z_0\.x_{20}x_{40}^2)^{\De-d+1}}{(-x_{24}^2)^{\frac{\tl{\De}_2+\tl{\De}_4+\De-J-2d+2}{2}}(x_{02}^2)^{\frac{\tl{\De}_2+\De+J-\tl{\De}_4}{2}}(x_{02}^2)^{\frac{\tl{\De}_4+\De+J-\tl{\De}_2}{2}}}.
\ee

In the second line of \eqref{eq:scalarfunctional_definition}, $\<\tl \cP_{\de_1}(z_1) \tl \cP_{\de_3}(z_3) \cP_{\de}(z_0)\>$ is a three-point structure of a fictitious $(d-2)$-dimensional Euclidean CFT on the celestial sphere, whose conformal symmetry is the Lorentz symmetry $\SO(d-1,1)$ of the original CFT.  The polarizations $z_i$ should be viewed as  embedding space coordinates of this $\mathrm{CFT}_{d-2}$, and $\cP_{\de}(z)$ is a primary with scaling dimension $\de$ on the celestial sphere. In order for the integral \eqref{eq:scalarfunctional_definition} to be conformally-invariant, we must have $\de=\frac{d-2}{2}+i\nu$. We also use the notation $\tl{\cP}_{\de_i} = \cP_{d-2-\de_i}$, and $\de_i=\De_i-1$, while $\tl{\de_i} = d-2 -\de_i$ is the shadow dimension with respect to the $(d-2)$-dimensional fictitious CFT. The celestial three-point structure is given by
\be
\<\tl \cP_{\de_1}(z_1) \tl \cP_{\de_3}(z_3) \cP_{\de}(z_0)\> = \frac{1}{(-2z_1\.z_0)^{\frac{\de+\tl{\de}_1-\tl{\de}_3}{2}}(-2z_3\.z_0)^{\frac{\de+\tl{\de}_3-\tl{\de}_1}{2}}(-2z_1\.z_3)^{\frac{\tl{\de}_1+\tl{\de}_3-\de}{2}}}.
\ee

Originally, the superconvergence sum rule \eqref{eq:L_commutator} is parametrized by the positions and polarizations of the operators, subject to conformal invariance. By integrating it against the dual structure and celestial structure described above, we have transformed the statement of superconvergence into $\nu$-space, which parametrizes the quantum numbers of the structures. In fact, one can also show that the condition $\Psi_{k,\nu}[\cG]=0$ is equivalent to \cite{Kologlu:2019mfz}
\be
C^{-}(\tfrac{d}{2}+i\nu,J=-1) = 0, \quad J_0<-1,
\ee
where $C^{-}(\De,J)$ is the coefficient function computed by the Lorentzian inversion formula \cite{Caron-Huot:2017vep, Simmons-Duffin:2017nub}, and encodes the analytic continuation of the CFT data from odd spin.

To further simplify the functional, let us focus on one term in the commutator $[\wL[\f_3],\wL[\f_1]]$ and consider
\be\label{eq:scalarfunctional_plus}
\Psi_{k,\nu} &= \Psi^{+}_{k,\nu} - (1\leftrightarrow 3), \nn \\
\Psi^{+}_{k,\nu}[\cG] &=  A_{k,\nu} \int_{\substack{2>4 \\ 0\approx 2,4}} \frac{d^d x_2 d^d x_4 d^d x_0 D^{d-2}z_0}{\vol(\tl \SO(d,2))} \<0|\f_4(x_4^{+})\wL[\cO](x_0,z_0)\f_2(x_2)|0\>^{-1} \nn \\
&\int D^{d-2}z_1 D^{d-2}z_3 \<\tl \cP_{\de_1}(z_1) \tl \cP_{\de_3}(z_3) \cP_{\de}(z_0)\> \<\O|[\f_4(x_4^{+}),\wL[\f_3](x_0,z_3)][\wL[\f_1](x_0,z_1),\f_2(x_2)]|\O\>.
\ee
Note that since light-transformed operators annihilate the vacuum, we can rewrite the four-point function as a double commutator. Moreover, in \eqref{eq:scalarfunctional_plus}, we can rewrite the integral over the two light transform directions $z_1,z_3$ as an integral over the positions of $\f_1,\f_3$ with additional delta function constraints \cite{Chang:2020qpj}. Explicitly, this gives 
\be\label{eq:scalarfunctional_plus_deltafunctions}
\Psi^{+}_{k,\nu}[\cG] &=  4A_{k,\nu} \int_{\substack{2>4 \\ 0\approx 2,4}} \frac{d^dx_1 d^d x_2 d^d x_3 d^d x_4 d^d x_0 D^{d-2}z_0}{\vol(\tl \SO(d,2))} \<0|\f_4(x_4^{+})\wL[\cO](x_0,z_0)\f_2(x_2)|0\>^{-1} \nn \\
& \<\tl \cP_{\de_1}(x_{10}) \tl \cP_{\de_3}(x_{30}) \cP_{\de}(z_0)\>\de(x_{10}^2)\de(x_{30}^2)\th(x_{10})\th(x_{30}) \<\O|[\f_4(x_4^{+}),\f_3(x_3)][\f_1(x_1),\f_2(x_2)]|\O\>,
\ee
where the theta-functions $\th(x_{10})\th(x_{30})$ indicate that $x_{10}$ and $x_{30}$ must be timelike and future-pointing. Finally, let us make a change of variables $x_1 \to x_{1}^{+},x_{2}\to x_{2}^{+}, x_{4} \to x_{4}^{-}$. This makes the causality configuration equivalent to $4>3>0>1>2$ (see figure \ref{fig:causality_config_2}), and we arrive at the expression
\be\label{eq:scalarfunctional_plus_final}
\Psi^{+}_{k,\nu}[\cG] &=  4A_{k,\nu} \int_{4>3>0>1>2} \frac{d^dx_1 d^d x_2 d^d x_3 d^d x_4 d^d x_0 D^{d-2}z_0}{\vol(\tl \SO(d,2))} \<0|\f_4(x_4)\wL[\cO](x_0,z_0)\f_2(x_2^{+})|0\>^{-1} \nn \\
& \<\tl \cP_{\de_1}(x_{1^{+}0}) \tl \cP_{\de_3}(x_{30}) \cP_{\de}(z_0)\>\de(x_{1^{+}0}^2)\de(x_{30}^2)\th(x_{1^{+}0})\th(x_{30}) \<\O|[\f_4(x_4),\f_3(x_3)][\f_1(x_1^{+}),\f_2(x_2^{+})]|\O\>.
\ee

\subsection{Action on conformal blocks}
Let us now apply the functional $\Psi^{+}_{k,\nu}$ to an s-channel conformal block $G^s_{\De,J}$ and study the action in the large $\De$ limit. In position space, the most useful expression for the block is the Lorentzian shadow representation \cite{Polyakov:1974gs}, given by integrating a fifth point $x_5$ over a causal diamond $4>5>3$: 
\be\label{eq:block_shadowrep0}
G^{s}_{\De,J} = \frac{1}{\b_{\De,J}} \int_{4>5>3}d^dx_5 |\<\f_1\f_2\cO^{\mu_1\cdots\mu_J}(x_5)\>| |\<\tl{\cO}^{\dag}_{\mu_1\cdots\mu_J}(x_5)\f_3\f_4\>|.
\ee
The external points should satisfy the causality constraint $1>2,4>3$ with all other pairs of points spacelike-separated. On the right-hand side, the operator $\cO$ has quantum numbers $(\De,J)$, and the quantum numbers of $\tl{\cO}^{\dag}$ are related to those of $\cO$ by a shadow transform and Hermitian conjugate:\footnote{In the case where $\rho$ is a traceless symmetric tensor representation with spin $J$, we have $\rho^R=\rho=(\rho^R)^*$.}
\be
\tl{\cO}&: (\De,\r) \mapsto (d-\De,\r^{R}),\nn \\
\cO^{\dag}&: (\De,\r) \mapsto (\De,\r^{\dag} = (\r^{R})^{*}).
\ee

In \eqref{eq:block_shadowrep0}, we contract the indices of the exchanged operator $\cO_5$. This is more natural for our purpose because the spin of the inserted block corresponds to spin of the exchanged massive particle in flat space, and hence we are only interested in the integer-spin case.\footnote{This is in contrast to the continuous-spin version of the Lorentzian shadow representation given in \cite{Kravchuk:2018htv} that involves integrating over the polarization vector of the exchanged operator.} Moreover, later we will find a correspondence between blocks and flat space partial waves, which also come from contracting the indices of spin-$J$ tensors. The shadow coefficient $\b_{\De,J}$ in \eqref{eq:block_shadowrep0} in the scalar case is given by
\be\label{eq:shadow_coeff_scalar}
\b_{\De,J} = \frac{(-1)^J 2^{J-1}\pi^{\frac{d-2}{2}}\G(\De+2-d)\G(\tfrac{J+\De}{2})^2\G(\tfrac{\De+2-d-J}{2})^2}{\G(\De+\tfrac{2-d}{2})\G(J+\De)\G(\De+2-d-J)}.
\ee

For the three-point structures in \eqref{eq:block_shadowrep0}, the notation $|\<\ldots\>|$ means that all the $x_{ij}$'s in the denominator should come with absolute values (see appendix \ref{app:conventions} for definitions). In our case, the correct interpretation of the absolute-valued structures is in terms of a double commutator,
\be
&\<\O|[\f_4(x_4),\f_3(x_3)][\f_1(x_1),\f_2(x_2)]|\O\>^{s}_{\De,J} \nn \\
&= \frac{-2(2\sin^2(\pi\tfrac{\De-J-2\De_\f}{2}))}{\b_{\De,J}} \int_{4>5>3}d^dx_5 |\<\f_1\f_2\cO^{\mu_1\cdots\mu_J}(x_5)\>| |\<\tl{\cO}^{\dag}_{\mu_1\cdots\mu_J}(x_5)\f_3\f_4\>|,
\ee
where the left hand side denotes the contribution from the $G^s_{\De,J}$ block to the double commutator. Each commutator gives $2i\sin(\pi\tfrac{\De-J-2\De_\f}{2})$ from the difference of two phase factors, so we have an additional $-2(2\sin^2(\pi\tfrac{\De-J-2\De_\f}{2}))$ factor, which is accounted for on the right-hand side.

In our calculation, we will use absolute-valued structures as convenient notation in intermediate steps, and replace them with commutators at the end as appropriate. For example, for the structure $|\<\f_1(x_1)\f_2(x_2)\cO(x_5)\>|$, which has causality relation $5\approx (1>2)$, we have the following identity:
\be\label{eq:abs_struct_relationtoWightman}
\<0|\cO(x_5)[\f_1(x_1),\f_2(x_2)]|0\> = 2i\sin(\pi\tfrac{\De-J-2\De_\f}{2})|\<\f_1(x_1)\f_2(x_2)\cO(x_5)\>|,\quad 5\approx (1>2),
\ee
where the left-hand side is a commutator of the standard Wightman structure. For the other structure $|\<\tl{\cO}^{\dag}(x_5)\f_3\f_4\>|$, this identity is not well-defined due to the additional phase factors from $x_{45}^2,x_{35}^2$. However, in the final formula for the heavy action, we will be able to apply the above identity to both structures, enabling us to remove absolute values and write the final formula in terms of standard Wightman structures.

We can now plug the Lorentzian shadow representation into our functional. In the functional \eqref{eq:scalarfunctional_plus_final}, the external points of the four-point function are $1^{+},2^{+},3,4$ with the condition $4>3>0>1>2$, so the causality configuration agrees with the Lorentzian shadow representation \eqref{eq:block_shadowrep0}. See also figure \ref{fig:causality_config_2}. Therefore, using \eqref{eq:block_shadowrep0}, the action of $\Psi^{+}_{k,\nu}$ on a conformal block can be written
\be
\Psi^{+}_{k,\nu}[G^s_{\De,J}] =& \frac{-8(2\sin^2(\pi\tfrac{\De-J-2\De_\f}{2}))A_{k,\nu}}{\b_{\De,J}} \int_{4>5>3>0>1>2} \frac{d^dx_1 d^d x_2 d^d x_3 d^d x_4 d^d x_5 d^d x_0 D^{d-2}z_0}{\vol(\tl \SO(d,2))}  \nn \\
&\x \<0|\f_4\wL[\cO](x_0,z_0)\f_{2^{+}}|0\>^{-1}\<\tl \cP_{\de_1}(x_{1^{+}0}) \tl \cP_{\de_3}(x_{30}) \cP_{\de}(z_0)\> \nn \\
&\x \de(x_{1^{+}0}^2)\de(x_{30}^2)\th(x_{1^{+}0})\th(x_{30})|\<\f_{1^{+}}\f_{2^{+}}\cO^{\mu_1\cdots\mu_J}(x_5)\>| |\<\tl{\cO}^{\dag}_{\mu_1\cdots\mu_J}(x_5)\f_3\f_4\>|.
\ee
For brevity, we have used the short-hand notation $\f_i=\f_i(x_i), \f_{i^{+}}=\f_i(x_i^{+})$.

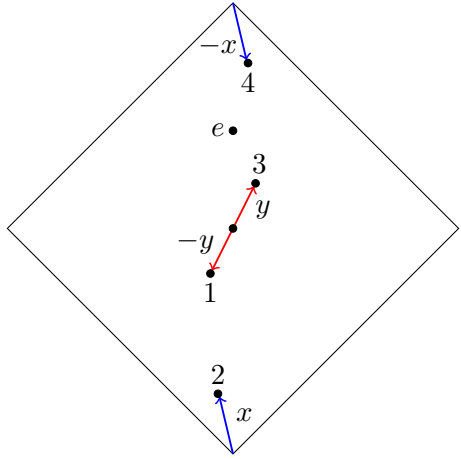
\begin{figure}[t!]
	\centering
	\begin{tikzpicture}	
	\draw[] (-3,0) -- (0,3) -- (3,0) -- (0,-3) -- cycle;
	\draw[->, line width=0.25mm, draw=red] (0+0.01,0+0.02) -- (0.3-0.02,0.6-0.04);
	\draw[->, line width=0.25mm, draw=red] (0-0.01,0-0.02) -- (-0.3+0.02,-0.6+0.04);
	\draw[->, line width=0.25mm, draw=blue] (0,3) -- (0.2-0.02,3-0.8+0.04);
	\draw[->, line width=0.25mm, draw=blue] (0,-3) -- (-0.2+0.02,-3+0.8-0.04);
	
	\draw[fill=black] (0,0) circle (0.05);
	\draw[fill=black] (0,1.3) circle (0.05);
	\node[above] at (-0.2,1.1) {$e$};
	\draw[fill=black] (0.3,0.6) circle (0.05);
	\node[above] at (0.3+0.05,0.6) {$3$};
	\node[above] at (0.4,0) {$y$};
	\draw[fill=black] (-0.3,-0.6) circle (0.05);
	\node[below] at (-0.3,-0.6) {$1$};
	\node[below] at (-0.5,0.1) {$-y$};
	\draw[fill=black] (0+0.2,3-0.8) circle (0.05);
	\node[below] at (0+0.2,3-0.8) {$4$};
	\node[below] at (-0.2,2.7) {$-x$};
	\draw[fill=black] (0-0.2,-3+0.8) circle (0.05);
	\node[above] at (0-0.2,-3+0.8) {$2$};
	\node[above] at (0.15,-2.7) {$x$};
	\end{tikzpicture}
	\caption{Our gauge fixing condition is $x_3=-x_1=y, x_4=-x_2=\frac{x}{-x^2}, x_5=e$. The causality constraint is $4>5>3>0>1>2$.}
	\label{fig:gaugefixing}
\end{figure}

Now we use conformal symmetry to gauge fix the integral. We will choose the gauge fixing in \cite{Caron-Huot:2021enk} and fix $x_1+x_3$ to the origin, $x_{2^{+}}+x_{4^-}$ to spatial infinity, and $x_5$ to the unit time vector $e$. More precisely, we choose (see figure \ref{fig:gaugefixing})
\be\label{eq:gaugefixing}
x_3=-x_1=y,\quad x_4=-x_2=\frac{x}{-x^2},\quad x_5=e,\quad x,y>0.
\ee
For this gauge fixing, the stabilizer group is $\SO(d-1)$, and the Faddeev-Popov determinant is \cite{Caron-Huot:2021enk}
\be\label{eq:FP_Det}
2^d(1+2x\.y+x^2y^2)(1-2x\.y+x^2y^2)(1-x^2y^2)^{d-2}.
\ee
After the gauge fixing, we obtain
\be\label{eq:scalarfunctional_plus_gaugefixed}
\Psi^{+}_{k,\nu}[G^s_{\De,J}] =& \frac{-8(2\sin^2(\pi\tfrac{\De-J-2\De_\f}{2}))A_{k,\nu}}{\b_{\De,J}\vol(\SO(d-1))}\int \frac{d^dx d^dy d^dx_0D^{d-2}z_0}{(-x^2)^{2d}}\nn \\
&\x 2^d(1+2x\.y+x^2y^2)(1-2x\.y+x^2y^2)(1-x^2y^2)^{d-2} \nn \\
&\x \<0|\f_4\wL[\cO](x_0,z_0)\f_{2^{+}}|0\>^{-1}\<\tl \cP_{\de_1}(x_{1^{+}0}) \tl \cP_{\de_3}(x_{30}) \cP_{\de}(z_0)\> \nn \\
&\x \left.\de(x_{1^{+}0}^2)\de(x_{30}^2)\th(x_{1^{+}0})\th(x_{30})|\<\f_{1^{+}}\f_{2^{+}}\cO^{\mu_1\cdots\mu_J}(x_5)\>| |\<\tl{\cO}^{\dag}_{\mu_1\cdots\mu_J}(x_5)\f_3\f_4\>|\right|_{\mathrm{gauge-fixed}},
\ee
where the last two lines should be evaluated in the gauge-fixed configuration \eqref{eq:gaugefixing}.

The integral over $z_0$ can also be fixed using $\SO(d-1)$ invariance. After integrating over $x,y,x_0$, the remaining vectors in the integrand of \eqref{eq:scalarfunctional_plus_gaugefixed} are $z_0$ and $e$. Homogeneity of $z_0$ then implies the $z_0$-integral must be of the form
\be
\int D^{d-2}z_0\ (-2z_0\.e)^{2-d} = \frac{\pi^{\frac{d-2}{2}}\G(\tfrac{d-2}{2})}{\G(d-2)}.
\ee
Thus, we can eliminate the $z_0$-integral by setting $z_0$ to be a fixed null vector $z_0^{*}$, and then introduce a factor 
\be
(-2z_0^{*}\.e)^{d-2}\frac{\pi^{\frac{d-2}{2}}\G(\tfrac{d-2}{2})}{\G(d-2)}.
\ee
Let us introduce lightcone coordinate $x=(u,v,\vec x_{\perp})$, where $x^2=-uv+\vec x_\perp^2$, and choose $z_0^{*}=(1,0,\vec 0)$. Then \eqref{eq:scalarfunctional_plus_gaugefixed} becomes
\be\label{eq:scalarfunctional_plus_gaugefixed_nozintegral}
\Psi^{+}_{k,\nu}[G^s_{\De,J}] =& \frac{-8(2\sin^2(\pi\tfrac{\De-J-2\De_\f}{2}))A_{k,\nu}}{\b_{\De,J}\vol(\SO(d-1))}\frac{\pi^{\frac{d-2}{2}}\G(\tfrac{d-2}{2})}{\G(d-2)}\int \frac{d^dx d^dy d^dx_0}{(-x^2)^{2d}}\nn \\
&\x 2^d(1+2x\.y+x^2y^2)(1-2x\.y+x^2y^2)(1-x^2y^2)^{d-2} \nn \\
&\x \<0|\f_4\wL[\cO](x_0,z_0^{*})\f_{2^{+}})|0\>^{-1}\<\tl \cP_{\de_1}(x_{1^{+}0}) \tl \cP_{\de_3}(x_{30}) \cP_{\de}(z_0^{*})\> \nn \\
&\x \left.\de(x_{1^{+}0}^2)\de(x_{30}^2)\th(x_{1^{+}0})\th(x_{30})|\<\f_{1^{+}}\f_{2^{+}}\cO^{\mu_1\cdots\mu_J}(x_5)\>| |\<\tl{\cO}^{\dag}_{\mu_1\cdots\mu_J}(x_5)\f_3\f_4\>|\right|_{\mathrm{gauge-fixed}}.
\ee

\subsection{Saddle point analysis}\label{sec:saddle_point_analysis}
We now study the action $\Psi^{+}_{k,\nu}[G^s_{\De,J}]$ in the bulk-point limit, where we take both $\nu$ and $\De$ to be large (and $\De\sim m$). To study the large $\nu$, large $\De$ limit of \eqref{eq:scalarfunctional_plus_gaugefixed_nozintegral}, we will consider the factors that depend exponentially on $\nu$ and $\De$ and look for a saddle point. The important factors in \eqref{eq:scalarfunctional_plus_gaugefixed_nozintegral} are given by
\be\label{eq:bulkpoint_integrand_factor}
&\left.\p{\frac{(2z_0\.x_{40}x_{20}^2-2z_0\.x_{20}x_{40}^2)^2(2x_{10}\.x_{30})}{(-x^2_{24})(-x^2_{40})(-x^2_{20})(2z_0\.x_{10})(-2z_0\.x_{30})}}^{\frac{i\nu}{2}}\p{\frac{(-x_{35}^2)(-x_{45}^2)(-x_{12}^2)}{(-x_{15}^2)(-x_{25}^2)(-x_{34}^2)}}^{\frac{m}{2}}\right|_{\mathrm{gauge-fixed},z_0=z_0^*} \nn \\
&=e^{f_{\nu,m}(x,y,x_0)},
\ee
where the function $f_{\nu,m}(x,y,x_0)$ is given by
\be
f_{\nu,m}(x,y,x_0) &= \frac{i\nu}{2}\log\p{\frac{4(z_0^{*}\.x(1+x^2x_0^2)-2z_0^{*}\.x_0 x^2x\.x_0)^2(-y^2)}{(-x^2)(1-2x\.x_0+x^2x_0^2)(1+2x\.x_0+x^2x_0^2)(z_0^{*}\.(y-x_0))(z_0^{*}\.(y+x_0))}} \nn \\
&+ \frac{m}{2}\log\p{\frac{(e-x)^2(e-y)^2}{(e+x)^2(e+y)^2}}.
\ee
The integral we have to consider in the bulk-point limit then takes the form
\be\label{eq:generalint_bulkpointlimit}
\int d^dx d^dy d^dx_0 \de((y+x_0)^2)\de((y-x_0)^2)\th(y-x_0)\th(y+x_0) e^{f_{\nu,m}(x,y,x_0)} g(x,y,x_0),
\ee
where $g(x,y,x_0)$ are all the factors that do not depend exponentially on $\nu$ and $m$ in the integrand of \eqref{eq:scalarfunctional_plus_gaugefixed_nozintegral}.

\begin{figure}[t!]
	\centering
	\begin{tikzpicture}	
		\draw[fill=black] (0.3,1.2) circle (0.05);
		\node[above] at (0.3,1.2) {$3$};	
		\draw[fill=black] (-0.3,-1.2) circle (0.05);
		\node[below] at (-0.3,-1.2) {$1$};	
		\draw[dashed] (0.3,1.2) -- (0.3-2,1.2-2);
		\draw[dashed] (0.3,1.2) -- (0.3+1.5,1.2-1.5);
		\draw[dashed] (-0.3,-1.2) -- (-0.3-1.5,-1.2+1.5);
		\draw[dashed] (-0.3,-1.2) -- (-0.3+2,-1.2+2);
		\draw[thick, draw=blue] (1.2,0.3) to[out=-135, in=-10, distance=0.5cm] (-1.2,-0.3);
		\draw[dotted, thick, draw=blue] (1.2,0.3) to[out=165, in=50, distance=0.5cm] (-1.2,-0.3);
		\node[below] at (-0.1,-0.2) {$S^{d-2}$};
		
		\draw[->] (1.2+2,0.3) -- (1.2+2+0.8,0.3+0.2);
		\node[above] at (1.2+2+0.8+0.3,0.3) {$\tl x_0$};
		\draw[->] (1.2+2,0.3) -- (1.2+2-0.2,0.3+0.8);
		\node[above] at (1.2+2-0.2,0.3+0.8) {$\tl t_0$};
	\end{tikzpicture}
	\caption{The two delta functions $\de(x_{10}^2)\de(x_{30}^2)$ force $x_0$ to be on an $S^{d-2}$ (the blue curves). The directions of the variables $\tl{t_0},\tl{x}_0$ defined in \eqref{eq:t0x0tilde_definition} are also indicated.}
	\label{fig:13lightcone}
\end{figure}
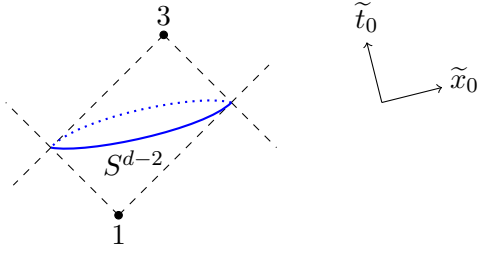

To find the location of the saddle, it is convenient to consider lightcone coordinates
\be
x=(u_x,v_x,\vec x_\perp),\quad y=(u_y,v_y,\vec y_\perp),\quad x_0=(u,v,\vec y_0).
\ee
Let us also define
\be\label{eq:t0x0tilde_definition}
\tl t_0=\frac{1}{2}\p{\frac{u}{u_y}+\frac{v}{v_y}-2\frac{\vec y_0\.\vec y_\perp}{u_yv_y}},\quad \tl x_0=\frac{1}{2}\p{\frac{u}{u_y}-\frac{v}{v_y}}.
\ee
In our functional, there are two delta functions $\de(x_{1^{+}0}^2)\de(x_{30}^2)$ in the integrand due to the light transforms. These delta functions restrict $x_0$ to lie on the intersection of the past lightcone of $x_3$ and the future lightcone of $x_1$. So, the $x_0$-integral is forced to be on an $S^{d-2}$, as shown by figure \ref{fig:13lightcone}. In terms of the variables defined above, the delta functions will localize $\tl t_0$ and $|\vec y_0|$ to be
\be
\tl t_0 =0,\quad |\vec y_0|^2 = \frac{u_yv_y(u_yv_y(1-\tl x_0^2)-\vec y_\perp^2)}{u_yv_y-(\vec n_0\.\vec y_\perp)^2},
\ee
where $\vec n_0=\vec y_0/|\vec y_0|$ is the unit vector in the $\vec y_0$ direction. The remaining integral over $S^{d-2}$ then becomes an integral over $\tl x_0\in [-(1-\tfrac{\vec y_\perp^2}{u_yv_y})^{\frac{1}{2}},(1-\tfrac{\vec y_\perp^2}{u_yv_y})^{\frac{1}{2}}]$ and $\vec n_0\in S^{d-3}$.

After going to lightcone coordinates and removing the delta functions, \eqref{eq:generalint_bulkpointlimit} becomes
\be\label{eq:generalint_bulkpointlimit_nodelta}
&\frac{(u_yv_y)^{\frac{d-2}{2}}}{32}\int d\tl x_0 du_x dv_x du_y dv_y d\O_{\vec n_0} d^{d-2}\vec x_\perp d^{d-2}\vec y_\perp (u_yv_y-(\vec n_0\.y_\perp)^2)^{\frac{2-d}{2}}(u_yv_y(1-\tl x_0^2)-\vec y_\perp^2)^{\frac{d-4}{2}}\nn \\
&\x e^{f_{\nu,m}(x,y,x_0)} g(x,y,x_0).
\ee
The large $\nu$ and large $m$ limit localizes all the variables in the integral, except for $\vec n_0$. In fact, the $f_{\nu,m}$ function leads to sixteen different saddles at
\be
x_0=(0,0,\vec y_0),\quad x=\p{-i\tfrac{m\pm\sqrt{m^2-\nu^2}}{\nu},i\tfrac{m\pm\sqrt{m^2-\nu^2}}{\nu},\vec 0}&,\ y=\p{i\tfrac{m\pm\sqrt{m^2-\nu^2}}{\nu},-i\tfrac{m\pm\sqrt{m^2-\nu^2}}{\nu},\vec 0},
\ee
where $|\vec y_0|^2=-y^2$. Each saddle corresponds to a locus $S^{d-3}$ parametrized by $\vec n_0$. To find the correct saddle approximation for the integral \eqref{eq:generalint_bulkpointlimit_nodelta}, we need to deform the integration contour into steepest descendant flows and see which saddle locus the contour goes through (see \cite{Witten:2010cx} for a pedagogical introduction). The analysis is essentially identical to \cite{Caron-Huot:2021enk}, which also studies saddle point of $x,y$ in the bulk-point limit. The result is that the dominant saddle depends on the relative size of $\nu$ and $m$. The case we will be particularly interested in is the region related to the flat space functional, where one considers $\nu\in[0,\De_{\mathrm{gap}}]$ and $m>\De_{\mathrm{gap}}$. In this case ($\nu<m$), the saddle locus that dominates the integral is given by
\be\label{eq:bulkpoint_saddle}
x_0=(0,0,\vec y_0),\quad x=\p{-i\tfrac{m-\sqrt{m^2-\nu^2}}{\nu},i\tfrac{m-\sqrt{m^2-\nu^2}}{\nu},\vec 0}&,\ y=\p{i\tfrac{m-\sqrt{m^2-\nu^2}}{\nu},-i\tfrac{m-\sqrt{m^2-\nu^2}}{\nu},\vec 0}.
\ee

Expanding the function $f_{\nu,m}$ around the dominant saddle \eqref{eq:bulkpoint_saddle}, we find
\be
f_{\nu,m} = -\pi\nu + i\nu \log2 - \frac{1}{2}\bpm \tl x_0 & u_x & v_x & u_y & v_y & \vec x_\perp & \vec y_\perp \epm M_{\nu,m,\vec y_0} \bpm \tl x_0 \\ u_x \\ v_x \\ u_y \\ v_y \\ \vec x_\perp \\ \vec y_\perp \epm + \ldots,
\ee
where $\ldots$ are higher-order terms in the expansion. The Hessian $M_{\nu,m,\vec y_0}$ has determinant
\be
\mathrm{Det} M_{\nu,m,\vec y_0} = -\frac{i\nu^{6d+1}(m^2-\nu^2)^2(m-\sqrt{m^2-\nu^2})^{-4d}}{16m^4}.
\ee
Therefore, the large $\nu$, large $m$ limit of the integral \eqref{eq:generalint_bulkpointlimit} can be written as
\be
&\lim_{\substack{\nu,m\gg 1 \\ \nu<m}}\int d^dx d^dy d^dx_0 \de((y+x_0)^2)\de((y-x_0)^2)\th(y-x_0)\th(y+x_0) e^{f_{\nu,m}(x,y,x_0)} g(x,y,x_0) \nn \\
=&\frac{\nu^{4-d}(m-\sqrt{m^2-\nu^2})^{d-4}}{32}\sqrt{\frac{(2\pi)^{2d+1}}{\mathrm{Det} M_{\nu,m,\vec y_0}}}e^{-\pi\nu}2^{i\nu} \int d\O_{\vec n_0} \left.g(x,y,x_0)\right|_{\mathrm{saddle}},
\ee
where saddle stands for the configuration \eqref{eq:bulkpoint_saddle}. Furthermore, the function $g(x,y,x_0)$ evaluated at the saddle does not include any $(d-2)$-dimensional vectors other than $\vec n_0$, and hence the remaining integral over the $S^{d-3}$ locus must be trivial. The final expression of the integral is then given by
\be\label{eq:generalint_bulkpointlimit_result}
&\lim_{\substack{\nu,m\gg 1 \\ \nu<m}}\int d^dx d^dy d^dx_0 \de((y+x_0)^2)\de((y-x_0)^2)\th(y-x_0)\th(y+x_0) e^{f_{\nu,m}(x,y,x_0)} g(x,y,x_0) \nn \\
=&\frac{e^{i\frac{\pi}{4}} (2\pi)^{d+\frac{1}{2}} m^2\nu^{-4d+\frac{7}{2}}(m-\sqrt{m^2-\nu^2})^{3d-4}}{8(m^2-\nu^2)}\vol(S^{d-3}) \left.e^{f_{\nu,m}(x,y,x_0)}g(x,y,x_0)\right|_{\mathrm{saddle}},
\ee
where we have plugged in the Hessian determinant and used the fact that $e^{-\pi\nu}2^{i\nu}$ comes from evaluating $e^{f_{\nu,m}(x,y,x_0)}$ at the saddle.

Now, we can compare \eqref{eq:generalint_bulkpointlimit_result} and the functional \eqref{eq:scalarfunctional_plus_gaugefixed_nozintegral} and find the expression for $\Psi^{+}_{k,\nu}[G^s_{\De,J}]$ in the bulk-point limit. From the above saddle analysis, we see that the integral in the bulk-point limit can be obtained by simply evaluating the integrand at the saddle and multiplying by the additional factors coming from the Jacobian and Hessian. This implies that in the bulk-point limit, the calculation of the heavy action of our functional can be simplified to evaluating conformally-invariant structures in the saddle configuration \eqref{eq:bulkpoint_saddle}, which leads to
\be\label{eq:scalarfunctional_bulkpoint_resultwithshadowcoeff}
&\lim_{\substack{\nu,m\gg 1 \\ \nu<m}}\frac{\Psi^{+}_{k,\nu}[G^s_{\De,J}]}{-2(2\sin^2(\pi\tfrac{\De-J-2\De_\f}{2}))} \nn \\
&= A_{k,\nu}\frac{2^{\frac{5}{2}+3d}\pi^{\frac{3d-1}{2}}e^{i\frac{\pi}{4}}m^{d+2}\nu^{\frac{7}{2}-4d}(m^2-\nu^2)^{\frac{d-2}{2}}(m-\sqrt{m^2-\nu^2})^{d-4}}{\b_{\De,J}\G\p{\frac{d-2}{2}}\vol(\SO(d-2))} \nn \\
&\x \<0|\f_4\wL[\cO](x_0,z_0^{*})\f_{2^{+}}|0\>^{-1}\<\tl \cP_{\de_1}(x_{1^{+}0}) \tl \cP_{\de_3}(x_{30}) \cP_{\de}(z_0^{*})\>\nn \\
&\x \left.|\<\f_{1^{+}}\f_{2^{+}}\cO^{\mu_1\cdots\mu_J}(x_5)\>| |\<\tl{\cO}^{\dag}_{\mu_1\cdots\mu_J}(x_5)\f_3\f_4\>|\right|_{\mathrm{saddle}}.
\ee

The saddle calculation done above involves computing the determinant of a $(2d+1)\x (2d+1)$ Hessian matrix. An alternative derivation of the above formula that can avoid this technically involved calculation is by first studying the large $\nu$ limit of the functional, and then use the result of the large $m$ saddle analysis that was already done in \cite{Caron-Huot:2021enk}. We describe this calculation in appendix \ref{app:derivation_largenu}.

\subsubsection{The shadow coefficient}\label{sec:shadow_coeff}
Finally, the remaining task is to study the shadow coefficient $\b_{\De,J}$ in the large $\De$ limit. For the scalar case, its explicit expression is known and given by \eqref{eq:shadow_coeff_scalar}. For the spinning case, the shadow transform will involve mixing of different tensor structures and therefore the shadow coefficients become a matrix. In principle it can be computed using e.g., weight-shifting operators \cite{Karateev:2017jgd}. However, as we discuss below, it turns out that the shadow transform at large $\De$ also gets localized to a saddle. Hence, the shadow transform at large $\De$ becomes algebraic and the coefficients can again be computed by evaluating structures at the saddle configuration.

Let us demonstrate this idea by studying the shadow transform in the scalar case. We will derive a formula for $\b_{\De,J}$ that reproduces the known answer \eqref{eq:shadow_coeff_scalar}. However, the advantage of this formula is that its generalization to the spinning case is almost trivial.

Considering the OPE limit of \eqref{eq:block_shadowrep0}, we see that the shadow coefficient $\b_{\De,J}$ should satisfy
\be\label{eq:shadowcoeff_definition}
&\frac{1}{\b_{\De,J}} \int_{4>5>3} d^d x_5 \<\cO^{\dag}_{\nu_1\cdots\nu_J}(x_6)\cO^{\mu_1\cdots\mu_J}(x_5)\>|\<\tl{\cO}^{\dag}_{\mu_1\cdots\mu_J}(x_5)\f_3\f_4\>| \nn \\
&= |\<\cO^{\dag}_{\nu_1\cdots\nu_J}(x_6)\f_3\f_4\>|.
\ee
Let us choose the external points $x_3,x_4,x_6$ to be
\be
x_3=\left.y\right|_{\mathrm{saddle}},\quad x_4=\left.\frac{x}{-x^2}\right|_{\mathrm{saddle}},\quad x_6=(-e)^{+},
\ee
where $x,y$ are evaluated at the saddle configuration \eqref{eq:bulkpoint_saddle}. With this choice of external points, we find that at large $\De$, the integrand in \eqref{eq:shadowcoeff_definition} has a saddle point at exactly $x_5=e$, which agrees with the saddle configuration of the original integral of the functional $\Psi^{+}_{k,\nu}$. After computing the determinant factor coming from the Gaussian integral, we obtain that in the large $\De$ limit, \eqref{eq:shadowcoeff_definition} becomes
\be\label{eq:shadowcoeff_largeDelta_v0}
&\lim_{\De\gg 1}\frac{2^d\pi^{\frac{d}{2}}\De^{-\frac{d}{2}}}{\b_{\De,J}} \left.\<\cO^{\dag}_{\nu_1\cdots\nu_J}(x_6)\cO^{\mu_1\cdots\mu_J}(x_5)\>|\<\tl{\cO}^{\dag}_{\mu_1\cdots\mu_J}(x_5)\f_3\f_4\>| \right|_{\substack{\mathrm{saddle} \\ x_6=(-e)^{+}}} \nn \\
&=  \left.|\<\cO^{\dag}_{\nu_1\cdots\nu_J}(x_6)\f_3\f_4\>|\right|_{\substack{\mathrm{saddle} \\ x_6=(-e)^{+}}}.
\ee

To recap, we see that at large $\De$, the shadow transform gets localized to a saddle point. Furthermore, with a good choice of external points $x_6=(-e)^{+}$, we can make the saddle point location agree with the saddle configuration for the functional \eqref{eq:bulkpoint_saddle}. We have also checked that \eqref{eq:shadowcoeff_largeDelta_v0} is consistent with the known scalar result \eqref{eq:shadow_coeff_scalar}.

To simplify the formula of the functional in the bulk-point limit given by \eqref{eq:scalarfunctional_bulkpoint_resultwithshadowcoeff}, we need the combination $\frac{1}{\b_{\De,J}}|\<\f_{1^{+}}\f_{2^{+}}\cO^{\mu_1\cdots\mu_J}(x_5)\>||\<\tl{\cO}^{\dag}_{\mu_1\cdots\mu_J}(x_5)\f_3\f_4\>|$, which is different from the structure appearing in \eqref{eq:shadowcoeff_largeDelta_v0}. To get the correct structure, let us first note that the two-point structure at the saddle $\<\cO^{\dag\nu_1\cdots\nu_J}((-e)^{+})\cO^{\mu_1\cdots\mu_J}(e)\>$ can be viewed as an invertible matrix for the spin-$J$ representation of the Lorentz group.\footnote{More generally, if $\cO$ has representation $\r$, it should be a map from $\r^{*}$ to $\r^{\dag}=(\r^{*})^{R}$.} Therefore, we can define an inverse ``$r$-tensor" that satisfies
\be\label{eq:rtensor_def_symmetric_traceless}
r_{\De,J;\r_1\cdots\r_J}{}^{\nu_1\cdots\nu_J}\<\cO^{\dag}_{\nu_1\cdots\nu_J}((-e)^{+})\cO^{\mu_1\cdots\mu_J}(e)\> = \de_{\{\r_1}{}^{\{\mu_1}\cdots\de_{\r_J\}}{}^{\mu_J\}} - \mathrm{traces}.
\ee 
Here, $\{\mu_1\cdots\mu_J\}$ means we symmetrize the indices, and ``$-\mathrm{traces}$" means we subtract terms proportional to $\de^{\mu_i\mu_j}$ and $\de_{\r_i\r_j}$ to make the result traceless. Concretely, in this case $r_{\De,J}$ is given by a reflection in the time direction,
\be
r_{\De,J;\r_1\cdots\r_J}{}^{\nu_1\cdots\nu_J} = (-1)^J 2^{-J+2\De}\p{\de_{\{\r_1}{}^{\{\nu_1}+2e_{\{\r_1}e^{\{\nu_1}}\cdots\p{\eta_{\r_J\}}{}^{\nu_J\}}+2e_{\r_J\}}e^{\nu_J\}}} - \mathrm{traces}.
\ee

By contracting both sides of \eqref{eq:shadowcoeff_largeDelta_v0} with $|\<\f_{1^{+}}\f_{2^{+}}\cO^{\r_1\cdots\r_J}(x_5)\>|r_{\De,J;\r_1\cdots\r_J}{}^{\nu_1\cdots\nu_J}$, we obtain
\be\label{eq:shadowcoeff_largeDelta_v1}
&\lim_{\De\gg 1}\frac{1}{\b_{\De,J}} \left.|\<\f_{1^{+}}\f_{2^{+}}\cO^{\mu_1\cdots\mu_J}(x_5)\>| |\<\tl{\cO}^{\dag}_{\mu_1\cdots\mu_J}(x_5)\f_3\f_4\>| \right|_{\substack{\mathrm{saddle} \\ x_6=(-e)^{+}}} \nn \\
&= 2^{-d}\pi^{-\frac{d}{2}}\De^{\frac{d}{2}} \left.|\<\f_{1^{+}}\f_{2^{+}}\cO^{\mu_1\cdots\mu_J}(x_5)\>|r_{\De,J;\mu_1\cdots\mu_J}{}^{\nu_1\cdots\nu_J}|\<\cO^{\dag}_{\nu_1\cdots\nu_J}(x_6)\f_3\f_4\>|\right|_{\substack{\mathrm{saddle} \\ x_6=(-e)^{+}}}.
\ee
The left hand side of the above equation now agrees with the structure in the formula \eqref{eq:scalarfunctional_bulkpoint_resultwithshadowcoeff}. We can then plug this equation into \eqref{eq:scalarfunctional_bulkpoint_resultwithshadowcoeff} and find a formula without the shadow coefficient. We write down this final formula in the next subsection.

Finally, let us discuss a natural formula for the $r_{\De,\r}$-tensor that will make it easier for us to generalize to the spinning case. It also makes it clear that the tensor depends on a choice of two-point structure convention. For a general representation $\r$, the corresponding $r$-tensor is defined as
\be
r_{\De,\r;a'}{}^{\bar b}\<\cO^{\dag}_{\bar b}((-e)^{+})\cO^{a}(e)\> = \de^{a}{}_{a'},
\ee
where the operator $\cO$ has quantum numbers $(\De,\r)$. Here, $a,a'$ are the indices of the representations $\r,\r^{*}$, and $\bar b$ labels the indices of the reflected representation $\r^{R}$ and its dual $\r^{\dag}$. On the right-hand side, $\de^{a}{}_{a'}$ is the identity matrix of the $\r$ representation (i.e., $\de^{a}{}_{a'}T^{a'}=T^{a}$ for any tensor $T$ with representation $\r$).

Motivated by symmetry, we can write down an ansatz for the $r_{\De,\r}$-tensor using the two-point function of the shadow operator $\tl \cO$,
\be
r_{\De,\r;a'}{}^{\bar b} = C_{\De,\r}\<\tl{\cO}^{\dag}_{a'}(e)\tl{\cO}^{\bar b}((-e)^{+})\>.
\ee
Plugging this ansatz into the definition of $r_{\De,\r}$ and taking the trace, we obtain
\be\label{eq:Rcoeff_eq}
C_{\De,\r} \<\tl{\cO}^{\dag}_{a}(e)\tl{\cO}^{\bar b}((-e)^{+})\>\<\cO^{\dag}_{\bar b}((-e)^{+})\cO^{a}(e)\> =  \de^{a}{}_{a}= \mathrm{dim}(\r).
\ee
We see that the unknown coefficient $C_{\De,\r}$ in the ansatz can be expressed in terms of the dimension of the representation and a pairing of the two-point functions. 

We define a natural Euclidean two-point pairing \cite{Karateev:2018oml},\footnote{Compared to the definition in \cite{Karateev:2018oml}, we have absorbed an infinite $\vol(\SO(1,1))$ factor to make the pairing finite.}
\be\label{eq:Euclidean_2pt_pairing}
\frac{\p{\<\tl\cO^{\dag}\tl\cO\>,\<\cO^{\dag}\cO\>}}{\vol(\SO(1,1))}&= \int \frac{d^dx d^dy}{\vol(\SO(d+1,1))} \<\tl{\cO}^{\dag}_{a}(x)\tl{\cO}^{\bar b}(y)\>\<\cO^{\dag}_{\bar b}(y)\cO^{a}(x)\>, \nn \\
&=\frac{1}{2^d \vol(\SO(d))\vol(\SO(1,1))}\<\tl{\cO}^{\dag}_{a}(0)\tl{\cO}^{\bar b}(\oo)\>\<\cO^{\dag}_{\bar b}(\oo)\cO^{a}(0)\>.
\ee
Originally, in \eqref{eq:Rcoeff_eq} we want to compute the pairing between Lorentzian two-point structures. Since the two points $e,(-e)^{+}$ are spacelike separated, the structures can be thought of as Euclidean two-point structures with distance $\sqrt{-(2e)^2}=2$, and we can replace the pairing with a Euclidean two-point pairing \eqref{eq:Euclidean_2pt_pairing}. Using the two-point pairing, one can then show that \eqref{eq:Rcoeff_eq} gives
\be
C_{\De,\r} = \frac{2^d \mathrm{dim}(\r)}{\vol(\SO(d))\p{\<\tl\cO^{\dag}\tl\cO\>,\<\cO^{\dag}\cO\>}},
\ee
and the $r$-tensor can be written as
\be\label{eq:rtensor_naturalformula}
r_{\De,\r;a'}{}^{\bar b} = \frac{2^d \mathrm{dim}(\r)}{\vol(\SO(d))\p{\<\tl\cO^{\dag}\tl\cO\>,\<\cO^{\dag}\cO\>}}\<\tl{\cO}^{\dag}_{a'}(e)\tl{\cO}^{\bar b}((-e)^{+})\>.
\ee

\subsection{Scalar sum rules}\label{sec:scalar_sum_rules}

\begin{figure}[t!]
	\centering
	\begin{tikzpicture}	
	\draw[dashed] (-3,-4) -- (-3,4);
	\draw[dashed] (3,-4) -- (3,4);
	\draw[] (-3,0) -- (0,3) -- (3,0) -- (0,-3) -- cycle;
	
	\draw[fill=black] (0,0) circle (0.05);
	\draw[fill=black] (0,1.5) circle (0.05);
	\node[left] at (0,1.5) {$5$};
	\draw[fill=black] (0,-1.5) circle (0.05);
	\node[left] at (0,-1.5) {$6$};
	\draw[fill=black] (0,1) circle (0.05);
	\node[left] at (0,1) {$3$};
	\draw[fill=black] (0,-1) circle (0.05);
	\node[left] at (0,-1) {$1$};
	\draw[fill=black] (0,2) circle (0.05);
	\node[left] at (0,2) {$4$};
	\draw[fill=black] (0,-2) circle (0.05);
	\node[left] at (0,-2) {$2$};
	
	\draw[fill=black] (0+3,-1.5+3) circle (0.05);
	\node[right] at (0+3,-1.5+3) {$6^+$};
	
	\draw[fill=black] (0+3,-1+3) circle (0.05);
	\node[right] at (0+3,-1+3) {$1^+$};
	
	\draw[fill=black] (0+3,-2+3) circle (0.05);
	\node[right] at (0+3,-2+3) {$2^+$};
	
	\draw[dashed] (0,1) -- (0+1,1-1);
	\draw[dashed] (0,1) -- (0-1,1-1);
	\draw[dashed] (0,-1) -- (0+1,-1+1);
	\draw[dashed] (0,-1) -- (0-1,-1+1);
	
	\draw[thick, draw=red] (1,0) to[out=-150, in=-30, distance=0.5cm] (-1,0);
	\draw[dotted, thick, draw=red] (1,0) to[out=150, in=30, distance=0.5cm] (-1,0);
	
	\draw[fill=red] (0,-0.18) circle (0.05); 
	\draw[fill=red] (0,0.18) circle (0.05);
	\node[below] at (0,-0.18) {$0$};
	
	\end{tikzpicture}
	\caption{The bulk-point limit localizes the spacetime integral to a saddle locus that we call the scattering-crystal. The entire saddle satisfies the causality constraint $4>5>3>0>1>6>2$. The two delta functions restrict $x_0$ to be on an $S^{d-2}$ shown by the red circle in the figure. The saddle locus further restricts $x_0$ to an $S^{d-3}$, which is represented by the two red dots ($0$-sphere). Points $5$ and $6$ are localized at $\pm e$. Points $1,2,3,4$ are restricted to be timelike vectors, although in practice they have purely imaginary spatial components (see \eqref{eq:scattering_crystal}).}
	\label{fig:scattering_crystal}
\end{figure}
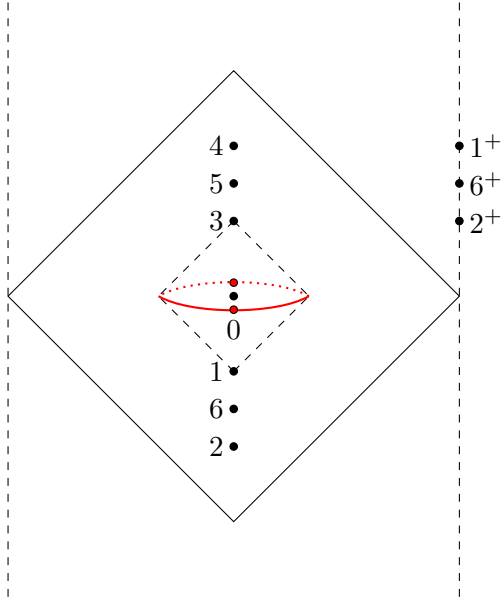

Let us briefly summarize what we have done and write down the final formula for the scalar functional in the bulk-point limit. We started with the commutator of two light-transformed operators and wrote down a functional $\Psi_{k,\nu}$ (\eqref{eq:scalarfunctional_definition}) whose action on any physical four-point function vanishes. The functional integrates the four-point function against a specific kernel over the external points $x_{1,2,3,4}$ and internal variables $x_0,z_0$. Using the Lorentzian shadow representation of conformal blocks, we wrote down the action of the functional on blocks, with an additional internal point $x_5$ from the shadow representation. We then found that in the bulk-point limit ($\nu,m\gg 1$ with $\nu<m$), the integral gets completely localized to a saddle configuration \eqref{eq:bulkpoint_saddle}. Furthermore we removed the shadow coefficient by introducing a final internal point $x_6$. 

In the end, we obtain a formula for the action of the functional on a conformal block in the bulk-point limit. The formula is given by a known coefficient and a product of conformally-invariant structures evaluated at a configuration that fixes all the positions and polarizations, $x_{0,1,2,3,4,5,6},z_0$. In lightcone coordinates, the configuration is given by (see figure \ref{fig:scattering_crystal})
\be\label{eq:scattering_crystal}
x_3&=-x_1=y,\quad x_4=-x_2=\frac{x}{-x^2},\quad x_5=e=(1,1,\vec 0),\quad x_6=-e, \nn \\
x&=\p{-i\tfrac{m-\sqrt{m^2-\nu^2}}{\nu},i\tfrac{m-\sqrt{m^2-\nu^2}}{\nu},\vec 0},\quad y=\p{i\tfrac{m-\sqrt{m^2-\nu^2}}{\nu},-i\tfrac{m-\sqrt{m^2-\nu^2}}{\nu},\vec 0},\nn \\
z_0&=(1,0,\vec 0),\quad x_0=(0,0,\vec y_0),\quad |\vec y_0|^2=\frac{(m-\sqrt{m^2-\nu^2})^2}{\nu^2}.
\ee
As we will show later, evaluating the conformal structures in this configuration reproduces the heavy action of a flat space sum rule and allows us to study the bulk scattering process. This is a generalization of the saddle point that leads to the \textit{spacelike scattering} phenomenon in \cite{Caron-Huot:2021enk}. We will call the configuration \eqref{eq:scattering_crystal} the ``scattering-crystal."

By plugging \eqref{eq:shadowcoeff_largeDelta_v1} in \eqref{eq:scalarfunctional_bulkpoint_resultwithshadowcoeff} to remove the shadow coefficient, we find
\be\label{eq:scalarfunctional_bulkpoint_withrtensor_v0}
&\lim_{\substack{\nu,m\gg 1 \\ \nu<m}}\frac{\Psi^{+}_{k,\nu}[G^s_{\De,J}]}{-2(2\sin^2(\pi\tfrac{\De-J-2\De_\f}{2}))} \nn \\
&=A_{k,\nu}\frac{2^{\frac{5}{2}+2d}\pi^{d-\frac{1}{2}}e^{i\frac{\pi}{4}}m^{\frac{3d}{2}+2}\nu^{\frac{7}{2}-4d}(m^2-\nu^2)^{\frac{d-2}{2}}(m-\sqrt{m^2-\nu^2})^{d-4}}{\G\p{\frac{d-2}{2}}\vol(\SO(d-2))} \nn \\
&\x \<0|\f_4\wL[\cO](x_0,z_0)\f_{2^{+}}|0\>^{-1}\<\tl \cP_{\de_1}(x_{1^{+}0}) \tl \cP_{\de_3}(x_{30}) \cP_{\de}(z_0)\>\nn \\
&\x \left.|\<\f_{1^{+}}\f_{2^{+}}\cO^{\mu_1\cdots\mu_J}(x_5)\>|r_{\De,J;\mu_1\cdots\mu_J}{}^{\nu_1\cdots\nu_J}|\<\cO^{\dag}_{\nu_1\cdots\nu_J}(x_6^{+})\f_3\f_4\>|\right|_{\textrm{scattering-crystal}}.
\ee

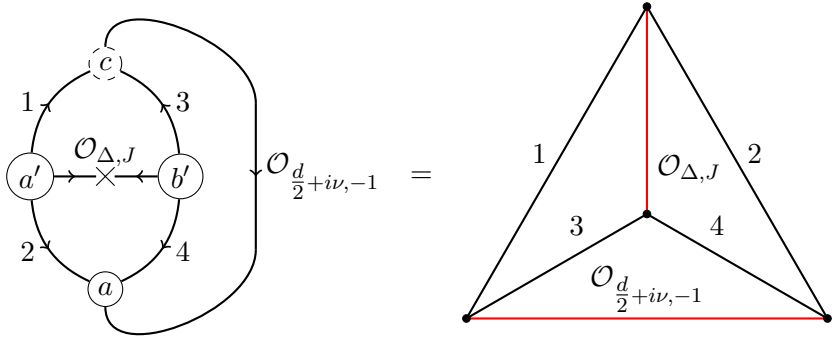
\begin{figure}[t!]
	\centering
	\begin{tikzpicture}
		\node at (-6,0) {$=$};
	\begin{scope}[xshift=-3cm, yshift=-0.5cm]	
		\draw[thick, draw =red] (-2.4,-1.39) -- (2.4,-1.39);
		\draw[thick] (-2.4,-1.39) -- (0,0);
		\draw[thick] (0,0) -- (2.4,-1.39);
		\draw[thick] (-2.4,-1.39) -- (0,2.76);
		\draw[thick] (2.4,-1.39) -- (0,2.76);
		\draw[thick, draw=red] (0,0) -- (0,2.76);
		\draw[fill=black] (2.4,-1.39) circle (0.05);
		\draw[fill=black] (-2.4,-1.39) circle (0.05);
		\draw[fill=black] (0,0) circle (0.05);
		\draw[fill=black] (0,2.76) circle (0.05);
		
		\node[left] at (-1.2,0.8) {$1$}; 
		\node[right] at (1.2,0.8) {$2$}; 
		\node[left] at (-0.7,-0.15) {$3$}; 
		\node[right] at (0.7,-0.15) {$4$}; 
		\node[right] at (0,0.6) {$\cO_{\De,J}$}; 
		\node at (0,-1) {$\cO_{\tfrac{d}{2}+i\nu,-1}$}; 
	\end{scope}
	
	\begin{scope}[xshift=-10.2cm]
		\node [style=threePoint] (12) at (-1, 0) {$a'$};
		\node [style=threePoint] (34) at (1, 0) {$b'$};
		\node [circle, draw=black, dashed, inner sep=2pt] (13) at (0, 1.5) {$c$};
		\node [style=threePoint] (24) at (0, -1.5) {$a$};
		\node [style=cross] (0) at (0, 0) {};
		\node [style=none] (133) at (2,1) {};
		\node [style=none] (244) at (2,-1) {};

		\draw [style=op] (12) to (0);
		\draw [style=op] (34) to (0);
		\draw [style=op, bend left] (12) to (13);
		\draw [style=op, bend right] (12) to (24);
		\draw [style=op, bend right] (34) to (13);
		\draw [style=op, bend left] (34) to (24);
		\draw[style=thick, out=90, in=90] (13.north) to (133.south);
		\draw[style=thick, out=-90, in=-90] (244.north) to (24.south);
		\draw[style=op] (133) to (244);
		
		\node[above] at (0,0) {$\cO_{\De,J}$};
		\node[left] at (-0.8,1) {$1$};
		\node[right] at (0.8,1) {$3$};
		\node[left] at (-0.8,-1) {$2$};
		\node[right] at (0.8,-1) {$4$};
		\node[right] at (2,0) {$\cO_{\tfrac{d}{2}+i\nu,-1}$};
	\end{scope}
	\end{tikzpicture}
	\caption{Our functional is a special $6j$ symbol (or a tetrahedron) where one of the three-point structures is a celestial structure. The cross in the left figure stands for the contraction of the $\<\tl{\cO}^{\dag}\tl{\cO}\>$ tensor. The flat space limit is the limit where the two red legs have large quantum numbers. (More precisely, the two legs have large scaling dimensions.)}
	\label{fig:6jsymbol}
\end{figure}

A nice feature of the scattering-crystal is that both structures with absolute values satisfy the identity \eqref{eq:abs_struct_relationtoWightman} in this configuration. Therefore, we can replace those structures with commutators of standard Wightman structures, and remove the $-2((2\sin^2(\pi\tfrac{\De-J-2\De_\f}{2}))$ factor on the left-hand side. Moreover, by applying the natural formula for the $r$-tensor given by \eqref{eq:rtensor_naturalformula}, we arrive at the final formula for the action of the scalar functional in the bulk-point limit:\footnote{To keep track of $(-1)^J$ factors, we should clarify our convention for $\<0|\cO^{\dag}(x_6)[\f_4,\f_3]|0\>$. Our convention is that it should be the analytic continuation of \eqref{eq:3pt_convention_scalar} with $1\to 3,2\to 4,3\to 6$.}
\be\label{eq:scalarfunctional_bulkpoint_withrtensor}
&\lim_{\substack{\nu,m\gg 1 \\ \nu<m}} \Psi^{+}_{k,\nu}[G^s_{\De,J}] \nn \\
&=A_{k,\nu}\frac{2^{\frac{5}{2}+2d}\pi^{d-\frac{1}{2}}e^{i\frac{\pi}{4}}m^{\frac{3d}{2}+2}\nu^{\frac{7}{2}-4d}(m^2-\nu^2)^{\frac{d-2}{2}}(m-\sqrt{m^2-\nu^2})^{d-4}}{\G\p{\frac{d-2}{2}}\vol(\SO(d-2))}\frac{2^d \mathrm{dim}(\r)}{\vol(\SO(d))} \nn \\
&\x \<0|\f_4\wL[\cO](x_0,z_0)\f_{2^{+}}|0\>^{-1}\<\tl \cP_{\de_1}(x_{1^{+}0}) \tl \cP_{\de_3}(x_{30}) \cP_{\de}(z_0)\>\nn \\
&\x \left.\frac{\<0|[\f_{1^{+}},\f_{2^{+}}]\cO^{\mu_1\cdots\mu_J}(x_5)|0\>\<\tl{\cO}^{\dag}_{\mu_1\cdots\mu_J}(x_5)\tl{\cO}^{\nu_1\cdots\nu_J}(x_6^{+})\>\<0|\cO^{\dag}_{\nu_1\cdots\nu_J}(x_6^{+})[\f_4,\f_3]|0\>}{\p{\<\tl{\cO}^{\dag}\tl\cO\>,\<\cO^\dag \cO\>}}\right|_{\textrm{scattering-crystal}}.
\ee

This formula can be viewed as a kind of $6j$ symbol, or equivalently a tetrahedron formed by gluing conformal three-point structures (see figure \ref{fig:6jsymbol}). Compared to the usual $6j$ symbol \cite{Liu:2018jhs}, in our case one of the three-point structures becomes a celestial structure, which originally comes from setting one of the legs to have $J=-1$. Furthermore, the bulk-point limit, which reproduces the flat space functional, corresponds to setting two of the legs to have large quantum numbers. In this limit, one of them has large and positive dimension $\De \gg 1$ and the other one has dimension $\frac{d}{2}+i\nu$ with $\nu \gg 1$. For usual $6j$ symbols in the classical limit, where all legs have large quantum number, they can be computed as the volume of a tetrahedron \cite{murakami2005volume}. It would be interesting to see if there is a similar argument for the bulk-point limit of our functional.

Before using this formula, we still need to address two more issues. First, note that the original functional $\Psi_{k,\nu}$ in \eqref{eq:scalarfunctional_definition} is given by
\be
\Psi_{k,\nu} = \Psi^{+}_{k,\nu} - (1\leftrightarrow 3).
\ee
Therefore, when we apply $\Psi_{k,\nu}$ to a four-point function of identical operators, which is manifestly symmetric under $(1\leftrightarrow 3)$, we will get a trivial sum rule since its action on each conformal block is simply zero. The second issue is about the convergence in the Regge limit discussed below \eqref{eq:L_commutator}. For external scalars, in order for the light transforms to converge, we need the Regge intercept to satisfy $J_0<-1$, while for non-perturbative CFTs we only know that $J_0\leq 1$ \cite{Hartman:2015lfa, Caron-Huot:2017vep, Kologlu:2019bco}.

It turns out that both issues can be resolved by introducing a ``subtraction factor" to the functional, that is, we consider a more general functional
\be
\Psi_{k,\nu,f}[\cG] = \Psi^{+}_{k,\nu}[f(u',v')\cG] - (1\leftrightarrow 3),
\ee
where $f(u',v')$ is a meromorphic function of the conformally-invariant cross-ratios $u',v'$ defined in \eqref{eq:cross_ratios}. Note that the kernel of $\Psi_{k,\nu,f}$ is again manifestly antisymmetric under $1\leftrightarrow 3$. On the other hand, points $1$ and $3$ are always spacelike in the range of integration of the functional, which comes from the light transforms. Therefore, causality/crossing guarantees that any physical correlator $\cG$ should be symmetric under $1\leftrightarrow 3$ in the integral, and the action of the general functional $\Psi_{k,\nu,f}$ should vanish on physical four-point functions. Additionally, by choosing subtraction factors that are antisymmetric under $1\leftrightarrow 3$, which swaps $u'$ and $v'$, we can get nontrivial sum rules for identical external scalars, meaning its action on each conformal block is nonzero. Furthermore, by restricting to subtraction factors with nice behaviors in the Regge limit, we can make sure the integral of the functional is convergent in the Regge limit.

For the scalar case, a nice choice of subtraction factors with the above properties is\footnote{The $k\equiv 2\text{ mod 4}$ case follows straightforwardly from the discussion in appendix D of \cite{Caron-Huot:2021enk}. The other case can be motivated by taking the large-$\nu$ limit of (D.12) in the same paper, and using the identity
\be
\p{(v'-u')^3 + \frac{3}{d-8}(v'-u')}G_{d-1,\tfrac{2-d}{2}+i\nu} \sim 8G_{d-5,\tfrac{2-d}{2}+i\nu},\quad \nu \to \oo.
\ee}
\be\label{eq:subtracction_scalars}
f^{\mathrm{scalar}}_{k}(u',v') &=
\begin{cases}
 \frac{v'-u'}{(u'v')^{\frac{k+2}{4}}} & k \equiv 2\text{ mod 4} \\
 \frac{(v'-u')^3-(v'-u')}{4(u'v')^{\frac{k+4}{4}}} & k \equiv 0\text{ mod 4}
\end{cases}, \qquad k=2,4,6,\ldots.
\ee
At the scattering-crystal locus, these subtraction factors become
\be\label{eq:subtraction_scalar_atsaddle}
\left.f^{\mathrm{scalar}}_k(u',v')\right|_{\textrm{scattering-crystal}} = -\nu^{2k+2}\frac{(2m^2-\nu^2)}{m^2(m^2-\nu^2)}\frac{1}{(m^2(m^2-\nu^2))^{\frac{k}{2}}}.
\ee
Note that this satisfies $f^{\mathrm{scalar}}_{k}(u',v')=-f^{\mathrm{scalar}}_{k}(v',u')$. In the Regge limit, the subtraction factor increases the effective Regge spin of the kernel by $k+1$. (One can see this from the large-$m$ scaling of \eqref{eq:subtraction_scalar_atsaddle}.) Therefore, the condition from convergence in the Regge limit becomes $J_0<k$, which is indeed true for any $k\geq 2$. 

Using the subtraction factors \eqref{eq:subtracction_scalars}, we can then write down nontrivial sum rules for identical scalars,
\be
\Psi_{k,\nu,f^{\mathrm{scalar}}_{k}}[\cG_{\f\f\f\f}] = 0.
\ee
Its action on a single $s$-channel block is given by
\be
\Psi_{k,\nu,f^{\mathrm{scalar}}_{k}}[G^s_{\De,J}]= 2\Psi^{+}_{k,\nu}[f^{\mathrm{scalar}}_k(u',v')G^s_{\De,J}].
\ee
In the bulk-point limit, since the subtraction factors do not have any $\nu$ and $m$ dependence, the formula \eqref{eq:scalarfunctional_bulkpoint_withrtensor} gets modified to
\be
&\lim_{\substack{\nu,m\gg 1 \\ \nu<m}}\Psi_{k,\nu,f^{\mathrm{scalar}}_{k}}[G^s_{\De,J}] \nn \\
&=A_{k,\nu}\frac{2^{\frac{7}{2}+2d}\pi^{d-\frac{1}{2}}e^{i\frac{\pi}{4}}m^{\frac{3d}{2}+2}\nu^{\frac{7}{2}-4d}(m^2-\nu^2)^{\frac{d-2}{2}}(m-\sqrt{m^2-\nu^2})^{d-4}}{\G\p{\frac{d-2}{2}}\vol(\SO(d-2))}\frac{2^d \mathrm{dim}(\r)}{\vol(\SO(d))} \nn \\
&\x \<0|\f_4\wL[\cO](x_0,z_0)\f_{2^{+}}|0\>^{-1}\<\tl \cP_{\de_1}(x_{1^{+}0}) \tl \cP_{\de_3}(x_{30}) \cP_{\de}(z_0)\>f^{\mathrm{scalar}}_k(u',v')\nn \\
&\x \left.\frac{\<0|[\f_{1^{+}},\f_{2^{+}}]\cO^{\mu_1\cdots\mu_J}(x_5)0|\>\<\tl{\cO}^{\dag}_{\mu_1\cdots\mu_J}(x_5)\tl{\cO}^{\nu_1\cdots\nu_J}(x_6^{+})\>\<0|\cO^{\dag}_{\nu_1\cdots\nu_J}(x_6^{+})[\f_4,\f_3]|0\>}{\p{\<\tl{\cO}^{\dag}\tl\cO\>,\<\cO^\dag \cO\>}}\right|_{\textrm{scattering-crystal}}.
\ee
We can then evaluate the structures and and the subtraction factor at the scattering-crystal locus to find an explicit expression. Moreover, for symmetric traceless tensors, the pairing $\frac{\vol(\SO(d))}{2^d \mathrm{dim}(\r)}\p{\<\tl{\cO}^{\dag}\tl\cO\>,\<\cO^\dag \cO\>}$ becomes $4^{-d}$ \cite{Karateev:2018oml}. After combining all the factors and writing the result in terms of the ``heavy action" defined in \eqref{eq:heavyaction_scalar_definition}, we obtain
\be\label{eq:scalar_sumrule_CFTtoflatspace}
&\lim_{\substack{\nu,m\gg 1 \\ \nu<m}}\Psi_{k,\nu,f^{\mathrm{scalar}}_{k}}[\De,J] = \frac{2m^2-\nu^2}{m^2-\nu^2}\frac{\cP_J(1-\tfrac{2\nu^2}{m^2})}{(m^2(m^2-\nu^2))^{\frac{k}{2}}}.
\ee
As reviewed in section \ref{sec:review_superconvergence}, this is precisely the heavy state contribution of the flat space sum rule for scalars.

\subsection{Spinning sum rules}\label{sec:spinning_sum_rules}

We are now ready to write down the sum rules for spinning operators. The above analysis for the bulk-point limit can be generalized straightforwardly. A spinning functional can be defined by
\be\label{eq:spinningfunctional_definition}
&\Psi^{(a),(c)}_{k,\nu,\l'}[\cG] \equiv \int_{\substack{2>4 \\ 0\approx 2,4}} \frac{d^d x_2 d^d x_4 d^d x_0 D^{d-2}z_0}{\vol(\tl \SO(d,2))} \p{\<0|\cO_4(x_4^{+})\wL[\cO](x_0,z_0)\cO_2(x_2)|0\>^{(a)}}^{-1} \nn \\
&\int D^{d-2}z_1 D^{d-2}z_3 \<\tl \cP^{\dag}_{\de_1,\l_1}(z_1) \tl \cP^{\dag}_{\de_3,\l_3}(z_3) \cP_{\de,\l'}(z_0)\>^{(c)} \<\O|\cO_4(x_4^{+})[\wL[\cO_3](x_0,z_3),\wL[\cO_1](x_0,z_1)]\cO_2(x_2)|\O\>.
\ee
Compared to the scalar case \eqref{eq:scalarfunctional_definition}, the main difference is that the operators $\cO$ and $\cP_{\de,\l'}$ can have more complicated representations, and there are multiple allowed dual structures and celestial structures. 

Moreover, in the spinning case we do not include any coefficient in front of the integral. In the scalar case, there is a natural choice for this coefficient $A_{k,\nu}$, which is the factor needed to exactly reproduce the action of flat space sum rule \eqref{eq:scalar_sumrule_CFTtoflatspace}. However, as we will see in the next section, for spinning operators, the CFT sum rules and flat space sum rules are related by a matrix. Since there is no natural choice for the overall factor, we do not include a factor in our definition of the spinning functional.

Let us introduce some notation to describe spinning operators. A spinning operator with $\SO(d-1,1)$ representation $\r$ can be described by a Young diagram with rows of length $(m_1,m_2,\ldots,m_n)$, where $n=\lfloor \frac{d}{2}\rfloor$. We will also use the notation $\r=(J,\l)$, where $J=m_1$ is the spin, and $\l=(m_2,\ldots,m_n)$ specifies an $\SO(d-2)$ representation (which we sometimes call the ``transverse" representation). We also define $j=m_2$ to be the transverse spin of the operator. 

In \eqref{eq:spinningfunctional_definition}, the external operators have quantum numbers $(\De_i,J_i,\l_i)$, and the operator $\cO$ in the first line has quantum numbers $(\tfrac{d}{2}+i\nu,J_1+J_3-1,\l')$. In the second line, the celestial operators, which are primary operators that live on the celestial sphere, also carry $\SO(d-2)$ representations. $\tl \cP^{\dag}_{\de_1,\l_1}$ and $\tl \cP^{\dag}_{\de_3,\l_3}$ have the same $\SO(d-2)$ representations as the transverse representations of the external operators $\cO_1,\cO_3$, and $\cP_{\de,\l'}$ have the same transverse representation as $\cO$. Note that the $\SO(d-1,1)$ indices of the $\r_2,\r_4$ representations, and the $\SO(d-2)$ indices of $\l_1,\l_3,\l'$ are implicitly contracted in \eqref{eq:spinningfunctional_definition}.

In addition to the $\nu$ variable, the spinning functional also depends on a choice of $\SO(d-2)$ representation $\l'$. Moreover, the functional has two tensor structure labels $(a),(c)$. The first label $(a)$ corresponds to the tensor structure of the spinning dual structure, defined as
\be\label{eq:dual_struct_definition_spinning}
\p{\p{\<0|\cO_4(x_4^{+})\wL[\cO](x_0,z_0)\cO_2(x_2)|0\>^{(a)}}^{-1},\<0|\cO_4(x_4^{+})\wL[\cO](x_0,z_0)\cO_2(x_2)|0\>_{(b)}}_L =\de^{a}_{b}. 
\ee
The allowed number of labels $(a)$ is the number of continuous-spin structures of $\<\cO_2\cO_4\cO\>$, which is given by \cite{Kravchuk:2016qvl,Kologlu:2019mfz}
\be\label{eq:dual_spinning_counting}
\p{\l' \otimes \p{\mathrm{Res}^{\SO(d-1,1)}_{\SO(d-2)}\r_2\otimes \r_4}}^{\SO(d-2)},
\ee
where $\Res^G_H$ denotes the restriction of a representation of $G$ to its subgroup $H$, and $\p{\ldots}^H$ denotes the number of $H$-singlets. The second tensor structure label $(c)$ corresponds to the structures of a celestial three-point function with $\SO(d-2)$ representation $\l_1,\l_3,\l'$. The number of those structures is \cite{Kravchuk:2016qvl}
\be\label{eq:celestial_spinning_counting}
\p{\mathrm{Res}^{\SO(d-2)}_{\SO(d-3)}\l_1\otimes\l_3\otimes\l'}^{\SO(d-3)}.
\ee
Combining \eqref{eq:dual_spinning_counting} and \eqref{eq:celestial_spinning_counting}, we see that given the representations of the external operators, the allowed choice of the transverse representation $\l'$ is
\be\label{eq:lambdaprime_selectionrule}
\left\{\l'| \l' \in \mathrm{Res}^{\SO(d-1,1)}_{\SO(d-2)}\r_2\otimes \r_4, \p{\mathrm{Res}^{\SO(d-2)}_{\SO(d-3)}\l_1\otimes\l_3\otimes\l'}^{\SO(d-3)} \neq 0\right\}.
\ee
In section \ref{sec:flat_space}, we will give the set of allowed $\l'$'s for some concrete examples.

Similar to the scalar case, commutativity of the light-transformed operators implies
\be
\Psi^{(a),(c)}_{k,\nu,\l'}[\cG] = 0,\quad J_1+J_3-1>J_0.
\ee
Again, this condition can be equivalently written in terms of the $C^{\pm}_{ab}(\De,\r)$ coefficient function,
\be
C^{(-1)^{J_1+J_3-1}}_{ab}(\tfrac{d}{2}+i\nu,J_1+J_3-1,\l') =0 ,\quad J_1+J_3-1>J_0,
\ee
where $a$ is the same tensor structure label of the functional \eqref{eq:spinningfunctional_definition}, which corresponds to the structures of the $\cO_2\x\cO_4$ OPE. On the other hand, $b$ is a subset of the structures of the $\cO_1\x\cO_3$ OPE that have the same counting rule as the celestial structures \eqref{eq:celestial_spinning_counting} (see \cite{Chang:2020qpj} for more details).

\subsubsection{The spinning heavy action}
We now derive a formula for the heavy action of the spinning functional in the bulk-point limit, similar to the scalar case \eqref{eq:scalarfunctional_bulkpoint_withrtensor}. For simplicity, in what follows we assume the external operators are symmetric traceless tensors with spin $J_e$, and they all have scaling dimensions $\De_e$. By rewriting the light transforms in \eqref{eq:spinningfunctional_definition}, the spinning functional can be written as
\be
&\Psi^{(a),(c)}_{k,\nu,\l'} = \Psi^{+(a),(c)}_{k,\nu,\l'} - (1\leftrightarrow 3),
\ee
and
\be
&\Psi^{+(a),(c)}_{k,\nu,\l'}[\cG] \nn \\
&= 4\int_{4>3>0>1>2}\frac{d^dx_1d^dx_2d^dx_3d^dx_4d^dx_0D^{d-2}z_0}{\vol(\tl{\SO}(d,2))} \de(x_{1^{+}0}^2)\de(x_{30}^2)\th(x_{1^{+}0})\th(x_{30}) \nn \\
&\x \p{\<0|\cO_4(x_4)\wL[\cO](x_0,z_0)\cO_2(x_2^{+})|0\>^{(a)}}^{-1}\<\tl \cP^{\dag}_{\de_1,\l_1}(x_{1^{+}0}) \tl \cP^{\dag}_{\de_3,\l_3}(x_{30}) \cP_{\de,\l'}(z_0)\>^{(c)} \nn \\
&\x \<\O|[\cO_4(x_4),\cO_3(x_3,x_{30})][\cO_1(x_1^{+},x_{1^{+}0}),\cO_2(x_2^{+})]|\O\>.
\ee
Note that the polarizations of $\cO_1,\cO_3$ are fixed to be $x_{30},x_{1^{+}0}$ (which are null vectors due to the delta functions in the integrand). This is because of the light transforms $\wL[\cO_3]\wL[\cO_1]$ in \eqref{eq:spinningfunctional_definition}.

To study the action of $\Psi^{+(a),(c)}_{k,\nu,\l'}$ on a conformal block, we again use the Lorentzian shadow representation of the block and consider the bulk-point limit, in which $\De\sim m$ and $\nu,m\gg 1,\nu<m$. To perform the saddle point analysis, we separate the integrand into a quickly-varying part and slowly-varying part. Crucially, in the bulk-point limit, the quickly-varying part of the integrand is the same as the scalar case! To see why, recall that the factor in the scalar case is given by \eqref{eq:bulkpoint_integrand_factor}, and it comes from powers of $x_{ij}^2,z_0\.x_{10},z_0\.x_{30},x_{10}\.x_{30}$, and $V_{0,24}$ in the conformally-invariant structures in the integrand. For spinning structures, the powers of these factors will only differ by integer values that depend on the representations and choice of tensor structures, and hence they do not matter in the $\nu,m \gg 1$ limit. Since the quickly-varying part in the scalar and spinning cases are identical, the saddle point analysis for the spinning case is identical to what we did in section \ref{sec:saddle_point_analysis}. As a result, the formula in the scalar case immediately generalizes to the spinning case, and by rewriting \eqref{eq:scalarfunctional_bulkpoint_withrtensor} we obtain that the formula in the spinning case is given by
\be\label{eq:spinningfunctional_bulkpoint_withrtensor}
&\lim_{\substack{\nu,m\gg 1 \\ \nu<m}} \Psi^{+(a),(c)}_{k,\nu,\l'}[G^s_{\De,\r,(a'b')}] \nn \\
&=\frac{2^{\frac{5}{2}+2d}\pi^{d-\frac{1}{2}}e^{i\frac{\pi}{4}}m^{\frac{3d}{2}+2}\nu^{\frac{7}{2}-4d}(m^2-\nu^2)^{\frac{d-2}{2}}(m-\sqrt{m^2-\nu^2})^{d-4}}{\G\p{\frac{d-2}{2}}\vol(\SO(d-2))}\frac{2^d \mathrm{dim}(\r)}{\vol(\SO(d))} \x \nn \\
&\p{\<0|\cO_4(x_4)\wL[\cO](x_0,z_0)\cO_2(x_2^{+})|0\>^{(a)}}^{-1}\<\tl \cP^{\dag}_{\de_1,\l_1}(x_{1^{+}0}) \tl \cP^{\dag}_{\de_3,\l_3}(x_{30}) \cP_{\de,\l'}(z_0)\>^{(c)} \x \nn \\
&\left.\frac{\<0|[\cO_1(x_{1^{+}},x_{1^{+}0}),\cO_{2}(x_{2^+})]\cO(x_5)|0\>_{(a')}\<\tl{\cO}^{\dag}(x_5)\tl{\cO}(x_6^{+})\>\<0|\cO^{\dag}(x_6^{+})[\cO_4(x_4),\cO_3(x_3,x_{30})]|0\>_{(b')}}{\p{\<\tl{\cO}^{\dag}\tl\cO\>,\<\cO^\dag \cO\>}}\right|_{\substack{\textrm{scattering-} \\ \textrm{crystal}}},
\ee
where $(a'),(b')$ are the three-point tensor structure labels of the block. The formula should be evaluated at the same ``scattering-crystal" as the scalar case, given by \eqref{eq:scattering_crystal}, and we have implicitly contracted all the $\SO(d-1,1)$ indices of $\r_2,\r_4,\r$ and the $\SO(d-2)$ indices of $\l_1,\l_3,\l'$.

\subsubsection{Subtraction factors}\label{sec:subtraction}
To finish the discussion of spinning sum rules, let us give the subtraction factors that should be used in the spinning case. We consider a more general functional
\be
\Psi^{(a),(c)}_{k,\nu,\l',f}[\cG] = \Psi^{+(a),(c)}_{k,\nu,\l'}[f(u',v')\cG] - (1\leftrightarrow 3).
\ee
As discussed in section \ref{sec:scalar_sum_rules}, to have nontrivial sum rules for identical external operators, the kernel of the $\Psi^{+(a),(c)}_{k,\nu,\l'}$ functional times the subtraction factor $f(u',v')$ should be antisymmetric under $(1\leftrightarrow 3)$. For the spinning functionals, the signature under swapping $1$ and $3$ depends on the transverse representation $\l'$. We will focus on the case where the external operators are symmetric traceless tensors, so the allowed $\l'$ satisfying the selection rule \eqref{eq:lambdaprime_selectionrule} should be a symmetric traceless tensor itself (as an $\SO(d-2)$ representation). Therefore, the spinning functionals in this case are labeled by an integer $j'$, which is the transverse spin. Due to the celestial structure $\<\tl{\cP}_1\tl{\cP}_3\cP_{\de,j'}\>$, the functional $\Psi^{+(a),(c)}_{k,\nu,j'}$ has signature $(-1)^{j'}$ under $(1\leftrightarrow 3)$. Therefore, the nontrivial spinning sum rules are given by
\be
\Psi^{+(a),(c)}_{k,\nu,j'}[f^{(-1)^{j'}}(u',v')\cG] = 0,
\ee
where the subtraction factors should satisfy
\be
f^{\pm}(v',u') = \mp f^{\pm}(u',v').
\ee

We find that for external operators with spin $J_e$, the subtraction factors with even $j'$ can be chosen to be
\be\label{eq:subtraction_factor_evenjprime}
f^{+}_{k;J_e}(u',v') &=
\begin{cases}
 \frac{v'-u'}{(u'v')^{\frac{k-2J_e+2}{4}}} & k \equiv 2J_e + 2\text{ mod 4} \\
 \frac{(v'-u')^3-(v'-u')}{4(u'v')^{\frac{k-2J_e+4}{4}}} & k \equiv 2J_e \text{ mod 4}
\end{cases},\quad &&k=2,4,6,\ldots,
\ee
which reduces to the scalar case \eqref{eq:subtracction_scalars} when $J_e=0$. On the other hand, for odd $j'$ we have
\be\label{eq:subtraction_factor_oddjprime}
f^{-}_{k;J_e}(u',v') &=
\begin{cases}
 -\frac{1}{(u'v')^{\frac{k-2J_e+1}{4}}} & k \equiv 2J_e+3 \text{ mod 4} \\
 -\frac{(u'-v')^2-1}{4(u'v')^{\frac{k-2J_e+3}{4}}} & k \equiv 2J_e+1 \text{ mod 4}
\end{cases},\quad &&k=3,5,7,\ldots.
\ee
When evaluating these subtraction factors at the scattering crystal \eqref{eq:scattering_crystal}, we find
\be
\left.f^{+}_{k;J_e}(u',v')\right|_{\textrm{scattering-crystal}} &= -\nu^{2(k-2J_e+1)}\frac{(2m^2-\nu^2)}{m^2(m^2-\nu^2)}\frac{1}{(m^2(m^2-\nu^2))^{\frac{k-2J_e}{2}}}, &&k=2,4,6,\ldots, \nn \\
\left.f^{-}_{k;J_e}(u',v')\right|_{\textrm{scattering-crystal}} &=  -\nu^{2(k-2J_e)}\frac{\nu^2}{m^2(m^2-\nu^2)}\frac{1}{(m^2(m^2-\nu^2))^{\frac{k-2J_e-1}{2}}}, &&k=3,5,7,\ldots.
\ee

From the flat space point of view, the expressions for the subtraction factors at the saddle are precisely what we need. We expect that functionals with even $j'$ are related to the $s\leftrightarrow t$ symmetric part of the flat space amplitude. Therefore, the corresponding flat space sum rule should have a factor
\be
\p{\frac{1}{s} + \frac{1}{s+u}}\frac{1}{(s(s+u))^{\frac{k}{2}}} = \frac{2s+u}{s(s+u)}\frac{1}{(s(s+u))^{\frac{k}{2}}}.
\ee
On the other hand, functionals with odd $j'$ should be related to the $s\leftrightarrow t$ antisymmetric part, and the corresponding flat space sum rule should have the factor
\be
\p{\frac{1}{s} - \frac{1}{s+u}}\frac{1}{(s(s+u))^{\frac{k}{2}}} = \frac{u}{s(s+u)}\frac{1}{(s(s+u))^{\frac{k}{2}}}.
\ee
Under the identification $s=m^2,u=-\nu^2$, we see that these expressions agree with $f^{\pm}_{k;J_e}(u',v')$ at the saddle. We make this connection more precise in section \ref{sec:flat_space}.

Finally, let us explain the allowed range of $k$ in \eqref{eq:subtraction_factor_evenjprime} and \eqref{eq:subtraction_factor_oddjprime}. In the Regge limit, the subtraction factor $f^{+}$ changes the Regge spin by $k-2J_e+1$, and $f^{-}$ changes the Regge spin by $k-2J_e$. Therefore, for the integral of the functional to be convergent, we must have $k>J_0$ for even $j'$ and $k-1>J_0$ for odd $j'$. For non-perturbative CFTs, we have $J_0\leq 1$. So, we should choose $k\geq 2$ for even $j'$ and $k\geq 3$ for odd $j'$. As we will show in the next section, $k$ also agrees with the Regge spin of the corresponding flat space sum rule.

Normally, introducing subtraction factors comes with a price that the action of the functional becomes non-vanishing on finitely many light double-traces \cite{Caron-Huot:2020adz}. However, if $J_e$ is large enough, sometimes we can have subtraction factors that make the Regge behavior of the kernel worse. For these unsubtracted sum rules, the action on all double-traces will remain zero. On the flat space side, the EFT series of the corresponding flat space sum rule should automatically truncate, since there is no denominator in the integrand. As an example, for $J_e=2$ (which corresponds to flat space gravitons), we have two such subtraction factors from $k=2,k=3$. Their expressions at the saddle are given by
\be
\left.f^{+}_{k=2;J_e=2}(u',v')\right|_{\textrm{scattering-crystal}} &= -\nu^{-2}(2m^2-\nu^2),\nn \\
\left.f^{-}_{k=3;J_e=2}(u',v')\right|_{\textrm{scattering-crystal}} &= 1.
\ee
These unsubtracted sum rules for gravitons will be given in section \ref{sec:flat_space}.

\section{Flat space interpretation}\label{sec:flat_space}
We now discuss the flat space interpretation of the CFT sum rules derived in the previous section. On the flat space side, we are mainly interested in sum rules of photons and gravitons, which are dual to conserved operators on the CFT side. Our goal is to relate the CFT heavy action formula \eqref{eq:spinningfunctional_bulkpoint_withrtensor} to the heavy state contribution of the flat space sum rule. The heavy action formula consists of a block part (fourth line of \eqref{eq:spinningfunctional_bulkpoint_withrtensor}) from the block insertion, and a kernel part (third line of \eqref{eq:spinningfunctional_bulkpoint_withrtensor}) from the kernel of the functional. We will be able to understand both of them in flat space. Throughout the discussion, we will study the photon case in detail and explain how to match the CFT result and the flat space result. For the graviton case, we simply give the final relation between CFT sum rules and flat space sum rules.

Hereafter, we use $D$ to denote the spacetime dimension of flat space, and $d=D-1$ is the spacetime dimension of the CFT.

\subsection{Conservation at large $\De$}\label{sec:conservation_largeDelta}

The flat space limit of a dispersive functional corresponds to an AdS scattering process where the incoming particles are boosted to high energies. Our main interest is in massless particles like photons and gravitons, which are dual to conserved CFT operators. In this highly boosted limit, conservation will substantially simplify the dictionary that we find. We expect that for external massive particles, it is most natural to use the goldstone equivalence theorem to break them into massless excitations, and use our dictionary for each excitation.

We first note that the conservation condition for CFT three-point structures simplifies when one of the operator dimensions $\De$ becomes large. For a three point function $\< \cO_1(x_1 , z_1) \cO_2(x_2 , z_2) \cO_3(x_3 , z_3) \> $, with $\cO_1$ conserved we can write its conservation condition as
\be\label{eq:conservation_usual}
0 &= \ptl_{\mu_1} \< \cO_1^{\mu_1\cdots \mu_J}(x_1) \cO_2(x_2 ,z_2) \cO_3(x_3 , z_3) \>.
\ee
We claim that at large $\De=\De_3$ this simplifies to
\be\label{eq:conservation_largeDelta}
v_{1,23}^{\mu_1}\<\cO_{1;\mu_1\cdots\mu_J}\cO_2\cO\> = 0,
\ee 
where 
\be
v_{1,23}^{\mu} &= \frac 1 2 \ptl^\mu_1 \log \frac{x_{12}^2}{x_{13}^2} = \frac{x_{12}^\mu x_{13}^2 - x_{13}^\mu x_{23}^2}{x_{12}^2 x_{13}^2}.
\ee
To see why, we can isolate the $\De$ dependence in the three-point structure:
\be
\label{eq:prefactor}
\< \cO_1^{\mu_1 \cdots \mu_J}(x_1) \cO_2(x_2 , z_2) \cO_3(x_3 , z_3) \> = \left( \frac{x_{12}^2}{x_{13}^2  x_{23}^2}  \right )^{\Delta /2} \x (\cdots),
\ee
where $(\cdots)$ does not depend exponentially on $\De$. Applying the conservation condition (\ref{eq:conservation_usual}), the leading contribution in the large-$\De$ limit comes from the prefactor in (\ref{eq:prefactor}), which gives (\ref{eq:conservation_largeDelta}).

Let us check this claim in the explicit example $\<JJ\cO\>$, where $J$ is a spin-1 conserved operator, and $\cO$ is a symmetric traceless tensor. Before imposing conservation, the general three-point function $\<JJ\cO\>$ in the embedding space (using the conventions in \eqref{eq:3pt_convention_spinning}) is given by
\be
&\<J(X_1,Z_1)J(X_2,Z_2)\cO(X_3,Z_3)\>
\nn\\
&=
\frac{c_1 (-2V_3)^{J-2} H_{23} H_{13} + c_2 H_{12}(-2V_3)^J +c_3 V_2 H_{13} (-2V_3)^{J-1} + c_4 V_1 H_{23} (-2V_3)^{J-1} +c_5 V_1 V_2 (-2V_3)^J}{X_{12}^{\frac{\De_1 +\De_2-\De +2-J}{2}}X_{13}^{\frac{\De_1 -\De_2+\De +J}{2}}X_{23}^{\frac{-\De_1 +\De_2+\De +J}{2}}}.
\ee

Let us now impose the conservation condition \eqref{eq:conservation_usual}. Starting with $\<JJ\cO\>$, we can see that $\partial_{X_1} \. \cD_{Z_1}\<JJ\cO\>$ transforms like $\<\phi J\cO\>$. It therefore has two independent structures; one proportional to $V_2$ and one proportional to $H_{23}$. These must independently vanish. Furthermore, we get two more equations from imposing that $J_2$ be conserved. If however we impose $\De_1 = d-1 = \De_2$ as required for a conserved spin 1 operator, we see that only 3 of these equations are independent. Imposing them and denoting $\De_3=\De$ we get
\be
(4-2d-J+\De)c_1 + 4J c_2 + 2 c_3 + 2(d-1+\De) c_4 &=0,
 \nn \\ (2J+2\De) c_2 + (2(d-1)+J-\De) c_3 + 2(d-1+\De) c_5 &=0,
  \nn \\
(4-2d-J+\De)c_1 + 4J c_2 + 2(d-1+\De) c_3 + 2 c_4 &=0,
 \nn \\ (2J+2\De) c_2 - (2(d-1)+J-\De) c_4 + 2(d-1+\De) c_5 &=0.
\ee

As mentioned above, of these four equations, only 3 are linearly independent. We can then take the large $\De$ limit to get 
\be 
c_1-2c_4 = 0\quad 2 c_2 +c_4+2c_5=0\quad c_3+c_4 = 0.
\ee

One can then verify that this is the same as \eqref{eq:conservation_largeDelta}, which is equivalent to demanding that in the embedding space
\be\label{eq:conservation_largeDelta_embedding}
&V_1\.D_{Z_1} \<\cO_1(X_1,Z_1)\cO_2(X_2,Z_2)\cO(X,Z)\> = 0. \nn \\
&V_2\.D_{Z_2} \<\cO_1(X_1,Z_1)\cO_2(X_2,Z_2)\cO(X,Z)\> = 0,
\ee
where $V^A_i$ is the usual embedding space structure $V_i$ defined in \eqref{eq:CFT_building_blocks_VH} with $Z^A_i$ stripped off, and $D^A_{Z_i} = \p{\frac{d}{2}-1+Z_i\.\ptl_{Z_i}}\ptl_{Z_i}^A-\tfrac{1}{2}Z_i^A \ptl^2_{Z_i}$ is the Todorov/Thomas operator \cite{Costa:2011mg}.

Importantly, the simplified conservation condition (\ref{eq:conservation_largeDelta}) is algebraic. It allows us to describe the conserved operator using a boundary polarization vector with one fewer degree of freedom, which will be necessary for writing a simple correspondence between boundary and bulk polarizations.

\subsection{CFT 3-point structures and flat space 3-point amplitudes}\label{sec:partial_waves}

Our goal now is to interpret the last line of (\ref{eq:spinningfunctional_bulkpoint_withrtensor}) in flat space language. The conformal three-point structures will become three-point amplitudes, and their contractions will become a partial wave.
 Recall that the partial wave decomposition of a spinning flat space amplitude is given by \cite{Caron-Huot:2022jli}
\be\label{eq:amp_partialwave_expansion}
\cM(s,u) = s^{\frac{4-D}{2}} \sum_{\r} n_{\r}^{(D)} \sum_{ab}(a_{\r}(s))_{ab}\pi_{\r,(ab)},
\ee
where $\r$ is an irrep of $\SO(d)$, $a,b$ represent the three-point structures (see below), and $n_{\r}^{(D)}=\frac{2^{d+1}(2\pi)^{d-1}\dim\r}{\vol S^{d-1}}$ is a normalization factor.  This is a generalization of the scalar partial wave decomposition \eqref{eq:amp_partialwave_expansion_scalar}, where instead of the Gegenbauer polynomial $\cP_J$ we now have a more general partial wave $\pi_{\r,(ab)}$.

The partial wave $\pi_{\r,(ab)}$ transforms in the representation $\r$ of the $\SO(d)$ group that stabilizes $P^{\mu} = p_1^{\mu}+p_2^{\mu}$. It can be obtained by gluing two vertices,
\be\label{eq:partial_wave_vertices_definition}
\pi_{\r,(ab)} \equiv \p{\bar v_b(n',e_3,e_4),v_{a}(n,e_1,e_2)},
\ee
where the pairing $(\ldots,\ldots)$ represents the contraction of $\SO(d)$ indices, and $\bar f(x) = f(x^{*})^{*}$ is  Schwarz reflection. The vertex $v_a(n,e_1,e_2)$ is a three-point amplitude of two massless particles and a massive particle. The massive particle has momentum $P^{\mu}$ and transforms in the representation $\r$ under the little group $\SO(d)$. We use $a$ to label different independent vertices, and we define $n^{\mu}$ and $e_i^{\mu}$ as
\be
n^{\mu} = \frac{p_1^{\mu}-p_2^{\mu}}{\sqrt{(p_1-p_2)^2}},\quad e_i^{\mu} = \e_i^{\mu} - p_i^{\mu}\frac{\e_i\.P}{p_i\.P}.
\ee
They satisfy $n\.e_i=e_i^2=0, n^2=1$. The other vertex can be defined in the same way with momenta $p_3^{\mu},p_4^{\mu}$.

Let us demonstrate how to obtain vertices $v_a(n,e_1,e_2)$ from CFT structures by studying the example of photon scattering. As we saw in the previous section, for a CFT three-point function between two conserved spin one currents and an operator with large $\De$, conservation places a simple constraint on the allowed three point structures. For example, when $\cO$ is a traceless symmetric tensor, we have
\be
&\<J_1(X_1,Z_1)J_2(X_2,Z_2)\cO(X_5,Z_5)\> \nn \\
&=\frac{c_1 (-2V_5)^{J-2} H_{25} H_{15} + c_2 H_{12}(-2V_5)^J +c_3 V_2 H_{15} (-2V_5)^{J-1} + c_4 V_1 H_{25} (-2V_5)^{J-1} +c_5 V_1 V_2 (-2V_5)^J}{X_{12}^{\frac{2d-\De -J}{2}}X_{15}^{\frac{\De +J}{2}}X_{25}^{\frac{\De +J}{2}}}.
\ee

At large $\De$, conservation demands that we have no terms proportional to $V_1$ or $V_2$. Moreover, at the saddle configuration of (\ref{eq:spinningfunctional_bulkpoint_withrtensor}) we find
\be\label{eq:vstructwithindex_saddle}
-v_{1,25}|_{\mathrm{saddle}}=v_{2,51}|_{\mathrm{saddle}}=v_{3,46}|_{\mathrm{saddle}}=-v_{4,63}|_{\mathrm{saddle}}=e,
\ee
where $e$ is the unit vector in the time direction. Therefore, if we choose the polarization vectors $z_1,z_2$ for the conserved currents to have no time component, the conserved structures of $\<JJ\cO\>$ at large $\De$ will satisfy
\be\label{eq:JJO_CFT3pt_saddle}
&\<J_1J_2\cO\>^{(1)}_{\mathrm{saddle},z_{i}=(0,z_i^x,\vec z_{i\perp})}= \left.\frac{(-2V_5)^{J-2}H_{15}H_{25}}{X_{12}^{\frac{2d-\De-J}{2}}X_{15}^{\frac{\De+J}{2}}X_{25}^{\frac{\De+J}{2}}}\right|_{\mathrm{saddle},z_{i}=(0,z_i^x,\vec z_{i\perp})}, \nn \\
&\<J_1J_2\cO\>^{(2)}_{\mathrm{saddle},z_{i}=(0,z_i^x,\vec z_{i\perp})}= \left.\frac{(-2V_5)^{J}H_{12}}{X_{12}^{\frac{2d-\De-J}{2}}X_{15}^{\frac{\De+J}{2}}X_{25}^{\frac{\De+J}{2}}}\right|_{\mathrm{saddle},z_{i}=(0,z_i^x,\vec z_{i\perp})}.
\ee

We would like to interpret these expressions in such a way that they give rise to flat-space three-point vertices. We can describe the possible vertices using the formalism of \cite{Chakraborty:2020rxf, Caron-Huot:2022jli}.
The two massless spin-1 particles have momenta $p_1$ and $p_2$, as well as polarizations $e_1$ and $e_2$. We additionally have the momentum of the third particle $p_3:=P= p_1+p_2$ and another Lorentz invariant quantity $n \propto p_1-p_2$. We have the freedom to normalize $n$, and to shift the polarizations $e_i$ by $p_i$. As the third particle is massive, we can go to its rest frame and parameterize its momentum as $P = (m,0,\cdots,0)$. We can then use the symmetry of our saddle point to find the flat space kinematics.

Since the vertices are tensors with $\SO(d)$ indices, we can introduce index free notation and contract the indices with polarization vectors $w_1,w_2,\ldots\in \C^d$ for each row of the Young diagram. For example, if $\r$ is a symmetric traceless tensor, we have a single polarization vector $w_1$, and from these quantities we can build two three-point amplitudes: 
\be\label{eq:JJO_flatspace3pt}
v_{\g\g J}^{(1)}(n,e_1,e_2) &= (n\.w_1)^{J-2}(e_1\.w_1)(e_2\.w_1) , \nn \\
v_{\g\g J}^{(2)}(n,e_1,e_2) &= (n\.w_1)^J (e_1\.e_2).
\ee
These two amplitudes should correspond to (\ref{eq:JJO_CFT3pt_saddle}) through the appropriate dictionary.

Our saddle point \eqref{eq:scattering_crystal} defines an $\SO(d-1) \subset \SO(d,2)$ subgroup that fixes the locations of the operators $J_1,J_2,\cO$. The vector $v_{5,12}^\mu$ points along the direction preserved by this $\SO(d-1)$ at the location of $\cO$. In the bulk, there is a corresponding $\SO(d-1)$ that preserves $n$. Thus it is natural to impose
\be 
(z_5 \. v_{5,12})^J  \leftrightarrow (w_1 \. n)^J.
\ee
The left-hand side is the projection of the polarization of $\cO$ along the $\SO(d-1)$-invariant direction, while the right-hand side is the projection of the polarization of the massive particle along the $\SO(d-1)$-invariant direction.

Therefore we should impose $p_1^{\mu} -p_2^{\mu} \propto (0,v_{5,12}^\mu)$. Evaluating this on the saddle \eqref{eq:scattering_crystal} we find $n^\mu = (0, \frac{\sqrt{m^2 - \nu^2}}{m},\frac{\nu}{m},0,\cdots,0)$, while we can choose the momenta of the scattering process to be:
\be\label{eq:momenta_for_partialwaves}
p_1=&\frac{1}{2}(m,\sqrt{m^2-\nu^2},\nu,0,\ldots,0),\quad\ \  p_2=\frac{1}{2}(m,-\sqrt{m^2-\nu^2},-\nu,0,\ldots,0), \nn \\
p_3=&\frac{1}{2}(m,\sqrt{m^2-\nu^2},-\nu,0,\ldots,0),\quad p_4=\frac{1}{2}(m,-\sqrt{m^2-\nu^2},\nu,0,\ldots,0).
\ee
Here, $p_1,p_2$ are incoming and $p_3,p_4$ are outgoing.

It now remains to parameterize the bulk polarization vectors $e_1,e_2,w_1$ in terms of CFT polarization vectors $z_1,z_2,z_5$ to get \eqref{eq:JJO_CFT3pt_saddle} to agree with \eqref{eq:JJO_flatspace3pt}. Our CFT polarizations can be expressed as 
\be
z_{1}=(0,z_1^x,\vec z_{1\perp}),\quad z_{2}=(0,z_2^x,\vec z_{2\perp}),\quad z_5=(z_5^t,z_5^x,\vec z_{5\perp}),
\ee
where we've explicitly set the time component of $z_1$ and $z_2$ to be zero, in accordance with (\ref{eq:vstructwithindex_saddle}). Meanwhile, we can express the bulk polarizations as
\be\label{eq:bulk_polarization_nperp}
e_1^{\mu} = e_1^{n} n_{\perp}^{\mu} + (0,0,0,\vec e_{1\perp}),\quad e_2^{\mu} = e_2^{n} n_{\perp}^{\mu} + (0,0,0,\vec e_{2\perp}),\quad w_1^{\mu}=(0,w_1^t,w_1^x,\vec w_{1\perp}),
\ee
where $n_{\perp}^{\mu}$ is a vector perpendicular to $n^{\mu}$, given by
\be
n_{\perp}^{\mu} = (0,-\tfrac{\nu}{m},\tfrac{\sqrt{m^2-\nu^2}}{m},0,\ldots,0).
\ee

As we argued above, the $\SO(d-1)$ singlets should match between CFT and flat space. In particular, this imposes
\be
&\left.\frac{H_{15}}{X_{15}}\right|_{\textrm{saddle}} \leftrightarrow e_1\.w_1,\quad \left.\frac{H_{25}}{X_{25}}\right|_{\textrm{saddle}} \leftrightarrow e_2\.w_1, \nn \\
&\left.\frac{H_{12}}{X_{12}}\right|_{\textrm{saddle}} \leftrightarrow e_1\.e_2,\quad \left.V_{5,12}\right|_{\textrm{saddle}} \leftrightarrow -i n\.w_1,
\ee
which we can achieve by setting 
\be\label{eq:polarvec_125_CFTtoflat}
z_1^x=e_1^{n}, \quad \vec z_{1\perp} = \vec e_{1\perp},\quad z_2^x=-e_2^{n},\quad \vec z_{2\perp} = \vec e_{2\perp},\quad z_5^t=-iw_1^t,\quad z_5^x=w_1^x,\quad \vec z_5= \vec w_{1\perp}.
\ee
Let us emphasize that even though we use a $t$ superscript for the $w_1^t$ component, it is actually a spatial direction in the bulk, while $z_5^t$ is the time direction in the CFT. This is why we need an $i$ factor in their relation. Note also that the relation maps the condition $z_i^2=0$ to $e_i^2=w_1^2=0$. If the Young diagram of the representation $\r$ has more than one row, then we need to introduce more polarization vectors for both the flat space vertex and the CFT three-point structure. The map between these polarizations is the same as the relation between $z_5$ and $w_1$ in \eqref{eq:polarvec_125_CFTtoflat}.

The polarization map \eqref{eq:polarvec_125_CFTtoflat} turns each CFT three-point structure building block into an amplitude building block, and thus the CFT three-point structures naturally become flat space vertices. 

Note that in the CFT sum rule \eqref{eq:spinningfunctional_bulkpoint_withrtensor}, the three-point structures appear as commutators of Wightman functions. So we will use the commutator to define vertices, and divide by the $\sin(\ldots)$ factor from the commutator by hand. For the cases we consider in this paper (photons and gravitons), the Young diagram of the exchanged representation $\r$ has at most three rows. We will often write $\r=(J,j,\tl{j})$ to denote the length of each row. For external operators $\cO_1,\cO_2$ with quantum numbers $(\De_i,J_i)$ and exchanged representation $\r$, we define the vertices as\footnote{Reference \cite{Li:2021snj} presents a similar relation between flat space 3-point amplitudes and CFT 3-point structures, but with a different configuration of CFT positions and polarizations $X_i,Z_i$. We expect that our relation is equivalent up to a choice of conformal frame.} 
\be\label{eq:CFT_vertices_definition}
&v_a(n,e_1,e_2,w_i) \nn \\
&\equiv \left.(-x_{12}^2)^{\frac{\De_1+\De_2}{2}}\p{\frac{x_{15}^2}{x_{25}^2}}^{\frac{\De_1-\De_2}{2}}\frac{\<0|[\cO_1(x_{1^+},z_1),\cO_2(x_{2^+},z_2)]\cO(x_5,z_5,\vec w_5)]|0\>_{(a)}}{2i(\sin(\pi\tfrac{\tl \tau_\r-\De_1-J_1-\De_2-J_2}{2}))}\right|_{\mathrm{saddle},\eqref{eq:polarvec_125_CFTtoflat}},
\ee
where $w_i$ on the left-hand side are the flat space polarizations, and $z_5,\vec w_5$ on the right-hand side are the CFT polarizations. The $2i(\sin(\pi\tfrac{\tl \tau_\r-\De_1-J_1-\De_2-J_2}{2}))$ factor comes from taking the commutator, and $\tl \tau_\r = \De-J+j+\tl{j}$ for $\r=(J,j,\tl{j})$.\footnote{We find this $\sin(\pi\tfrac{\tl \tau_\r-\De_1-J_1-\De_2-J_2}{2})$ factor by studying the commutator of \eqref{eq:3pt_convention_spinning}. It would be interesting to determine what $\tl \tau_\r$ should be for general $\r$.} We introduce the factor $(-x_{12}^2)^{\frac{\De_1+\De_2}{2}}\p{\frac{x_{15}^2}{x_{25}^2}}^{\frac{\De_1-\De_2}{2}}$ to remove factors of $\nu$.

As an example, we can apply the polarization map \eqref{eq:polarvec_125_CFTtoflat} to the $\<JJ\cO\>$ CFT three-point structures \eqref{eq:JJO_CFT3pt_saddle}. We get
\be
&\left.(-x_{12}^2)^{d-1}\left\{\frac{\<0|[J(x_{1^+},z_1),J(x_{2^+},z_2)]\cO(x_5,z_5)|0\>_{(1,2)}}{2i\p{\sin(\pi\tfrac{\De-J-2d}{2})}}\right\}\right|_{\mathrm{saddle},\eqref{eq:polarvec_125_CFTtoflat}} \nn \\
&=-(-i)^J2^{J}\left\{2^{-2}(n\.w_1)^{J-2}(e_1\.w_1)(e_2\.w_1),  (n\.w_1)^J (e_1\.e_2) \right\}.
\ee 
As expected, the result correctly reproduces the three-point amplitudes \eqref{eq:JJO_flatspace3pt}.

The other vertex $\bar v(n',e_3,e_4)$ can be defined using the $\<\cO_3\cO_4\cO_6\>$ CFT three-point structure in a similar way. The polarization map is given by
\be\label{eq:polarvec_346_CFTtoflat}
z_3^x = e_3^{n'},\quad \vec z_{3\perp} = \vec e_{3\perp},\quad z_4^x = -e_4^{n'},\quad z_{4\perp} = \vec e_{4\perp},\quad z_6^t = -iw_1^t,\quad z_6^x = w_1^x,\quad \vec z_6=\vec w_{1\perp}.
\ee
We again set the time component of $z_3,z_4$ to be zero, and
\be
e_3^{\mu} = e_3^{n'} n_{\perp}^{\prime\mu} + (0,0,0,\vec e_{3\perp}),\quad e_4^{\mu} = e_4^{n'} n_{\perp}^{\prime\mu} + (0,0,0,\vec e_{4\perp}),\quad w_1^{\mu}=(0,w_1^t,w_1^x,\vec w_{1\perp}),
\ee
where
\be
n_{\perp}^{\prime\mu} = (0,\tfrac{\nu}{m},\tfrac{\sqrt{m^2-\nu^2}}{m},0,\ldots,0),
\ee
which is perpendicular to $n^{\prime\mu}=(0,\tfrac{\sqrt{m^2-\nu^2}}{m},-\tfrac{\nu}{m},0,\ldots,0)\propto p_3^{\mu}-p_4^{\mu}$.

The definition of the $\bar v(n',e_3,e_4)$ is given by 
\be\label{eq:CFT_vertices_definition_bar}
&\bar v_b(-n',e_3,e_4,w_i) \nn \\
&\equiv (-x_{34}^2)^{\frac{\De_3+\De_4}{2}}\p{\frac{x_{36}^2}{x_{46}^2}}^{\frac{\De_3-\De_4}{2}}\mathcal{I}_e^{\r} \left.\frac{\<0|\cO^{\dag}(x_{6^+},z_6,\vec w_6)[\cO_4(x_{4},z_4),\cO_3(x_3,z_3)]|0\>_{(b)}}{2i(\sin(\pi\tfrac{\tl \tau_\r-\De_3-J_3-\De_4-J_4}{2}))}\right|_{\mathrm{saddle},\eqref{eq:polarvec_346_CFTtoflat}}.
\ee
On the right hand side, $\mathcal{I}_e^\r$ is a tensor that reflects all the indices in the time direction. More precisely, for any tensor $T$ with representation $\r$, in the index-free notation we have
\be
(\mathcal{I}_e^{\r}T)(z,\vec w) = T(I_e\.z,I_e\.\vec w),
\ee
where $I^{\mu}_e{}_{\nu} = \de^{\mu}{}_{\nu}+2e^{\mu}e_{\nu}$ is the usual reflection in time direction.

Note that \eqref{eq:CFT_vertices_definition_bar} gives a definition for $\bar v(-n',e_3,e_4)$ instead of $\bar v(n',e_3,e_4)$, which appears in the actual partial wave. This is to ensure that the two definitions \eqref{eq:CFT_vertices_definition} and \eqref{eq:CFT_vertices_definition_bar} are consistent. In particular, one can show that the partial waves are always positive in the forward limit. Furthermore, the definitions for $v(n,e_1,e_2)$ and $\bar v(-n',e_3,e_4)$ are related by CRT symmetry on the CFT side. We give more details in appendix \ref{app:partial_waves_more}, in which we also write down the relation between the partial waves coming from CFT and the Young tableaux basis of \cite{Chakraborty:2020rxf, Caron-Huot:2022jli}.

The appearance of $\mathcal{I}_e^{\r}$ in \eqref{eq:CFT_vertices_definition_bar} is due to the two-point structure $\<\tl{\cO}^{\dag}(e)\tl{\cO}((-e)^{+})\>$ in the block part of the CFT sum rule \eqref{eq:spinningfunctional_bulkpoint_withrtensor}. The effect of this two-point structure is a reflection of all the indices of $\cO$ in the time direction with some overall factor. For a representation $\r=(J,j,\tl{j})$, we find
\be\label{eq:rtensor_general_flatspace}
\left.\frac{2^d \mathrm{dim}(\r)}{\vol(\SO(d))}\frac{\<\tl{\cO}^{\dag}_a(x_5)\tl{\cO}^{\bar b}(x_6^{+})\>}{\p{\<\tl{\cO}^{\dag}\tl\cO\>,\<\cO^\dag \cO\>}}\right|_{\mathrm{saddle}} = 2^{-J+2\De}(-1)^{J-j+\tl{j}}R_\r \p{\mathcal{I}_e^{\r}}_a{}^{\bar b},
\ee
where the lower index $a$ and upper index $\bar b$ are indices in the dual $\r^{*}$ and reflected $\r^R$ representations respectively. As one can see from the left-hand side, the coefficient $R_{\r}$ depends on our convention of the two-point structure. For the two-point convention \eqref{eq:2pt_withtransverserep_convention}, we have
\be\label{eq:Rrhocoeff_definition}
R_{\r=(J,j,\tl{j})} = \frac{(j-\tl{j}+1)(J-j+1)(J-\tl{j}+2)}{(j+1)(J+1)(J+2)}.
\ee
This result can be obtained by using weight-shifting operators \cite{Karateev:2017jgd} to derive a recursion relation of $R_{\r}$. We give an example in appendix \ref{app:partial_waves_more}.

Using the momentum configuration and maps of polarizations discussed above, we can now write down the precise flat space interpretation of the block part of the CFT sum rule. Let us define
\be\label{eq:cB_definition}
&\cB_{\De,\r,(a'b')} = \frac{2^d \mathrm{dim}(\r)}{\vol(\SO(d))}\x \nn \\
&\left. \frac{\<0|[\cO_1(x_{1^{+}},z_1),\cO_{2}(x_{2^+},z_2)]\cO(x_5)|0\>_{(a')}\<\tl{\cO}^{\dag}(x_5)\tl{\cO}(x_6^{+})\>\<0|\cO^{\dag}(x_6^{+})[\cO_4(x_4,z_4),\cO_3(x_3,z_3)]|0\>_{(b')}}{\p{\<\tl{\cO}^{\dag}\tl\cO\>,\<\cO^\dag \cO\>}}\right|_{\substack{\mathrm{saddle}, \\ \eqref{eq:polarvec_125_CFTtoflat},\eqref{eq:polarvec_346_CFTtoflat}}}.
\ee
From the definitions \eqref{eq:CFT_vertices_definition}, \eqref{eq:CFT_vertices_definition_bar}, and \eqref{eq:rtensor_general_flatspace}, we find 
\be\label{eq:Blockvspartialwave}
&\frac{\cB_{\De,\r,(a'b')}}{-2(2\sin(\pi\tfrac{\tl{\tau}_{\r}-\De_1-J_1-\De_2-J_2}{2})\sin(\pi\tfrac{\tl{\tau}_{\r}-\De_3-J_3-\De_4-J_4}{2}))} \nn \\
&= 2^{-J+2\De-\De_1-\De_2-\De_3-\De_4}m^{-\De_1-\De_2-\De_3-\De_4}\nu^{2(\De_1+\De_3)}\p{m-\sqrt{m^2-\nu^2}}^{-\De_1+\De_2-\De_3+\De_4}R_{\r}\pi_{\r,(a'b')},
\ee
where $\pi_{\r,(a'b')}$ is the flat space partial wave \eqref{eq:partial_wave_vertices_definition} in the CFT three-point structure basis, and $R_{\r}$ is given by \eqref{eq:Rrhocoeff_definition}. Note that when turning the vertex $\bar v(-n',e_3,e_4)$ in \eqref{eq:CFT_vertices_definition_bar} into $\bar v(n',e_3,e_4)$ to get the partial wave, we get a $(-1)^{J-j+\tl{j}}$ factor, which exactly cancels with the same factor in \eqref{eq:rtensor_general_flatspace}.

Recall that in the scalar case, where the exchanged operators are symmetric traceless tensors, we define a heavy action $\Psi[\De,J]$ in \eqref{eq:heavyaction_scalar_definition} by rescaling the action of the functional on the block by a positive factor. Relation \eqref{eq:Blockvspartialwave} in fact suggests a way to generalize this definition to more general representation $\r$. Since the other parts of the sum rule should not know about the exchanged quantum numbers of the inserted block, the heavy action should remove the $m$- and $\r$-dependent factors that we don't expect to appear in the flat space answer, such as the $R_{\r}$ coefficient. For external operators with quantum numbers $(\De_i,J_i)$ and a general exchanged representation $\r$, we define the large $\De\sim m$ heavy action $\Psi[\De,\r]$ to be\footnote{In the scalar case \eqref{eq:heavyaction_scalar_definition}, the factor $q_{\De,J}$ in the heavy action definition comes from the OPE coefficient of Mean Field Theory. So, to find the analogous factor for more general external operators, one can consider the corresponding OPE coefficients in the Mean Field Theory. However, we do not do this in this paper, and the definition \eqref{eq:heavyaction_definition_generalrep} is good enough for our purpose.}
\be\label{eq:heavyaction_definition_generalrep}
\Psi[\De,\r]_{(a'b')} \equiv& \frac{2^{J-2\De-\De_1-\De_2-\De_3-\De_4}m^{\De_1+\De_2+\De_3+\De_4-2-\frac{d}{2}}}{R_{\r}}\nn \\
&\x\frac{\Psi[G^s_{\De,\r,(a'b')}]}{2\sin(\pi\tfrac{\tl{\tau}_{\r}-\De_1-J_1-\De_2-J_2}{2})\sin(\pi\tfrac{\tl{\tau}_{\r}-\De_3-J_3-\De_4-J_4}{2})}.
\ee

\subsection{Kernels and the shock frame}\label{sec:kernel_shock}

We now consider the kernel part of our spinning sum rules. It is given by evaluating a celestial structure $\<\tl{\cP}_1\tl{\cP}_3\cP_{\de,j'}\>$ and a dual structure $\<0|\cO_4\wL[\cO]\cO_2|0\>^{-1}$ at the scattering-crystal. In the heavy action formula, we first evaluate these structures at the scattering crystal \eqref{eq:scattering_crystal}, and then we can use the maps of CFT and flat space polarizations \eqref{eq:polarvec_125_CFTtoflat}, \eqref{eq:polarvec_346_CFTtoflat} to get the corresponding flat space answer. We claim that the resulting expression can be seen to correspond to a shockwave amplitude in flat space. The shockwave amplitude is defined as the amplitude of an elastic scattering of a probe particle $X$ and two shockwave gravitons $g^*$ \cite{Kologlu:2019bco},
\be
g^*(p_1)X(p_2) \to g^*(p_3)X(p_4).
\ee
In our case, we will take the probe particle $X$ to be a graviton. The two shockwave gravitons are dual to the two light transformed operators $\wL[\cO_1],\wL[\cO_3]$ in the CFT functional. A convenient choice of momenta for shockwave amplitudes is given by (in all incoming convention)
\be\label{eq:shock_frame_momenta}
p_1^{\mu} &= p_{\bar s}\bar s^{\mu} + q_1^{\mu},\quad ~~ p_2^{\mu} = p_s s^\mu - q_3^{\mu}\nn \\
p_3^{\mu} &= -p_{\bar s}\bar s^{\mu} + q_3^{\mu}, \quad p_4^{\mu}= -p_s s^{\mu} - q_1^{\mu},
\ee
where $s^{\mu}$ and $\bar s^{\mu}$ are two null directions, and the polarization of the shock gravitons should be in the $\bar s^{\mu}$ direction. The transverse momenta $q_1^{\mu},q_3^{\mu}$ are perpendicular to the null directions, and in order to make the external momenta on-shell, we must have $q_1^2=q_3^2=0$. We will call the momentum configuration \eqref{eq:shock_frame_momenta} the shock frame. More explicitly, let $D=d+1$ be the spacetime dimension of the bulk, in the shock frame we have two null vectors $s^{\mu},\bar s^{\mu}$ and $D-2$ transverse unit vectors $v_{1;\perp}^{\mu},v_{2;\perp}^{\mu},\ldots, v_{D-2;\perp}^{\mu}$ satisfying
\be
s^2=\bar s^2=s\.v_{i;\perp}=\bar s\.v_{i;\perp}=0,\quad s\.\bar s =-2,\quad v_{i;\perp}\.v_{j;\perp} = \de_{ij}.
\ee

In the previous subsection, we have shown that in order to match the blocks to flat space partial waves, we should choose the flat space momentum to be \eqref{eq:momenta_for_partialwaves}. The shock frame momenta \eqref{eq:shock_frame_momenta} should also agree with these center of mass frame momenta \eqref{eq:momenta_for_partialwaves}. By relating the two frames, we obtain that in terms of the bulk Minkowski coordinates used in \eqref{eq:momenta_for_partialwaves}, the vectors in the shock frame can be written as
\be\label{eq:shock_frame_vecs_inCMframe}
\bar s^{\mu} &= \p{\tfrac{\sqrt{m^2-\nu^2}}{m},1,0,\tfrac{i\nu}{m},0,\ldots,0} \nn \\
s^{\mu} &= \p{\tfrac{\sqrt{m^2-\nu^2}}{m},-1,0,\tfrac{i\nu}{m},0,\ldots,0} \nn \\
v^{\mu}_{1;\perp} &= \p{\tfrac{i\nu}{m},0,0,\tfrac{\sqrt{m^2-\nu^2}}{m},0,\ldots} \nn \\
v^{\mu}_{2;\perp} &= (0,0,1,0,0,\ldots,0) \nn \\
v^{\mu}_{3;\perp} &= (0,0,0,0,1,0,\ldots,0) \nn \\
&\vdots \nn \\
v^{\mu}_{D-2;\perp} &= (0,0,0,0,0,0,\ldots,1),
\ee
and
\be\label{eq:shock_momenta_inbasisvec}
p_{\bar s} =p_{s} = \frac{\sqrt{m^2-\nu^2}}{2},\quad q_1^{\mu}=-\frac{i\nu}{2}v^{\mu}_{1;\perp} + \frac{\nu}{2}v^{\mu}_{2;\perp},\quad q_3^{\mu}= \frac{i\nu}{2}v^{\mu}_{1;\perp} + \frac{\nu}{2}v^{\mu}_{2;\perp}.
\ee

One motivation for relating the CFT kernel to the shock frame is that we set the CFT polarization vectors $z_1,z_3$ to be $x_{1^{+}0},x_{30}$ in the CFT sum rule \eqref{eq:spinningfunctional_bulkpoint_withrtensor}. Using the map of CFT and flat space polarizations \eqref{eq:polarvec_125_CFTtoflat},\eqref{eq:polarvec_346_CFTtoflat} derived in the previous subsection, we find that in flat space the two polarizations become (up to $e_i\sim e_i+\a p_i$ gauge redundancy)
\be
x_{1^{+}0}|_{\eqref{eq:polarvec_125_CFTtoflat}} \sim i\frac{m(m-\sqrt{m^2-\nu^2})}{\nu^2} \bar s^{\mu},\qquad x_{30}|_{\eqref{eq:polarvec_346_CFTtoflat}} \sim i\frac{m(m-\sqrt{m^2-\nu^2})}{\nu^2} \bar s^{\mu},
\ee
where $\bar s^{\mu}$ is given by \eqref{eq:shock_frame_vecs_inCMframe}. We see that the polarizations of particles $1$ and $3$ should be set to the same null direction $\bar s^{\mu}$, which agrees with the fact that the two shock gravitons in the shockwave amplitude should have the same polarizations along the direction of the shock.

We can now discuss the CFT kernel in the shock frame. Explicitly, evaluating the celestial three-point structure at the saddle configuration gives
\be\label{eq:celestial_3pt_saddle}
&\left.\<\tl{\cP}^{\dag}_{\de_1}(x_{1^{+}0})\tl{\cP}^{\dag}_{\de_3}(x_{30})\cP_{\de,j'}(z_0,w_0)\>\right|_{\mathrm{saddle}} \nn \\
&= (-1)^{j'}(-i)^{-\de}2^{j'+\de-\tl{\de}_1-\tl{\de}_3}\nu^{\tl{\de}_1+\tl{\de}_3}(m-\sqrt{m^2-\nu^2})^{-\tl{\de}_1-\tl{\de}_3}(\vec n_0\.\vec w_{0\perp})^{j'},
\ee
where we have introduced a polarization $w_0=(0,0,\vec w_{0\perp})$ for the transverse indices of $\cP_{\de,j'}$, and $\vec n_0=\vec y_0/|\vec y_0|$ is the unit vector in the transverse $\vec y_0$ direction in the scattering crystal \eqref{eq:scattering_crystal}. 

For each transverse spin $j'$, the celestial three-point function has a unique structure which gives \eqref{eq:celestial_3pt_saddle} at the saddle. On the other hand, the dual structure $\<0|\cO_4\wL[\cO]\cO_2|0\>^{-1}$ can have multiple tensor structures. Moreover, they are subject to constraints from imposing conservation on $\cO_2$ and $\cO_4$. Interestingly, the space of allowed structures becomes easier to study when we just focus on the saddle configuration. This fact can be seen from the identity
\be\label{eq:kernel_identity_saddle}
 -\left.V_{2,40}V_{0,24}\right|_{\mathrm{saddle},z_{i}=(0,z_i^x,\vec z_{i\perp})}= \left.H_{20}\right|_{\mathrm{saddle},z_{i}=(0,z_i^x,\vec z_{i\perp})},
\ee
and a similar identity for $V_{4,02}$ and $H_{40}$. Note that we also set the time component of the polarizations to zero, which is due to the large-$\De$ conservation analysis in section \ref{sec:partial_waves}. This identity allows us to just remove all $V_{2,40}, V_{4,02}$ structures when writing down the basis for the spinning dual structures. Additionally, it can be shown that the basis constructed without $V_{2,40}, V_{4,02}$ gives the correct number of independent tensor structures of continuous-spin three-point function of two conserved operators, which should be given by \cite{Kravchuk:2016qvl, Kologlu:2019mfz}
\be\label{eq:spinning_dual_counting}
\p{\l'\otimes \mathrm{Res}^{\SO(d-1)}_{\SO(d-2)}\r_2\otimes \r_4}^{\SO(d-2)}.
\ee
We give the argument for this statement in appendix \ref{app:dual_large_nu}, and also explain how to compute these spinning dual structures.

As an example, let us consider $j'=0$. Focusing just on the spinning part of $\cO_2,\cO_4$, we see that the allowed structures should be
\be
H_{20}^{J_2-n}H_{40}^{J_4-n}H_{24}^n,
\ee
where $n=0,1,\ldots,\min(J_2,J_4)$. The number of structures is $\min(J_2,J_4)+1$, which agrees with \eqref{eq:spinning_dual_counting}.

We now study the four-photon example in detail, where all external operators should be conserved spin-$1$ currents. The selection rule \eqref{eq:lambdaprime_selectionrule} implies that the allowed values of transverse spin should be $j'=0,1,2$. Using the algorithm explained in appendix \ref{app:dual_large_nu}, we obtain that the dual structures for $j'=0$ in the large $\nu$ limit are given by (for $4>0>2^{-}$)
\be\label{eq:JJO_kernel_dual}
&\left\{\p{\<0|J(X_4,Z_4)\wL[\cO](X_0,Z_0)J(X_{2},Z_{2})|0\>^{(1)}}^{-1},\p{\<0|J(X_4,Z_4)\wL[\cO](X_0,Z_0)J(X_{2},Z_{2})|0\>^{(2)}}^{-1}\right\} \nn \\
&= 2^{\frac{3d-5}{2}-i\nu}e^{-\frac{i\pi}{4}}\sqrt{\frac{\nu}{\pi}}\vol\SO(d-2)\frac{1}{X_{24}^{\frac{4-\De_F-J_F}{2}}X_{02}^{\frac{\De_F+J_F}{2}}(-X_{04})^{\frac{\De_F+J_F}{2}}} \nn \\
&\x\left\{\frac{16(d-1)}{d-2}(-2V_{0,42})^{J_F-2}H_{20}H_{40}+\frac{4}{d-2}(-2V_{0,42})^{J_F}H_{24},\right.\nn \\
&\qquad\quad\left.\frac{4}{d-2}(-2V_{0,42})^{J_F-2}H_{20}H_{40}+\frac{1}{d-2}(-2V_{0,42})^{J_F}H_{24}\right\} + \ldots,
\ee
where $\De_F=d,J_F=\frac{2-d}{2}+i\nu$, and $\ldots$ are subleading terms at large $\nu$. The first structure is dual to the light-transformed structure $\<0|J\wL[\cO]J|0\>$ with $H_{20}H_{40}$ (see \eqref{eq:light_transform_largeDelta_spinning}), and the second structure is dual to the one with $H_{24}$.

Then, the next step is to evaluate \eqref{eq:JJO_kernel_dual} at the saddle (after applying a $\cT_2$ to \eqref{eq:JJO_kernel_dual} to make the causality configuration agree with the saddle). This will give us an expression that depends on the external polarizations $z_2,z_4$ (note that $z_0$ is fixed in \eqref{eq:scattering_crystal}). For example, for the $H_{24}$ structure, we have
\be
\left.\frac{H_{24}}{X_{24}}\right|_{\mathrm{saddle},z_{i}=(0,z_i^x,\vec z_{i\perp})} = -z_2^xz_4^x+\vec z_{2\perp}\.\vec z_{4\perp}.
\ee
Using \eqref{eq:polarvec_125_CFTtoflat} and \eqref{eq:polarvec_346_CFTtoflat} to map this to flat space, we find
\be
-z_2^xz_4^x+\vec z_{2\perp}\.\vec z_{4\perp} \to -e_2^{n}e_4^{n'} + \vec e_{2\perp}\.\vec e_{4\perp} = e_{2\mu_2}e_{4\mu_4}\p{-n_\perp^{\mu_2}n_\perp^{\prime\mu_4}+\de^{\mu_2\mu_4}_{\perp}}.
\ee
Our goal is to rewrite this expression in the shock frame \eqref{eq:shock_frame_vecs_inCMframe}. Because of the gauge redundancy, we can shift everything that is contracted with $e_2$ by $p_2$. Furthermore, thanks to the momentum configuration we choose in section \ref{sec:partial_waves}, we have $e_2\.p_1=0$, so we can also shift by $p_1$. Similarly, we can shift things that are contracted with $e_4$ by $p_3,p_4$. Using this, we can make a gauge choice to remove all the $s^{\mu}, \bar s^{\mu}$, and we find that the $H_{24}$ structure mapped to the shock frame can be written as
\be\label{eq:H24_kernel_toflat}
\left.\frac{H_{24}}{X_{24}}\right|_{\mathrm{saddle},\eqref{eq:polarvec_125_CFTtoflat},\eqref{eq:polarvec_346_CFTtoflat}} =e_{2\mu_2}e_{4\mu_4}\p{\de^{\mu_2\mu_4}_{\perp;D-4}-\frac{1}{\nu^2(m^2-\nu^2)^2}(\cV_2^{\mu_2}\cV_4^{\mu_4}+\cW_2^{\mu_2}\cW_4^{\mu_4})},
\ee
where we define
\be\label{eq:kernel_VWvec_definition}
\cV_2^{\mu_2} = m^2 q_1^{\mu_2} + (m^2-\nu^2) q_3^{\mu_2}, \nn \\
\cV_4^{\mu_4} = (m^2-\nu^2) q_1^{\mu_4} + m^2  q_3^{\mu_4}, \nn \\
\cW_2^{\mu_2} = m^2 q_1^{\mu_2} - (m^2-\nu^2) q_3^{\mu_2}, \nn \\
\cW_4^{\mu_4} = (m^2-\nu^2) q_1^{\mu_4} - m^2  q_3^{\mu_4}.
\ee
The momenta $q_1^{\mu},q_3^{\mu}$ are given by \eqref{eq:shock_frame_vecs_inCMframe}, \eqref{eq:shock_momenta_inbasisvec}, and $\de^{\mu_2\mu_4}_{\perp;D-4}$ is the metric in the $(D-4)$-dimensional transverse space spanned by $v^{\mu}_{3;\perp},\ldots,v^{\mu}_{D-2;\perp}$. Similarly, the structure with $H_{20}H_{40}$ becomes
\be\label{eq:H20H40_kernel_toflat}
\left.\frac{H_{20}H_{40}}{X_{20}X_{40}}\right|_{\mathrm{saddle},\eqref{eq:polarvec_125_CFTtoflat},\eqref{eq:polarvec_346_CFTtoflat}} =\frac{1}{4\nu^2(m^2-\nu^2)}e_{2\mu_2}e_{4\mu_4}\cV_2^{\mu_2}\cV_4^{\mu_4}.
\ee
Using \eqref{eq:JJO_kernel_dual}, \eqref{eq:H24_kernel_toflat}, and \eqref{eq:H20H40_kernel_toflat}, we can then obtain the flat space kernels for the two $j'=0$ dual structures. 

For nonzero $j'$, we also need to contract the transverse indices carried by $\cO$ in the dual structure and $\cP_{\de,j'}$ in the celestial structure. Let us define
\be\label{eq:cK_definition}
\cK^{4\g, (a)}_{\nu,j'} = \left.\p{\<0|J(x_4,z_4)\wL[\cO_{j'}](x_0,z_0)J(x_{2^+},z_2)|0\>^{(a)}}^{-1}\<\tl{\cP}^{\dag}_{\de_J}(x_{1^{+}0})\tl{\cP}^{\dag}_{\de_J}(x_{30})\cP_{\de,j'}(z_0)\>\right|_{\substack{\mathrm{saddle},\eqref{eq:polarvec_125_CFTtoflat},\eqref{eq:polarvec_346_CFTtoflat} \\  \nu \to \oo}},
\ee
where $\nu \to \oo$ indicates that we take the leading term at large $\nu$. The results for the four-photon kernels are given by
\be\label{eq:JJO_kernels_flatspace}
\cK^{4\g, (1)}_{\nu,j'=0}&=\frac{K^{4\g}_{\nu}}{2}e_{2\mu_2}e_{4\mu_4}\p{(\cV_2^{\mu_2}\cV_4^{\mu_4}+\cW_2^{\mu_2}\cW_4^{\mu_4}-\nu^2(m^2-\nu^2)^2\de^{\mu_2\mu_4}_{\perp;D-4})-(D-2)\cV_2^{\mu_2}\cV_4^{\mu_4}}, \nn \\
\cK^{4\g, (2)}_{\nu,j'=0}&= \frac{K^{4\g}_{\nu}}{8}e_{2\mu_2}e_{4\mu_4}\p{\cW_2^{\mu_2}\cW_4^{\mu_4}-\nu^2(m^2-\nu^2)^2\de^{\mu_2\mu_4}_{\perp;D-4}},  \nn \\
\cK^{4\g, (1)}_{\nu,j'=1}&=K^{4\g}_{\nu}e_{2\mu_2}e_{4\mu_4}\cW_2^{\mu_2}\cV_4^{\mu_4}, \nn \\
\cK^{4\g}_{\nu,j'=2}&=-\frac{4K^{4\g}_{\nu}}{(D-1)}e_{2\mu_2}e_{4\mu_4}\p{\cW_2^{\mu_2}\cW_4^{\mu_4}+\frac{\nu^2(m^2-\nu^2)^2}{D-4}\de^{\mu_2\mu_4}_{\perp;D-4}},
\ee
where
\be
K^{4\g}_{\nu} = \frac{2^{-\frac{d}{2}+i\nu}e^{-\pi \nu}m^{-d}\nu^{-4+3d}(m^2-\nu^2)^{-2-\frac{d}{2}}(m-\sqrt{m^2-\nu^2})^{2-d}}{d-2}.
\ee
Note that for $j'=1$ there are two tensor structures (see \eqref{eq:3pt_convention_spinning}). However, they are just related by $2\leftrightarrow 4$, and as we will see later, kernels that are related by $2\leftrightarrow 4$ will give the same sum rule, so we just include one of the structures here.

Finally, let us comment on the symmetry properties of the kernel. On the CFT side, the celestial three-point structure has a stabilizer group $\SO(d-3)$, which is also the stabilizer group of the scattering crystal \eqref{eq:scattering_crystal}. In the shock frame, this group becomes the $\SO(D-4)$ that fixes the shockwave amplitude momenta \eqref{eq:shock_frame_momenta}, and the group only rotates in the $v_{3;\perp}^{\mu},\ldots,v_{D-2;\perp}^{\mu}$ directions. On the other hand, for the continuous-spin dual structure we can choose a conformal frame to fix all three points to be on the same line, and fix the continuous-spin polarization vector in a null direction. The configuration should be fixed by an $\SO(d-2)$, and the transverse spin $j'$ is a representation of this group. In flat space, this should correspond to an $\SO(D-3)$. From the expressions of the kernels \eqref{eq:JJO_kernels_flatspace}, it is not clear what this $\SO(D-3)$ should be. It turns out that (at least in the Regge limit) the group can be understood as the rotation group that fixes the $s^{\mu},\bar s^{\mu},v_{2;\perp}^{\mu}$ shock frame vectors. To see this we will first need to study how to apply the kernels \eqref{eq:JJO_kernels_flatspace} to four-photon amplitudes and write down the full sum rules for photons. We will therefore explain this $\SO(D-3)$ group and the meaning of transverse spin in the bulk in section \ref{sec:transverse_spin}.

\subsection{Review: flat space sum rules for photons}\label{sec:review_photon_flatspace}

After understanding how to map each part of the CFT bulk-point limit formula \eqref{eq:spinningfunctional_bulkpoint_withrtensor} to flat space, we can now derive the dictionary between CFT and flat space sum rules for photons and gravitons. Let us first review flat space sum rules for photons \cite{simon-notes,Caron-Huot:2022jli}. The graviton sum rules will be considered in section \ref{sec:graviton_sum_rules}. 

The four-photon amplitude $\cM_{4\g}(p_i,\e_i)$ is a function of photon momenta $p_i$ and polarizations $\e_i$ for $i=1,2,3,4$. They should satisfy $p_i^2=p_i\.\e_i=\e_i^2=0$ and $\e_i \sim \e_i + \#p_i$. In general, the amplitude can be written as a sum of Lorentz-invariant polynomials of $p_i,\e_i$ times functions of Mandelstam variables defined in \eqref{eq:Mandelstams_definition}. Depending on the choice of the polynomials, sometimes the functions of Mandelstam variables can develop spurious poles. It turns out that there exists a ``local module" such that the amplitude won't have any spurious poles \cite{Chowdhury:2019kaq}. Namely, an amplitude that is a polynomial can always be written as generators of the local module times polynomials of Mandelstam variables.

To describe the generators of the local module, let us introduce some basic building blocks,\footnote{We are using $V$ and $H$ for the building blocks of both CFT tensor structures and flat space amplitudes. We hope the meaning of the notations will be clear from the context and this won't cause too much confusion.}
\be
H_{ij} &= F^{\mu}_{i\nu}F^{\nu}_{j\mu},\qquad\quad\quad H_{ijk}= F^{\mu}_{i\nu}F^{\nu}_{j\s}F^{\s}_{k\mu},\nn \\
H_{ijkl} &= F^{\mu}_{i\nu}F^{\nu}_{j\s}F^{\s}_{k\r}F^{\r}_{l\mu},\quad V_{i,jk} = p_{k\mu}F^{\mu}_{i\nu}p_j^{\nu},
\ee
where $F^{\mu}_{i\nu}= p_i^{\mu}\e_{i\nu} - \e_i^{\mu}p_{i\nu}$. We also define
\be
X_{ijkl} &= H_{ijkl} - \frac{1}{4}H_{ij}H_{kl} - \frac{1}{4}H_{ik}H_{jl} - \frac{1}{4}H_{il}H_{jk}, \nn \\
S &= V_{1,24}H_{234} + V_{2,31}H_{341} + V_{3,42}H_{412} + V_{4,13}H_{123}.
\ee
Then, the four-photon amplitude can be written as
\be\label{eq:4photon_amp_expr0}
\cM_{4\g}(s,u) = \p{H_{14}H_{23}\cM_{4\g}^{(1)}(s,u) + X_{1243}\cM_{4\g}^{(2)}(s,u) + \mathrm{permutations}} + S\cM_{4\g}^{(3)}(s,t),
\ee
where $\cM_{4\g}^{(1)},\cM_{4\g}^{(2)}$ are symmetric in their two arguments, and $\cM_{4\g}^{(3)}$ is symmetric under all permutations of $s,t,u$.

To write down dispersion relations for $\cM_{4\g}(s,u)$, we need to study how the polarization structures behave in the fixed-$u$ Regge limit, where we take $s\to \oo$. This will tell us how many subtractions we should have in the dispersion relations. Moreover, for fixed-$u$ sum rules it is useful to separately consider structures that are symmetric and antisymmetric under $s\leftrightarrow t$. The polarization structures in the Regge limit scale as \cite{Caron-Huot:2022jli}
\be\label{eq:amp_structure_Regge}
H_{12},H_{14} \sim s,\quad H_{13} \sim s^{0},\quad X_{1234}, X_{1243}-X_{1324} \sim s,\quad X_{1243}+X_{1324} \sim s^2,\quad S \sim s^2, 
\ee
where other structures can be obtained by permutations.

The amplitude itself $\cM_{4\g}$ satisfies boundedness and analyticity properties due to unitarity and causality \cite{Caron-Huot:2021rmr, Caron-Huot:2021enk, Caron-Huot:2022ugt, Haring:2022cyf}. As shown by \cite{Haring:2022cyf}, for a suitably smeared amplitude,
\be
\cM_{\Psi}(s) \equiv \int_0^{M} dp \Psi(p) \cM(s,-p^2),
\ee
along any complex direction of $s$ it should satisfy
\be\label{eq:amp_Regge_bound}
|\cM_{\Psi}(s)|_{s \to \oo} \leq s\times \mathrm{constant}.
\ee

The Regge behaviors \eqref{eq:amp_structure_Regge}, \eqref{eq:amp_Regge_bound} then combine to give boundedness conditions on each coefficient function $\cM^{(i)}$. For example, since $X_{1243}-X_{1324} \sim s$ in the Regge limit, $\cM^{(2)}(s,u) - \cM^{(2)}(t,u)$ can't grow faster than $s^0$ in the Regge limit. Performing this analysis for all the structures, we can then determine the subtractions needed for the dispersion relation of each coefficient $\cM^{(i)}$. The complete list of flat space fixed-$u$ sum rules for photons is given by
\be\label{eq:photon_flatspace_sumrules}
\left\{\cC^{+(1,2,3,4,5)}_{4\g;k,u}\right\} &=-\oint_{\oo}\frac{ds}{4\pi i}\frac{1}{(-st)^{\frac{k}{2}}}\left\{(s-t)\cM_{4\g}^{(1,2)+}(s,u),(s-t)\p{\tfrac{-u}{2}}\cM_{4\g}^{(3)}(s,t),\right. \nn \\
&\qquad\qquad\qquad\qquad\qquad \left.\cM^{(2)-}_{4\g}(s,u),\tfrac{(s-t)}{(-st)}\cM^{(1)}_{4\g}(s,t)\right\},\quad k=2,4,\ldots, \nn \\
\left\{\cC^{-(1,2)}_{4\g;k,u}\right\} &= -\oint_{\oo}\frac{ds}{4\pi i}\frac{1}{(-st)^{\frac{k-1}{2}}}\left\{\cM^{(1)-}_{4\g}(s,u),\tfrac{(s-t)}{(-st)}\cM^{(2)}_{4\g}(s,t)\right\},\quad\ \  k=3,5,\ldots,
\ee
where we have used the notation $\cM^{(i)\pm}(s,u) = \cM^{(i)}(s,u) \pm \cM^{(i)}(t,u)$. For later convenience, we introduce an additional $\frac{-u}{2}$ factor for $\cM_{4\g}^{(3)}$. This would also make all the sum rules have the same power counting since the $S$ structure has one more $p_i\.p_j$ compared to the other structures. The parameter $k$ is the Regge spin of the sum rule. A sum rule with Regge spin $k$ would be convergent if the amplitude $\cM$ satisfies $\cM/s^k\to 0$ in the Regge limit.

At low energy, the amplitude can be computed by an EFT with cutoff $M$. To derive bounds from the dispersion relations, one can deform the contour in \eqref{eq:photon_flatspace_sumrules} and separate the contributions from high energy $s\geq M^2$ and low energy $s< M^2$. Our main focus is the high energy part, where one decomposes the amplitude into $\SO(d)$ partial waves and imposes unitarity. 

When applying the dispersion relations \eqref{eq:photon_flatspace_sumrules} to the partial wave expansion \eqref{eq:amp_partialwave_expansion}, for each sum rule $\cC$ we can write its heavy contribution as
\be\label{eq:sumrules_partialwave_actions}
\left\<\cC[m^2,\r]_{(ab)}\right\>,
\ee
where $\<\cdots\>$ is a heavy average defined as
\be\label{eq:heavyaverage_flatspace}
\<\cdots\> = \frac{1}{\pi} \sum_{\r} n_{\r}^{(D)} \sum_{ab}\int_{M^2}^{\oo}  \frac{dm^2}{m^2}m^{4-D}  \Im(a_{\r}(s))_{ab} \p{\cdots}.
\ee
For example, for the $\cC^{+(1)}_{4\g;k,u}$ sum rule in \eqref{eq:photon_flatspace_sumrules}, we have
\be\label{eq:flat_space_heavy_contribution_example}
\cC^{+(1)}_{4\g;k,u}[m^2,\r]_{(ab)} = \frac{2m^2+u}{m^2+u}\frac{1}{[m^2(m^2+u)]^{\frac{k-2}{2}}} \pi^{(1)}_{\r,(ab)}(m^2,u),
\ee
where we have decomposed the partial wave into different polarization structures similar to \eqref{eq:4photon_amp_expr0}, so $\pi^{(1)}_{\r,(ab)}$ is the $H_{14}H_{23}$ component of the partial wave.

This dictionary we obtain below in sections \ref{sec:sumrules_forphotons} and \ref{sec:graviton_sum_rules} will be a relation between the action of CFT sum rules on conformal blocks in the bulk-point limit \eqref{eq:spinningfunctional_bulkpoint_withrtensor} and the action of flat space sum rules on partial waves \eqref{eq:sumrules_partialwave_actions}. This will allow us to find CFT sum rules with positive action on conformal blocks with large $\De$ straightforwardly from the flat space result.

\subsection{From CFT to flat space for photons}\label{sec:sumrules_forphotons}
Now, we combine the discussion about the block/partial wave in section \ref{sec:partial_waves} and the discussion about the kernel/shockwave amplitude in section \ref{sec:kernel_shock} to write down the relation between four-photon flat space sum rules and the bulk-point limit of our CFT sum rules. 

When combining the two parts, we first have to understand how to contract the indices of the external operators. Recall that in the CFT functional, the polarizations $z_1,z_3$ are set to be $z_1=x_{1^{+}0},z_3=x_{30}$. As discussed in \ref{sec:kernel_shock}, this implies that in the bulk, we should set $e_1^{\mu}=e_3^{\mu}=i\frac{m(m-\sqrt{m^2-\nu^2})}{\nu^2} \bar s^{\mu}$. For 2 and 4, we have to contract the indices of the kernel and the block. Note that we have chosen the CFT polarization vectors to have no time component. In the bulk, this means that $e_2$ has no time and $n^{\mu}$ components (see \eqref{eq:bulk_polarization_nperp}), and similarly $e_4^{\mu}$ has no time and $n^{\prime\mu}$ components. Let us consider the following index contraction in the CFT sum rule:
\be\label{eq:contraction_ex_CFT}
(z_2^x K^x + \vec z_{2\perp}\.\vec K_{\perp})\odot(z_2^x G^x + \vec z_{2\perp}\.\vec G_{\perp}),
\ee
where $K^x,\vec K_{\perp}$ are from the kernel and $G^x,\vec G_{\perp}$ are from the block, and the symbol $\odot$ represents the index contraction. For spin-1, it is simply stripping off the $z_2$'s and contracting both sides, and we see that after the index contraction, we should get $K^x G^x+\vec K_{\perp}\.\vec G_{\perp}$. On the other hand, mapping \eqref{eq:contraction_ex_CFT} to the bulk, we have
\be
&(-e_2^n K^x + \vec e_{2\perp}\.\vec K_{\perp})\odot(-e_2^n G^x + \vec e_{2\perp}\.\vec G_{\perp}) \nn \\
=&e_{2\mu_2}(-n_{\perp}^{\mu_2}K^x + \de^{\mu_2 i}_{\perp} \vec K_{\perp i})\odot e_{2\r_2}(-n_{\perp}^{\r_2}G^x + \de^{\r_2 i}_{\perp} \vec G_{\perp i}) \nn \\
=&e_{2\mu_2} K^{\mu_2}\odot e_{2\r_2} G^{\r_2},
\ee
where $K^{\mu_2}$ is $-n_{\perp}^{\mu_2}K^x + \de^{\mu_2 i}_{\perp} \vec K_{\perp i}$ up to a shift by $p_1^{\mu_2}, p_2^{\mu_2}$ and similarly for $G^{\r_2}$. For example, in the $j'=2$ kernel given by \eqref{eq:JJO_kernels_flatspace} we have $K^{\mu_2} \propto m^2q_1^{\mu_2}-(m^2-\nu^2)q_3^{\mu_2}$. To correctly reproduce $K^x G^x+\vec K_{\perp}\.\vec G_{\perp}$, we should perform a polarization sum in $e_2$ using\footnote{For a general polarization $\e_2$, $p_1$ can be replaced with any null vector $q$. However, here we have chosen the polarization to satisfy the condition $e_2\.p_1=e_2\.p_2=0$, so the correct choice here is $q=p_1$.}
\be
e_{2\mu_2}e_{2\r_2} \to \eta_{\mu_2\r_2}-\frac{p_{1\mu_2}p_{2\r_2}+p_{1\r_2}p_{2\mu_2}}{p_1\.p_2}.
\ee
Then, we obtain
\be
K^{\mu_2} G^{\r_2}\p{\eta_{\mu_2\r_2}-\frac{p_{1\mu_2}p_{2\r_2}+p_{1\r_2}p_{2\mu_2}}{p_1\.p_2}} = K^x G^x+\vec K_{\perp}\.\vec G_{\perp}
\ee

In summary, for the flat space sum rules we should set the polarizations $e_1,e_3$ to be proportional to $\bar s$, and perform polarization sums for $e_2,e_4$. More precisely,
\be\label{eq:sumrule_bulk_polarizationmapall}
e_{1\mu}&=e_{3\mu}=i\frac{m(m-\sqrt{m^2-\nu^2})}{\nu^2} \bar s_{\mu}, \nn \\
e_{2\mu_2}e_{2\r_2} &\to \eta_{\mu_2\r_2}-\frac{p_{1\mu_2}p_{2\r_2}+p_{1\r_2}p_{2\mu_2}}{p_1\.p_2},\quad e_{4\mu_4}e_{4\r_4} \to \eta_{\mu_4\r_4}-\frac{p_{3\mu_4}p_{4\r_4}+p_{3\r_4}p_{4\mu_4}}{p_3\.p_4}.
\ee

To study how the spinning CFT sum rules are mapped to flat space, we first start with the heavy action formula \eqref{eq:spinningfunctional_bulkpoint_withrtensor}. Then, we further decompose the conformal block $G^{s}_{\De,\r}$ into independent four-point tensor structures. Namely, we want to consider
\be
\Psi^{+(a)}_{k,\nu,j'}[G^s_{\De,\r,(a'b'),I}Q^I],
\ee
where $Q^I$ is a basis of CFT four-point structures. We will choose the CFT four-point structures such that under the polarization maps \eqref{eq:polarvec_125_CFTtoflat},\eqref{eq:polarvec_346_CFTtoflat}, they get mapped to the generators of the local module of the four-photon amplitude. Namely, the four-point structures satisfy
\be\label{eq:CFT_4pt_toflatspace}
\left.Q^{I}\right|_{\mathrm{saddle},\eqref{eq:polarvec_125_CFTtoflat},\eqref{eq:polarvec_346_CFTtoflat}} = \frac{16}{\nu^4}\left\{H_{14}H_{23},H_{13}H_{24},H_{12}H_{34},X_{1243},X_{1234},X_{1324},\frac{2}{\nu^2}S\right\},
\ee
where the structures on the right hand side is the basis of polarization structures of the four-photon amplitude introduced in section \ref{sec:review_photon_flatspace}. We give the explicit expressions of $Q^{I}$'s in appendix \ref{app:4pt_structures}.

We can now apply the polarization map \eqref{eq:polarvec_125_CFTtoflat},\eqref{eq:polarvec_346_CFTtoflat} to the right hand side of the heavy action formula \eqref{eq:spinningfunctional_bulkpoint_withrtensor}. From the previous discussions, we know that the kernel part should become $\cK^{(a)}_{\nu,j'}$ defined in \eqref{eq:cK_definition} and it acts on the four-point tensor structures $Q^I$'s through \eqref{eq:sumrule_bulk_polarizationmapall}. For the block part, by \eqref{eq:cB_definition}, \eqref{eq:Blockvspartialwave}, it should turn into four-photon partial waves which are further decomposed in four-point structures. In summary, we have
\be
&\lim_{\substack{\nu,m\gg 1 \\ \nu<m}} \frac{\Psi^{+(a)}_{k,\nu,j'}[G^s_{\De,\r,(a'b')}]}{-2(2\sin^2(\pi\tfrac{\tl{\tau}_{\r}-2d}{2}))} \nn \\
&=\frac{2^{\frac{5}{2}+2d}\pi^{d-\frac{1}{2}}e^{i\frac{\pi}{4}}m^{\frac{3d}{2}+2}\nu^{\frac{7}{2}-4d}(m^2-\nu^2)^{\frac{d-2}{2}}(m-\sqrt{m^2-\nu^2})^{d-4}}{\G\p{\frac{d-2}{2}}\vol(\SO(d-2))} \nn \\
&\x \cK^{(a)}_{\nu,j'}[Q^I] \nn \\
&\x 2^{-J+2m-4d+4}m^{-4d+4}\nu^{4d-4}R_{\r}\pi_{\r,(a'b')}^{I}.
\ee

Our remaining task is to evaluate the above expression explicitly. We perform the polarization sum of $\cK^{(a)}_{\nu,j'}[Q^I]$ using \eqref{eq:JJO_kernels_flatspace}, \eqref{eq:sumrule_bulk_polarizationmapall}, and also insert the subtraction factors $f_{k;J_e}^{(-1)^{j'}}$ given by \eqref{eq:subtraction_factor_evenjprime}, \eqref{eq:subtraction_factor_oddjprime} with $J_e=1$. Finally, by comparing the result with the heavy state contribution of the four-photon flat space sum rules \eqref{eq:photon_flatspace_sumrules}, we obtain the dictionary between the four-photon CFT and flat space sum rules.
\be\label{eq:final_dictionary_photons}
\lim_{\substack{\De,\nu \gg 1 \\ \nu< m}}\bpm \Psi^{(1)}_{4\g;k,\nu,j'=0;f^{+}_{k;J_e=1}}[\De,\r] \\  \Psi^{(2)}_{4\g;k,\nu,j'=0;f^{+}_{k;J_e=1}}[\De,\r] \\  \Psi_{4\g;k,\nu,j'=2;f^{+}_{k;J_e=1}}[\De,\r] \epm =&\frac{1}{A^{4\g}_{k,\nu}}M^{+}_{4\g} \bpm \cC^{+(1)}_{4\g;k,-\nu^2}[m^2,\r] \\ \cC^{+(2)}_{4\g;k,-\nu^2}[m^2,\r] \\ \cC^{+(3)}_{4\g;k,-\nu^2}[m^2,\r] \epm, \quad k=2,4,=\ldots, \nn \\
\lim_{\substack{\De,\nu \gg 1 \\ \nu< m}} \bpm \Psi^{(1)}_{4\g;k,\nu,j'=1;f^{-}_{k;J_e=1}}[\De,\r] \epm =& \frac{1}{A^{4\g}_{k,\nu}}M^{-}_{4\g} \bpm \cC^{-(1)}_{4\g;k,-\nu^2}[m^2,\r] \epm,\quad k=3,5,\ldots,
\ee
where $A^{4\g}_{k,\nu}$ is
\be
A^{4\g}_{k,\nu} = 2^{5d-3-2j'}e^{\pi\nu}\pi^{1-d}\nu^{6-3d-2k}\G(\tfrac{d}{2}).
\ee
The two matrices $M^{+}_{4\g},M^{-}_{4\g}$ are our main results. They are given by
\be\label{eq:final_dictionary_photons_matrices}
M^{+}_{4\g}&=\left(
\begin{array}{ccc}
 -4 (D-2) & -2 (D-4) & 0 \\
 -1 & -\frac{D-4}{2} & (D-3) \\
 \frac{2}{(D-1)} & -\frac{1}{(D-1)} & 0 \\
\end{array}
\right), \nn \\
M^{-}_{4\g} &= \bpm 2 \epm.
\ee
The left-hand side of the dictionary is written as the heavy action defined by \eqref{eq:heavyaction_definition_generalrep}, and the right hand side is the heavy state contribution given by e.g., \eqref{eq:flat_space_heavy_contribution_example}. The dictionary should be true for all tensor structures of the block/partial wave, so we suppress their labels $(a'),(b')$ in \eqref{eq:final_dictionary_photons}. Moreover, since $(-1)^{j'}$ gives the signature of the integrand under $1\leftrightarrow 3$, or equivalently $s\leftrightarrow t$, we can separately consider the CFT sum rules with even/odd $j'$ and the flat space sum rules that are symmetric/antisymmetric under $s\leftrightarrow t$. On the CFT side, $k$ is a parameter of the functional and the subtraction factor, which controls the behavior of the integrand in the Regge limit. On the flat space side, the same $k$ gives the Regge spin of the sum rule.

We also notice that the all the CFT sum rules correspond to the ``lowest-subtracted" flat space sum rules, meaning that they have the minimal number of subtractions (for a given Regge spin) among all the sum rules in \eqref{eq:photon_flatspace_sumrules}. (In other words, their corresponding polarization structures grow the fastest in the Regge limit.) Interestingly, upon setting $e_1=e_3$ to be along the $\bar s^{\mu}$ direction following \eqref{eq:sumrule_bulk_polarizationmapall}, we find that the only four-photon amplitude structures that are still non-vanishing are the ones that appear in the lowest subtracted sum rules. It would be nice to better understand why this happens.

\subsubsection{Transverse spin of sum rules}\label{sec:transverse_spin}
By construction, our CFT sum rules are labeled by a transverse spin $j'$. As discussed below the dictionary \eqref{eq:final_dictionary_photons}, the signature $(-1)^{j'}$ tells us if the flat space sum rule is symmetric or antisymmetric under $s\leftrightarrow t$. However, $j'$ is a representation of $\SO(d-2)$, so it should encode more information than just a signature. On the CFT side, the transverse spin appears in the dual structure $\p{\<0|\cO_4(x_4)\wL[\cO](x_0,z_0)\cO_2(x_{2}^+)\>^{(a)}}^{-1}$, and $j'$ can be realized as a representation of the $\SO(d-2)$ group that stabilizes the saddle configuration of $x_{2^+},x_4,x_0,z_0$. On the flat space side, this should correspond to an $\SO(D-3)$. What is this $\SO(D-3)$ in flat space? Moreover, in the dictionary \eqref{eq:final_dictionary_photons}, \eqref{eq:final_dictionary_photons_matrices}, each CFT sum rule with a fixed $j'$ gets mapped to a linear combination of the coefficient functions $\cM^{(i)}_{4\g}$. How can we interpret this linear combination as having a transverse spin $j'$ directly in flat space?

We find that the simplest way to understand transverse spin in flat space is by considering the fixed-$u$ Regge limit in the shock frame \eqref{eq:shock_frame_vecs_inCMframe}. First, note that in the Regge limit, the $\cV, \cW$ vectors defined in \eqref{eq:kernel_VWvec_definition} become $\cV^{\mu} \sim q_1^{\mu}+q_3^{\mu}= \nu v_{2;\perp}^{\mu},\cW^{\mu}\sim q_1^{\mu}-q_3^{\mu}= -i\nu v_{1;\perp}^{\mu}$. Therefore, the four kernels in \eqref{eq:JJO_kernels_flatspace} become
\be\label{eq:kernels_photon_Regge}
\left.\cK^{4\g,(1)}_{j'=0}\right|_{\mathrm{Regge}} \propto& e_{2\mu_2}e_{4\mu_4}\p{(D-3) v_{2;\perp}^{\mu_2}v_{2;\perp}^{\mu_4}+v_{1;\perp}^{\mu_2}v_{1;\perp}^{\mu_4}+\de_{\perp;D-4}^{\mu_2\mu_4}}, \nn \\
\left.\cK^{4\g,(2)}_{j'=0}\right|_{\mathrm{Regge}} \propto& e_{2\mu_2}e_{4\mu_4}\p{v_{1;\perp}^{\mu_2}v_{1;\perp}^{\mu_4}+\de_{\perp;D-4}^{\mu_2\mu_4}}, \nn \\
\left.\cK^{4\g,(1)}_{j'=1}\right|_{\mathrm{Regge}} \propto& e_{2\mu_2}e_{4\mu_4}v_{1;\perp}^{\mu_2}v_{2;\perp}^{\mu_4}, \nn \\
\left.\cK^{4\g}_{j'=2}\right|_{\mathrm{Regge}} \propto& e_{2\mu_2}e_{4\mu_4}\p{v_{1;\perp}^{\mu_2}v_{1;\perp}^{\mu_4}-\frac{v_{1;\perp}^{\mu_2}v_{1;\perp}^{\mu_4}+\de_{\perp;D-4}^{\mu_2\mu_4}}{D-3}}.
\ee

From the expression of the kernels in the Regge limit, we see that the $\SO(D-3)$ that realizes the transverse spin in flat space should be the stabilizer group of $s,\bar s,v_{2;\perp}$, and $v_{1;\perp}^{\mu_2}v_{1;\perp}^{\mu_4}+\de_{\perp;D-4}^{\mu_2\mu_4}$ is the metric of the corresponding $(D-3)$-dimensional space. In fact, we can see this more clearly by studying the four-photon amplitude in the Regge limit. To make the action of the $\SO(D-3)$ more manifest, we can parametrize the polarizations as
\be\label{eq:e24_shock_parametrization_forjprime}
e_2^{\mu} = e_{2s}s^{\mu} + e_{2\bar s}\bar s^{\mu} + e_{2+}v_{2;\perp}^{\mu} + z_{2}^{\mu}, \nn \\
e_4^{\mu} = e_{4s}s^{\mu} + e_{4\bar s}\bar s^{\mu} + e_{4+}v_{2;\perp}^{\mu} + z_{4}^{\mu},
\ee
where $z_2^{\mu},z_4^{\mu}$ are $(D-3)$-dimensional vectors that transform under the $\SO(D-3)$ group. Using transversality $p_2\.e_2=0$ and the gauge redundancy $e_2\sim e_2+\# p_2$, we can express $e_{2s},e_{2\bar s}$ in terms of $e_{2+},z_2$ and the shock frame momenta (and similarly for $e_4$). Therefore, we can parametrize the polarizations using $e_{2+},e_{4+}$ and $z_2,z_4$.

We now take the four-point amplitude and set $e_1=e_3=i\frac{m(m-\sqrt{m^2-\nu^2})}{\nu^2}\bar s$ as required by our functional. As mentioned above, after this step only the polarization structures that appear in the lowest-subtracted sum rules are non-vanishing. We then take the Regge limit and consider the leading term. With the parametrization \eqref{eq:e24_shock_parametrization_forjprime}, we find
\be
\cM^{\mathrm{Regge}}_{4\g} &= \frac{m^2\nu^2(D-1)}{8}\p{\tfrac{2}{D-1}\cM^{(1)+}(s,u)-\tfrac{1}{D-1}\cM^{(2)+}(s,u)} E^{4\g}_{24,j'=2} \nn \\
&+\frac{im^2\nu^2}{8}\p{2\cM^{(1)-}(s,u)} E^{4\g}_{24,j'=1} \nn \\
&-\frac{m^2\nu^2}{4(D-3)}\p{-\cM^{(1)+}(s,u)-\tfrac{D-4}{2}\cM^{(2)+}(s,u)+(D-3)\tfrac{\nu^2}{2}\cM^{(3)}}E^{4\g,(1)}_{24,j'=0} \nn \\
&-\frac{m^2\nu^2}{16(D-3)}\p{-4(D-2)\cM^{(1)+}(s,u)-2(D-4)\cM^{(2)+}(s,u)}E^{4\g,(2)}_{24,j'=0},
\ee
where each $E^{4\g}_{24,j'}$ is a tensor built from $e_2,e_4$ that transforms as a spin-$j'$ representation under the $\SO(D-3)$ group. They are defined as
\be\label{eq:E24_tensors_photons}
E^{4\g}_{24,j'=2} &= z_{2}\.v_{1;\perp} z_{4}\.v_{1;\perp} - \frac{z_2\.z_4}{D-3} ,\qquad E^{4\g}_{24,j'=1} = e_{4+}z_{2}\.v_{1;\perp} - e_{2+}z_{4}\.v_{1;\perp}, \nn \\
E^{4\g,(1)}_{24,j'=0} &= z_2\.z_4-e_{2+}e_{4+},\qquad\qquad\qquad E^{4\g,(2)}_{24,j'=0}= e_{2+}e_{4+}.
\ee

We see that by decomposing $\cM^{\mathrm{Regge}}_{4\g} $ into different representations under $\SO(D-3)$, the linear combinations of the coefficient functions exactly agree with the ones in the dictionary \eqref{eq:final_dictionary_photons_matrices}. In other words, we can write $\cM^{\mathrm{Regge}}_{4\g}$ as
\be
\cM^{\mathrm{Regge}}_{4\g} =& \bpm -\frac{m^2\nu^2}{4(D-3)} E^{4\g,(1)}_{24,j'=0} &&-\frac{m^2\nu^2}{16(D-3)}  E^{4\g,(2)}_{24,j'=0} && \frac{m^2\nu^2(D-1)}{8} E^{4\g}_{24,j'=2}  \epm M^{+}_{4\g}\bpm \cM^{(1)+}(s,u) \\ \cM^{(2)+}(s,u) \\ \frac{\nu^2}{2}\cM^{(3)}(s,t) \epm \nn \\
&+ \bpm \frac{im^2\nu^2}{8}  E^{4\g}_{24,j'=1}  \epm M_{4\g}^{-} \bpm \cM^{(1)-}(s,u) \epm,
\ee
where $M^{+}_{4\g}, M^{-}_{4\g}$ are the matrices in the dictionary \eqref{eq:final_dictionary_photons_matrices}.

Another way to see why we get the same matrices that appear in the dictionary is that under the polarization sum, the pairing between the Regge limit kernels \eqref{eq:kernels_photon_Regge} and the spin-$j'$ tensors \eqref{eq:E24_tensors_photons} form a diagonal matrix. 

For the $j'=0$ case, there are two different tensors structures. By comparing $E^{(1)}_{j'=0},E^{(2)}_{j'=0}$ and the CFT structures that give the two kernels for $j'=0$, we can even write down a dictionary between the embedding space CFT structures and spin-$j'$ tensors:
\be\label{eq:CFT_spinjtensor}
H_{20}H_{40}\leftrightarrow e_{2+}e_{4+},\quad H_{24} \leftrightarrow z_2\.z_4-e_{2+}e_{4+}.
\ee
A similar dictionary between CFT structures and polarization structures in a conformal frame is also given in \cite{Kravchuk:2016qvl}.

\subsection{Sum rules for gravitons}\label{sec:graviton_sum_rules}

In this section we give the dictionary between the CFT sum rules and flat space sum rules for gravitons. We first review the flat space graviton sum rules.

The local module of the graviton amplitude has $29$ generators, and $28$ of them can be constructed from taking products of the photon generators defined in section \ref{sec:review_photon_flatspace}. The remaining generator $\cG$ can be written as the Gram determinant of all dot products between $(p_1,p_2,p_3,e_1,e_2,e_3,e_4)$. The general four-graviton amplitude in generic spacetime dimension ($D\geq 8$) takes the form \cite{Chowdhury:2019kaq}
\be
\cM_{4g}(s,u) =& \cG \cM^{(1)}_{4g}(s,u) + S^2\cM^{(10)}_{4g}(s,u) \nn \\
+&\left(H_{14}^2H_{23}^2\cM^{(2)}_{4g}(s,u) + H_{12}H_{13}H_{24}H_{34}\cM^{(3)}_{4g}(s,u)\right. \nn \\
&+H_{14}H_{23}(X_{1243}-X_{1234}-X_{1324})\cM^{(4)}_{4g}(s,u) + X_{1243}^2 \cM^{(6)}_{4g}(s,u) \nn \\
&+X_{1234}X_{1324}\cM^{(7)}_{4g}(s,u) + H_{14}H_{23}S\cM^{(8)}_{4g}(s,u) \nn \\
&\left.+X_{1243}S\cM^{(9)}_{4g}(s,u) + \textrm{triplet permutations}\right) \nn \\
+&H_{12}H_{34}X_{1243}\cM^{(5)}_{4g}(s,u) + \textrm{sextuplet permutations}.
\ee
The functions that multiply the polarization structures $\cG$ and $S^2$ are symmetric under all permutations of $s,t,u$, and the ones with ``triplet permutations" are symmetric in their two arguments. For the $\cM^{(5)}$ function one should include all six permutations.

By repeating the analysis of the Regge limit behavior reviewed in section \ref{sec:review_photon_flatspace}, one can obtain the dispersion relations for the graviton amplitude. There are $19$ independent fixed-$u$ sum rules with even Regge spin $k$ and are symmetric under $s\leftrightarrow t$ \cite{Caron-Huot:2022jli}. Similar to the photon case, the number of subtractions for a given Regge spin of these sum rules can be different.
In the dictionary for photons, we see that our CFT sum rules all correspond to the lowest-subtracted flat space sum rules. This turns out to be true for gravitons as well. For graviton sum rules with even Regge spin, $7$ of them are the lowest-subtracted. They are given by
\be\label{eq:graviton_flatspace_sumrules_evenk}
\left\{\cC^{+(1-7)}_{4g;k,u}\right\}= -\oint_{\oo}\frac{ds}{4\pi i}\frac{(s-t)}{(-st)^{\frac{k-2}{2}}}&\left\{\cM_{4g}^{(3)}(s,t),\p{\tfrac{-u}{2}}^2\cM_{4g}^{(10)}(s,t),\cM_{4g}^{(2,5)+}(s,u),\right. \nn \\
&\quad\left.\p{\tfrac{-u}{2}}\cM_{4g}^{(8,9)+}(s,u), \cM_{4g}^{(6)+}(s,u)+\cM_{4g}^{(7)}(s,t)\right\},\quad k=2,4,\ldots.
\ee
For the remaining $12$ sum rules that have higher number of subtractions, see appendix A of \cite{Caron-Huot:2022jli}. We again use the notation $\cM^{(i)\pm}(s,u)=\cM^{(i)}(s,u)\pm\cM^{(i)}(t,u)$, and rescale some of the sum rules by $\p{\tfrac{-u}{2}}$ to make all of them have the same power counting.

For graviton sum rules with odd Regge spin, there are $10$ independent sum rules, and $3$ of them are the lowest-subtracted,
\be\label{eq:graviton_flatspace_sumrules_oddk}
\left\{\cC^{-(1-3)}_{4g;k,u}\right\}= -\oint_{\oo}\frac{ds}{4\pi i}\frac{1}{(-st)^{\frac{k-3}{2}}}\left\{\cM_{4g}^{(2,5)-}(s,u),\p{\tfrac{-u}{2}}\cM_{4g}^{(8)-}(s,u)\right\},\quad k=3,5,\ldots.
\ee

When performing the CFT analysis for gravitons, all the discussions in section \ref{sec:partial_waves} and \ref{sec:kernel_shock} can be generalized straightforwardly. In particular, we can also obtain a simple relation similar to \eqref{eq:Blockvspartialwave} that relates the block part of the four-stress-tensor functional to the four-graviton partial waves, which include $20$ different cases in total \cite{Chakraborty:2020rxf,Caron-Huot:2022jli}. Moreover, we can obtain the graviton kernels that are similar to \eqref{eq:JJO_kernels_flatspace}. The number of independent kernels are $3$ for $j'=0$, $2$ for $j'=1$, $3$ for $j'=2$, $1$ for $j'=3$, $1$ for $j'=4$. Therefore, we have $7$ even $j'$ and $3$ odd $j'$ sum rules in total, and they indeed agree with the number of ``lowest-subtracted" flat space graviton sum rules that are symmetric or antisymmetric under $s\leftrightarrow t$ given in \eqref{eq:graviton_flatspace_sumrules_evenk} and \eqref{eq:graviton_flatspace_sumrules_oddk}.

Finally, the only remaining difference in the graviton case is that instead of \eqref{eq:sumrule_bulk_polarizationmapall}, the rules for polarization sum of $e_2,e_4$ become
\be
e_i^{\mu}e_i^{\nu}e_i^{\a}e_i^{\b} \to \frac{1}{2}\p{P_{e_i}^{\mu\a}P_{e_i}^{\nu\b}+P_{e_i}^{\nu\a}P_{e_i}^{\mu\b}} - \frac{1}{D-2}P_{e_i}^{\mu\nu}P_{e_i}^{\a\b},\quad i=2,4,
\ee
where
\be
P_{e_2}^{\mu\nu} = \eta^{\mu\nu}-\frac{p_{1}^{\mu}p_{2}^{\nu}+p_{1}^{\nu}p_{2}^{\mu}}{p_1\.p_2}, \nn \\
P_{e_4}^{\mu\nu} = \eta^{\mu\nu}-\frac{p_{3}^{\mu}p_{4}^{\nu}+p_{3}^{\nu}p_{4}^{\mu}}{p_3\.p_4}.
\ee

Our final dictionary for gravitons is given by
\be
\lim_{\substack{\De,\nu \gg 1 \\ \nu< m}}\bpm \Psi^{(1)}_{4g;k,\nu,j'=0;f^{+}_{k;J_e=2}}[\De,\r] \\  \Psi^{(2)}_{4g;k,\nu,j'=0;f^{+}_{k;J_e=2}}[\De,\r] \\ \Psi^{(3)}_{4g;k,\nu,j'=0;f^{+}_{k;J_e=2}}[\De,\r] \\  \Psi^{(1)}_{4g;k,\nu,j'=2;f^{+}_{k;J_e=2}}[\De,\r] \\ \Psi^{(2)}_{4g;k,\nu,j'=2;f^{+}_{k;J_e=2}}[\De,\r] \\ \Psi^{(3)}_{4g;k,\nu,j'=2;f^{+}_{k;J_e=2}}[\De,\r] \\ \Psi_{4g;k,\nu,j'=4;f^{+}_{k;J_e=2}}[\De,\r] \epm &= \frac{1}{A^{4g}_{k,\nu}}M^{+}_{4g}\bpm \cC^{+(1)}_{4g;k,-\nu^2}[m^2,\r] \\ \cC^{+(2)}_{4g;k,-\nu^2}[m^2,\r] \\ \cC^{+(3)}_{4g;k,-\nu^2}[m^2,\r] \\ \cC^{+(4)}_{4g;k,-\nu^2}[m^2,\r] \\ \cC^{+(5)}_{4g;k,-\nu^2}[m^2,\r] \\ \cC^{+(6)}_{4g;k,-\nu^2}[m^2,\r] \\ \cC^{+(7)}_{4g;k,-\nu^2}[m^2,\r] \epm, \quad k=2,4,=\ldots, \nn \\
\lim_{\substack{\De,\nu \gg 1 \\ \nu< m}} \bpm \Psi^{(1)}_{4g;k,\nu,j'=1;f^{-}_{k;J_e=2}}[\De,\r] \\  \Psi^{(2)}_{4g;k,\nu,j'=1;f^{-}_{k;J_e=2}}[\De,\r] \\ \Psi^{(1)}_{4g;k,\nu,j'=3;f^{-}_{k;J_e=2}}[\De,\r] \epm &= \frac{1}{A^{4g}_{k,\nu}}M^{-}_{4g}\bpm \cC^{-(1)}_{4g;k,-\nu^2}[m^2,\r] \\ \cC^{-(2)}_{4g;k,-\nu^2}[m^2,\r] \\ \cC^{-(3)}_{4g;k,-\nu^2}[m^2,\r] \epm, \quad k=3,5,=\ldots,
\ee
where the coefficient $A^{4g}_{k,\nu}$ is
\be
A^{4g}_{k,\nu} = 2^{5d+3-2j'}e^{\pi\nu}\pi^{1-d}\nu^{2-3d-2k}\G(\tfrac{d+2}{2}),
\ee
and the two matrices are given by
\be\label{eq:graviton_matrices_final}
M^{+}_{4g} =&\left(\scalemath{0.7}{
\begin{array}{ccccccc}
 16 (D-4) (D-2) & 0 & 16 D (D+2) & 8 (D-4) D & 0 & 0 & 4 (D-4) (D-2) \\
 16 & 0 & 16 D & 2 \left(D^2-3 D-2\right) & -4 (D-2) (D-1) & -2 (D-4) (D-1) & 2 \left(D^2-6 D+7\right) \\
 2 & (D-3) (D-1) & 2 & \frac{D-3}{2} & 1-D & -\frac{1}{2} (D-4) (D-1) & \frac{1}{4} \left(D^2-6 D+7\right) \\
 -\frac{8 D}{D+1} & 0 & \frac{8 (D+2)}{D+1} & -\frac{4}{D+1} & 0 & 0 & \frac{2}{D+1} \\
 -\frac{32}{D+1} & 0 & -\frac{32 (D+2)}{D+1} & \frac{16}{D+1} & 0 & 0 & \frac{4 (D-1)}{D+1} \\
 -\frac{8}{D+1} & 0 & -\frac{8}{D+1} & -\frac{D-3}{D+1} & 2 & -1 & \frac{D-1}{D+1} \\
 \frac{24}{(D+1) (D+3)} & 0 & \frac{24}{(D+1) (D+3)} & -\frac{12}{(D+1) (D+3)} & 0 & 0 & \frac{6}{(D+1) (D+3)} \\
\end{array}
}
\right), \nn \\
M^{-}_{4g} =&  \left(
\begin{array}{ccc}
 -16 (D+2) & 4 (D-4) & 0 \\
 -8 & (D-3)& 2(D-1) \\
 \frac{24}{D+1} & \frac{6}{D+1} & 0 \\
\end{array}
\right).
\ee

Finally, the transverse spin analysis done in section \ref{sec:transverse_spin} can also be performed in the graviton case. In this case we have transverse spin up to $j'=4$. As an example, let us consider the spin-$4$ tensor which is defined as
\be
E^{4g}_{24,j'=4}&= (z_2\.v_{1;\perp})^2(z_4\.v_{1;\perp})^2 -\mathrm{traces}.
\ee
When we isolate the spin-$4$ part of the four-graviton amplitude in the Regge limit (and after setting $e_1,e_3$ to be in the $\bar s$ direction), we obtain
\be
\cM_{4g}^{\mathrm{Regge}} = \frac{m^4\nu^4}{64}&\left(4\cM^{(3)}(s,t)+4\cM^{(2)+}(s,u)-2\cM^{(5)+}(s,u)\right. \nn \\
&\qquad\left.+\cM^{(6)+}(s,u) + \cM^{(7)}(s,t)\right)E^{4g}_{24,j'=4} + \ldots,
\ee
where $\ldots$ are other terms with $j'< 4$. Indeed, we see that the linear combination of the coefficient functions agrees with the dictionary \eqref{eq:graviton_matrices_final} (the last row of $M^{+}_{4g}$).

\section{Discussion}\label{sec:discussion}

In this work, we studied a basis of CFT dispersive sum rules for spinning operators. The basis was constructed using the fact that the commutator of two null-integrated operators on the same null plane vanishes, also known as superconvergence. Using the Lorentzian shadow representation of conformal blocks, we expressed the action of our sum rule on a conformal block as an integral over spacetime. We showed that in the bulk point limit, where $\nu$ (parameter of the sum rule) and $\De$ (dimension of the block) both become large, the spacetime integral gets completely localized to a ``scattering-crystal." This enabled us to derive a simple formula for the block action of our sum rule in the bulk point limit, generalizing the results in \cite{Caron-Huot:2021enk} to the spinning case.

The main result of this paper is a dictionary between the block action of CFT sum rules in the bulk point limit and the heavy state contribution of flat space sum rules. We fixed the dictionary by exploring flat space interpretations for the inserted block and the kernel in the simple formula of the CFT sum rule. We showed that CFT three-point structures at the saddle form a natural basis for the flat space three-point amplitudes, allowing the inserted block to be directly related to flat space partial waves in this basis. On the other hand, the kernels of the sum rule can be interpreted as actions on  shockwave amplitudes in flat space. Combining the two results, we derived the dictionary between our CFT sum rules and the flat space sum rules for photons and gravitons.

By construction, the CFT sum rules are labeled by a transverse spin $j'$. Through the dictionary, this gives a basis of flat space sum rules each labeled by a transverse spin (an $\SO(D-3)$ representation). This $\SO(D-3)$ stabilizes the shock directions and the momentum transfer $p_1+p_3$ of the shockwave amplitude, and the transverse spin of each sum rule can be realized by studying the amplitude in the Regge limit. However, from the flat space perspective, it remains unclear how this ``transverse spin basis" is useful in practice. It would be interesting to explore this further.

So far, we have only found a dictionary that relates CFT sum rules to some of the known flat space sum rules. In particular, we only get sum rules that have the fewest subtractions. As discussed in \ref{sec:sumrules_forphotons}, to get other sum rules we have to leave the configuration where the polarizations of the shock gravitons are along the shock direction. In CFT this condition comes from setting the polarizations of the light-transformed operators $\cO_1,\cO_3$ to be the null vectors $x_{10},x_{30}$. Therefore, it is conceivable that new sum rules can be obtained by studying null-integrated operators that are not integrated along their polarizations. Equivalently, we can consider taking derivatives of the light transformed operators. Schematically, we would like to find a differential operator $\cD$ such that
\be
\cD_{x,z_1,z_3}\wL[\cO_1](x,z_1)\wL[\cO_3](x,z_3) 
\ee
is a conformal primary at $x$. These conformally-invariant differential operators have interesting connections to reducible representations of the conformal group \cite{Penedones:2015aga}. We hope to address this problem in future.

Our construction of spinning CFT sum rules is an important step toward lifting the flat space gravitational bounds in \cite{Caron-Huot:2022jli} to AdS/CFT. It provides functionals with positive action on conformal blocks with large $\De$ and fixed $J$. In order to complete the bootstrap argument, there are still several remaining tasks. First, as discussed above, we have to construct the additional CFT sum rules that give rise to the other flat space sum rules which cannot be written in terms of the shock amplitudes. Second, one also needs to check the positivity of the sum rules in the Regge limit, where $\De$ and $J$ are both large. Finally, one has to derive a similar dictionary that relates the low energy EFT contribution of the CFT and flat space sum rules. In the scalar case \cite{Caron-Huot:2021enk}, the light contribution analysis also involves a saddle point. Therefore we expect we should be able to generalize it to the spinning case by arguments similar to section \ref{sec:heavyaction_CFT}. Once these calculations are done, one can then derive bounds on the higher derivative corrections of gravitational AdS EFTs in terms of the large CFT gap. These bounds would give a sharp form of the HPPS conjecture \cite{Heemskerk:2009pn} at the level of four-point functions of stress tensors. 

It would be interesting to also test the HPPS conjecture beyond four-point functions using superconvergence. For instance, one could consider commutators of light-transformed operators in a multi-point correlation function. In flat space, we expect this should correspond to commutators of multiple shockwaves. Another interesting direction is studying superconvergence in a thermal state by using the folded light transform contour from \cite{Caron-Huot:2022lff}.

In \cite{Caron-Huot:2021enk}, it was shown that superconvergence sum rules also exist at $J=-2,-4,\ldots$. In these sum rules, one integrates a commutator against derivatives of delta functions, which again constrains the two operators in the commutator to be spacelike separated. It would be interesting to find spinning versions of these sum rules.

In flat space, unitarity additionally imposes an upper bound on the spectral density. This condition has been used in the context of EFT amplitudes through primal bootstrap methods \cite{Chiang:2022ltp, Chen:2022nym}. However, it is not clear what the corresponding CFT statement is. To understand how this upper bound is realized in CFT, our dictionary will be an important ingredient.

Crossing-symmetric dispersion relations in flat space \cite{Sinha:2020win, Chowdhury:2021ynh} have also proven valuable in studying dispersive bounds on EFTs. They have the advantage of making low-energy crossing almost trivial to impose, and are more well-behaved when including loop EFT corrections \cite{Li:2023qzs}. Analogous CFT dispersion relations have been established in Mellin space \cite{Gopakumar:2021dvg}, and crossing-symmetric dispersion relations in position space were recently studied in \cite{Bissi:2022fmj}. We expect that position space methods will be more straightforward to generalize to spinning case.

Including loop EFT corrections often makes the forward limit in flat space divergent \cite{Bellazzini:2020cot, Bellazzini:2021oaj}. Consequently, one has to consider a limited set of sum rules that give weaker positivity bounds. On the other hand, loop corrections in AdS are automatically regularized by $R_{\mathrm{AdS}}$. By computing the loop contributions from summing over double-trace exchanged operators \cite{Aharony:2016dwx, Meltzer:2019nbs, Meltzer:2020qbr}, one should be able to obtain finite loop corrections to the tree-level AdS bounds.

\section*{Acknowledgements}
We thank Simon Caron-Huot, Yue-Zhou Li, Dalimil Maz\'{a}\v{c}, Julio Parra-Martinez, and Leonardo Rastelli for helpful discussions. We are
supported by Simons Foundation grant 488657 (Simons Collaboration on the Nonperturbative
Bootstrap) and a DOE Early Career Award under grant no.\ DE-SC0019085. YL is additionally supported by the National Science Foundation Graduate Research Fellowship under grant no.\ DGE‐1745301
\newpage

\appendix

\section{Conventions for two-point and three-point structures}\label{app:conventions}
In this appendix, we summarize the conventions for the conformally-invariant structures we use in the main text.

For standard two-point and three-point structures, we use
\be
\<\cO(x_1,z_1)\cO(x_2,z_2)\> &= \frac{(2z_1\.I(x_{12})\.z_2)^J}{x_{12}^{2\De}},\nn \\
I^{\mu}{}_{\nu}(x) &= \de^{\mu}{}_{\nu} - \frac{2x^{\mu}x_{\nu}}{x^2},
\ee
and
\be\label{eq:3pt_convention_scalar}
\<\f_1(x_1)\f_2(x_2)\cO(x_3,z_3)\> = \frac{(2z_3\.x_{23}x_{13}^2-2z_3\.x_{13}x_{23}^2)^J}{(x_{12}^2)^{\frac{\De_1+\De_2-\De+J}{2}}(x_{13}^2)^{\frac{\De_1+\De+J-\De_2}{2}}(x_{23}^2)^{\frac{\De_2+\De+J-\De_1}{2}}}.
\ee

In the embedding space, the standard structures can be written as
\be\label{eq:standard_struct_conventions}
\<\cO(X_1,Z_1)\cO(X_2,Z_2)\>&=\frac{(2H_{12})^J}{X_{12}^{\De+J}}, \nn \\
\<\f_1(X_1)\f_2(X_2)\cO(X_3,X_3)\> &= \frac{(-2V_{3,12})^J}{X_{12}^{\frac{\De_1+\De_2-\De-J}{2}}X_{13}^{\frac{\De_1+\De+J-\De_2}{2}}X_{23}^{\frac{\De_2+\De+J-\De_1}{2}}},
\ee
where
\be\label{eq:CFT_building_blocks_VH}
X_{ij} &= -2X_i\.X_j,\nn \\
V_{i,jk} &= \frac{Z_i\.X_j X_i\.X_k - Z_i\.X_k X_i\.X_j}{X_j\.X_k}, \nn \\
H_{ij} &= -2(Z_i\.Z_j X_i\.X_j - Z_i\.X_j Z_j\.X_i).
\ee

We also need the structures with more complicated representations. In the main text, the most complicated case will be when $\cO_1,\cO_2$ are symmetric traceless tensors with spins $J_1,J_2$, and $\cO$ has representation $\r$ whose Young diagram has three rows of length $(J,j,\tl{j})$. Our convention for the two-point structure is
\be\label{eq:2pt_withtransverserep_convention}
&\<\cO(X_1,Z_1,W_1,\tl W_1)\cO(X_2,Z_2,W_2,\tl W_2)\> = \frac{(2H_{12})^{J-j}(2Y_{12})^{j-\tl j}(2\tl{Y}_{12})^{\tl j}}{X_{12}^{\De+J}}.
\ee
For the additional two rows of the Young diagram, we introduce two null polarization vectors $W_i,\tl W_i$ with the conditions
\be\label{eq:embeddingvec_conditions}
X_i\.W_i&=Z_i\.W_i=X_i\.\tl W_i=Z_i\.\tl W_i=W_i\. \tl W_i =0, \nn \\
W_i &\sim W_i + \# X_i + \# Z_i, \nn \\
\tl W_i &\sim \tl W_i + \# X_i + \# Z_i + \# W_i.
\ee
The structures $Y_{12},\tl Y_{12}$ are index contractions of the antisymmetrization of the embedding space vectors $X_i,Z_i,W_i,\tl{W}_i$. Let us introduce the notation
\be\label{eq:commutator_contraction_definition}
[V_1,V_2,\ldots,V_n]\.[W_1,W_2,\ldots,W_n] \equiv \sum_{\s \in S_n} \mathrm{sgn}(\s) V_{1\a_1}V_{2\a_2}\ldots V_{n\a_n} W_1^{\a_{\s(1)}}W_2^{\a_{\s(2)}}\ldots W_n^{\a_{\s(n)}}.
\ee
Then, the structures $Y_{12},\tl Y_{12}$ are defined as
\be
Y_{ij} &= -2[X_i,Z_i,W_i]\.[X_j,Z_j,W_j], \nn \\
\tl Y_{ij} &=-2[X_i,Z_i,W_i,\tl{W}_i]\.[X_j,Z_j,W_j,\tl{W}_j].
\ee
One can check that subject to the conditions \eqref{eq:embeddingvec_conditions}, \eqref{eq:2pt_withtransverserep_convention} is the only conformally-invariant structure with the correct homogeneity of the embedding space vectors. We have chosen the two-point convention such that \eqref{eq:2pt_withtransverserep_convention} itself is Rindler positive.

For three-point structures, we have
\be\label{eq:3pt_convention_spinning}
&\<\cO_1(X_1,Z_1)\cO_2(X_2,Z_2)\cO(X_3,Z_3,W_3,\tl W_3)\>^{(a)} \nn \\
&= \frac{(-2V_{3,12})^{m_3}V_{1,23}^{m_1}V_{2,31}^{m_2}H_{12}^{n_{12}}H_{13}^{n_{13}}H_{23}^{n_{23}}(-2U_{3,12})^{k_{31}}(-2U_{3,21})^{k_{32}}(-2\tl U_{3,12})^{\tl {j}}}{X_{12}^{\frac{\De_1+J_1+\De_2+J_2-\De-J-j-\tl{j}}{2}}X_{13}^{\frac{\De_1+J_1+\De+J+j+\tl{j}-\De_2-J_2}{2}}X_{23}^{\frac{\De_2+J_2+\De+J+j+\tl{j}-\De_1-J_1}{2}}},
\ee
where the tensor structure is labeled by the nonnegative integers $m_i,n_{ij},k_{ij}$, subject to the constraints
\be
m_1+n_{12}+n_{13}+k_{31} &= J_1-\tl{j}, \nn \\
m_2+n_{12}+n_{23}+k_{32} &= J_2-\tl{j}, \nn \\
m_3+n_{13}+n_{23}&=J-j, \nn \\
k_{31} + k_{32} &= j-\tl{j}.
\ee
The structures $U_{i,jk},\tl{U}_{i,jk}$ are defined as
\be
U_{i,jk}&=\frac{X_{ij}}{X_{jk}} [X_i,Z_i,W_i]\.[X_j,Z_j,X_k], \nn \\
\tl{U}_{i,jk} &=\frac{X_{ij}X_{ik}}{X_{jk}} [X_i,Z_i,W_i,\tl{W}_i]\.[X_j,Z_j,X_k,Z_k].
\ee

When writing the Lorentzian shadow representation of the block (e.g., \eqref{eq:block_shadowrep0}), we need to use the three-point structures with absolute values. Also, note that the two structures in \eqref{eq:block_shadowrep0} have different causality configurations. Their explicit expressions are given by
\be
|\<\f_1(x_1)\f_2(x_2)\cO(x_5,z_5)\>| &= \frac{(2z_5\.x_{25}x_{15}^2-2z_5\.x_{15}x_{25}^2)^J}{(-x_{12}^2)^{\frac{\De_1+\De_2-\De+J}{2}}(x_{15}^2)^{\frac{\De+J+\De_1-\De_2}{2}}(x_{25}^2)^{\frac{\De+J+\De_2-\De_1}{2}}},\quad 1>2, 5\approx 1,2 \nn \\
|\<\f_3(x_3)\f_4(x_4)\cO(x_5,z_5)\>| &= \frac{(2z_5\.x_{45}x_{35}^2-2z_5\.x_{35}x_{45}^2)^J}{(-x_{43}^2)^{\frac{\De_4+\De_3-\De+J}{2}}(-x_{45}^2)^{\frac{\De+J+\De_4-\De_3}{2}}(-x_{35}^2)^{\frac{\De+J+\De_3-\De_4}{2}}},\quad 4>5>3.
\ee 
From the expression of $|\<\f_1(x_1)\f_2(x_2)\cO(x_5,z_5)\>|$, one can explicitly verify the identity \eqref{eq:abs_struct_relationtoWightman}.

In the definition of the functional \eqref{eq:spinningfunctional_definition}, we introduce a celestial three-point structure. This structure is a standard three-point structure in a Euclidean $(d-2)$-dimensional CFT, where $z_i$ are viewed as the embedding space coordinates. Therefore, we can get this structure by taking the $d$-dimensional three-point structure in \eqref{eq:standard_struct_conventions} and make the replacement $X_i\to z_i,Z_i\to w_i$. This gives
\be
\<\cP_{\de_1}(z_1)\cP_{\de_2}(z_2)\cP_{\de,j}(z,w)\> = \frac{(4w\.z_1 z\.z_2-4w\.z_2 z\.z_1)^j}{(-2z_1\.z)^{\frac{\de_1+\de+j-\de_2}{2}}(-2z_2\.z)^{\frac{\de_2+\de+j-\de_1}{2}}(-2z_1\.z_2)^{\frac{\de_1+\de_2-\de+j}{2}}}.
\ee

Finally, in the functional we also use a dual structure $\<0|\cO_4\wL[\cO]\cO_{2^{+}}|0\>^{-1}$, which should be a continuous-spin structure. In particular, we are interested in the structure with the causality constraint $4>0>2$. We define the standard continuous-spin structure with this configuration as
\be
&\<0|\f_4(x_4)\cO(x_0,z_0)\f_2(x_2^{+})|0\> \nn \\
&= \frac{(2z_0\.x_{40}x_{20}^2-2z_0\.x_{20}x_{40}^2)^{J}}{(-x_{24}^2)^{\frac{\De_2+\De_4-\De+J}{2}}(-x_{02}^2)^{\frac{\De+J+\De_2-\De_4}{2}}(-x_{04}^2)^{\frac{\De+J+\De_4-\De_2}{2}}},\quad 4>0>2.
\ee
In the embedding space, it is given by
\be\label{eq:continuous_spin_convention}
&\<0|\f_4(X_4)\cO(X_0,Z_0)\f_2(X_2^{+})|0\> \nn \\
&= \frac{(-2V_{0,42}(-X_{24}))^{J}}{(-X_{24})^{\frac{\De_2+\De_4-\De+J}{2}}(-X_{02})^{\frac{\De+J+\De_2-\De_4}{2}}(-X_{04})^{\frac{\De+J+\De_4-\De_2}{2}}},\quad 4>0>2.
\ee
Note that the combination $-2V_{0,42}(-X_{24})$ is always positive due to the causality constraint $4>0>2$. The actual dual structure will be the above structure with the replacement $\De\to J+d-1,J\to \De-d+1,\De_i\to \tl \De_i$, multiplied by a computable prefactor. In the scalar case, it is given by \eqref{eq:dual_struct_scalar}. We discuss the spinning case in more detail in appendix \ref{app:dual_large_nu}.

\section{Heavy action formula from the large $\nu$ limit}\label{app:derivation_largenu}
In the main text of this paper we have taken the large $\nu$ limit of the kernel and large $\De$ limit of the block at the same time. However, in this appendix we compute the large $\nu$ limit independently. Taking the large $\nu$ limit of the kernel alone will slightly differ from taking $\De$ and $\nu$ limits simultaneously. Recall that when we perform the gauge fixing, we also fix a fifth point $x_5$ coming from the conformal block we want to act the functional on. If we just study the action of the functional on a general four-point function, we can again use conformal symmetry to fix $x_1+x_3$ to the origin and $x_{2^{+}}+x_{4^{-}}$ to spatial infinity, and since we don't have a fifth point, the stabilizer group is generated by Lorentz transformations and dilatations. We can then use Lorentz transformation to fix all four points to be on the same plane. After performing the large $\nu$ analysis (which will be done below), one can then obtain a kernel that only depends on the cross ratios (see \eqref{eq:superconvergence_integrandx0z_expr5}). Therefore, we can undo the gauge fixing and go back the the bulk-point gauge fixing again, and then the rest of the calculation is the same as \cite{Caron-Huot:2021enk}.

We now want to take the large $\nu$ limit of 
\be\label{eq:scalarfunctional_plus_final_app}
\Psi^{+}_{k,\nu}[\cG] &=  4A_{k,\nu} \int_{4>3>0>1>2} \frac{d^dx_1 d^d x_2 d^d x_3 d^d x_4 d^d x_0 D^{d-2}z_0}{\vol(\tl \SO(d,2))} \<0|\f_4(x_4)\wL[\cO](x_0,z_0)\f_2(x_2^{+})|0\>^{-1} \nn \\
& \<\tl \cP_{\de_1}(x_{1^{+}0}) \tl \cP_{\de_3}(x_{30}) \cP_{\de}(z_0)\>\de(x_{1^{+}0}^2)\de(x_{30}^2)\th(x_{1^{+}0})\th(x_{30}) \<\O|[\f_4(x_4),\f_3(x_3)][\f_1(x_1^{+}),\f_2(x_2^{+})]|\O\>.
\ee
The only $\nu$ dependence is in dual and celestial structures, and so we need only focus on
\be\label{eq:superconvergence_integrandx0z}
&\int d^dx_0D^{d-2}z \<0|\f_4(x_4)\wL[\cO](x_0,z_0)\f_2(x_2^{+})|0\>^{-1} \de(x_{1^{+}0}^2)\de(x_{30}^2)\nn \\
&\qquad\qquad \x\<\cP_{\de}(z)\cP_{\tl{\de}_1}(x_{1^{+}0})\cP_{\tl{\de}_3}(x_{30})\>(-x_{13}^2)^{\tl{\De}_\f}(-x_{24}^2)^{\tl{\De}_\f},
\ee
which we can rewrite using the explicit expression for the dual structure \eqref{eq:dual_struct_scalar} which gives us
\be
\int d^dx_0D^{d-2}z 
\x\<\cP_{\de}(z)\cP_{\tl{\de}_1}(x_{1^{+}0})\cP_{\tl{\de}_3}(x_{30})\>(-x_{13}^2)^{\tl{\De}_\f}(-x_{24}^2)^{\tl{\De}_\f}\de(x_{1^{+}0}^2)\de(x_{30}^2)\nn \\
\frac{(2z_0\.x_{40}x_{20}^2-2z_0\.x_{20}x_{40}^2)^{\De-d+1}}{(-x_{24}^2)^{\frac{\tl{\De}_2+\tl{\De}_4+\De-J-2d+2}{2}}(x_{02}^2)^{\frac{\tl{\De}_2+\De+J-\tl{\De}_4}{2}}(x_{02}^2)^{\frac{\tl{\De}_4+\De+J-\tl{\De}_2}{2}}},
\ee
where we have dropped the overall coefficient in \eqref{eq:dual_struct_scalar}. It will be useful to notice that the dual structure has the form $\< \tl{\f}(x_4) \cO^F(x_0, z_0) \tl{\f}(x_2) \>$, which is a three-point function of two scalars with shadow dimensions $\tl{\De}_\phi = d- \De_\phi$, and a third operator with $(\De_F,J_F) = (J+d-1,\De -d+1)$, where $J$ and $\De$ are the dimension and spin of the light transformed operator. In particular we will focus on the limit $J_F = 2-d+\de=\frac{2-d}{2}+i\nu$, $\nu\to \oo$.

From the analysis in the cross-ratio space \cite{Caron-Huot:2021enk}, we know that in the bulk-point limit $x$ and $y$ will get fixed to saddle points of the form
\be
x=(u_x,v_x,\vec 0),\qquad y=(u_y,v_y,\vec 0),
\ee
and therefore we can just study the large $\nu$ limit for this particular $x$ and $y$ using the gauge fixing procedure described above.

Let us choose the coordinates of $x_0$ and $z$ to be $x_0=(u,v,\vec y_0), z=(1,\vec y_z^2,\vec y_z)$. In the integrand, the $J_F$ dependent factors are (we replace $\de$ with $J_F+d-2$ as well as $\De$ with $J_F+d-1$)
\be
&\<0|\tl{\f}_4(x_4)\cO^F(x_0,z)\tl{\f}_2(x_2^{+})|0\>\<\cP_{\de}(z)\cP_{\tl{\de}_1}(x_{1^{+}0})\cP_{\tl{\de}_3}(x_{30})\>\nn \\
=&\frac{(-2V_{0,24})^{J_F}}{(x_{2^{+}4}^2)^{\frac{2\tl{\De}_\f-d+2-J_F}{2}}(-x_{40}^2)^{\frac{d-2+J_F}{2}}(x_{2^{+}0}^2)^{\frac{d-2+J_F}{2}}}\x \nn \\
&\qquad \frac{1}{(-2z\.x_{1^{+}0})^{\frac{d-2+J_F}{2}}(-2z\.x_{30})^{\frac{d-2+J_F}{2}}(-2x_{1^{+}0}\.x_{30})^{\frac{2\tl{\de}_\f-d+2-J_F}{2}}}.
\ee
Note that we have the delta functions $\de(x_{1^{+}0}^2)\de(x_{30}^2)$, and therefore $-2x_{1^{+}0}\.x_{30}=(x_{1^{+}0}-x_{30})^2 = x_{1^{+}3}^2$. 

The two delta functions can be written as
\be
\de(x_{1^{+}0}^2)\de(x_{30}^2) =  \de((u+u_y)(v+v_y)-\vec y_0^2)\de(-(u-u_y)(v-v_y)+\vec y_0^2).
\ee
In terms of $\tl t_0, \tl x_0$ defined in \eqref{eq:t0x0tilde_definition} (in this case we have $\vec y_\perp=0$), the delta functions fix $\tl t_0=0$ and $r_0 \equiv |\vec y_0| = (u_yv_y(1-\tl x_0^2))^{\frac 12}$. The Jacobian relating $\tl t_0,\tl x_0$ and $u,v$ is given by $du dv = 2u_y v_yd\tl t_0 d\tl x_0$. This gives (for any function $f(u,v,\vec y_0)$)
\be\label{eq:superconvergence_deltafunctionintegral}
&\frac{1}{2}\int dudv d^{d-2}\vec y_0\ \de(-(u-u_y)(v-v_y)+\vec y_0^2)\de((u+u_y)(v+v_y)-\vec y_0^2)f(u,v,\vec y_0)\nn \\
=&  \int d\tl x_0 d\O_{\vec n_0} \frac{(u_y v_y (1-\tl x_0^2))^{\frac{d-4}{2}}}{8} f(\tl t_0=0,\tl x_0,r_0=(u_yv_y(1-\tl x_0^2))^{\frac 12},\O_{\vec n_0}),
\ee
where $\vec n_0$ is the unit vector in the $\vec y_0$ direction.

After removing the delta functions, \eqref{eq:superconvergence_integrandx0z} becomes
\be\label{eq:superconvergence_integrandx0z_expr2}
&\int d\tl x_0 d\O_{\vec n_0} d^{d-2}\vec y_z \frac{(u_y v_y (1-\tl x_0^2))^{\frac{d-4}{2}}}{8}\frac{2^{-J_F}}{(-x_{2^{+}4^{-}}^2)^{\frac{2\tl{\De}_\f-d+2-J_F}{2}}(x_{4^{-}0}^2)^{\frac{d-2}{2}}(x_{2^{+}0}^2)^{\frac{d-2}{2}}}\x \nn \\
&\qquad \frac{1}{(-2z\.x_{1^{+}0})^{\frac{d-2}{2}}(-2z\.x_{30})^{\frac{d-2}{2}}(-2x_{1^{+}0}\.x_{30})^{\frac{2\tl{\de}_\f-d+2-J_F}{2}}}e^{J_F\x h(\tl x_0,\vec y_z,\O_{\vec n_0})}(4u_y v_y)^{(d-\De_\f)}(\tfrac{4}{u_x v_x})^{(d-\De_{\f})},
\ee
where the function $h(\tl x_0,\vec y_z,\O_{\vec n_0})$ is given by
\be
h(\tl x_0,\vec y_z,\O_{\vec n_0}) &= \log\p{\frac{1}{u_x} + \frac{\vec y_z^2}{v_x} - u_y v_y (1-\tl x_0^2)(v_x + u_x \vec y_z^2) - u_xv_y^2 \tl x_0^2 - u_y^2 v_x \tl x_0^2 \vec y_z^2 + 2(u_y v_x - v_y u_x)\tl x_0 \vec y_0 \.\vec y_z} \nn \\ 
&-\frac 12 \log \p{(v_y(1-\tl x_0) - 2\vec y_0 \.\vec y_z + u_y(1+\tl x_0)\vec y_z^2)(v_y(1+\tl x_0)+ 2\vec y_0 \.\vec y_z u_y(1-\tl x_0)\vec y_z^2)} \nn \\
&-\frac 12 \log \p{\frac{(1- u_yv_y u_x v_x - (u_y v_x - v_y u_x)\tl x_0)(1-u_yv_y u_x v_x + (u_y v_x - u_x v_y)(1-\tl x_0))}{u_x^2 v_x^2}}.
\ee
We find that at large $J_F$, the integral has two saddle loci: one at $\tl x_0=0,\vec y_z=0$ and one at $\tl x_0=0,\vec y_z=\oo$. Namely,
\be
&\left.\frac{\ptl}{\ptl \tl x_0}h(\tl x_0,\vec y_z,\O_{\vec n_0})\right|_{\tl x_0\to 0, \vec y_z\to 0} = \left.\frac{\ptl}{\ptl \vec y_z}h(\tl x_0,\vec y_z,\O_{\vec n_0})\right|_{\tl x_0\to 0, \vec y_z\to 0} = 0 \nn \\
&\left.\frac{\ptl}{\ptl \tl x_0}h(\tl x_0,\vec y_z,\O_{\vec n_0})\right|_{\tl x_0\to 0, \vec y_z\to \oo} = \left.\frac{\ptl}{\ptl \vec y_z}h(\tl x_0,\vec y_z,\O_{\vec n_0})\right|_{\tl x_0\to 0, \vec y_z\to \oo} = 0. 
\ee
Let us first consider the locus at $\tl x_0= \vec y_z=0$. We will see what the other saddle should give shortly. Expanding $h(\tl x_0,\vec y_z,\O_{\vec n_0})$ around $\tl x_0= \vec y_z=0$, we get
\be\label{eq:hfunc_aroundsaddle_afterdelta}
h(\tl x_0,\vec y_z,\O_{\vec n_0}) \approx& - \log\p{\frac{v_y}{v_x}} + \frac{(1+u_y^2 v_x^2 - 2u_x u_y v_x v_y)(1-u_x^2 v_y^2)}{2(1-u_y v_y u_x v_x)^2}\tl x_0^2 \nn \\
&+ \frac{2(1-u_x^2 v_y^2)}{v_y(1-u_y v_y u_x v_x)}\tl x_0 \vec y_0 \.\vec y_z  + \frac{2v_x (\vec y_0\.\vec y_z)^2 - v_y(u_y v_x-v_y u_x)\vec y_z^2}{v_x v_y^2}+ \ldots \nn \\
=& - \log\p{\frac{v_y}{v_x}} + \frac 12 \bpm \tl x_0 & \vec y_z \epm M_{\vec y_0} \bpm \tl x_0 \\ \vec y_z \epm.
\ee
Alternatively, one can also choose to not integrate $\tl t_0, |\vec y_0|$ and keep the delta functions. Then the function $h$ will also depend on $\tl t_0$ and $|\vec y_0|$, and one can show that there are saddle loci at $\tl t_0=\tl x_0=0,\vec y_z =0,\oo$. We can then expand $h$ around the saddle locus and use the delta functions to fix $\tl t_0$ and $|\vec y_0|$. This calculation should give the same result. 

Now let us compute the saddle integral in \eqref{eq:superconvergence_integrandx0z_expr2}. By rotational invariance, the determinant of the Hessian $M_{\vec y_0}$ in \eqref{eq:hfunc_aroundsaddle_afterdelta} does not depend on the direction of $\vec y_0$, so we can pick a direction that makes computing the determinant simple. For example, $\vec y_0 = |\vec y_0 |(1,0,\ldots,0)$. We then find
\be
\mathrm{Det} M_{\vec y_0} =\frac{2^{d-2}(1-u_y^2 v_x^2)(1-v_y^2 u_x^2)(v_y u_x - u_y v_x)^{d-2}}{(v_x v_y)^{d-2}(1-u_y v_y u_x v_x)^2}.
\ee
One also has to check the sign of each eigenvalue of $\mathrm{Det} M_{\vec y_0}$ in order to get the correct phase in the saddle integral. We find that for $v_yu_x - u_yv_x>0$, all eigenvalues are positive, and for $v_yu_x - u_yv_x<0$, one eigenvalue is positive and the other $d-2$ eigenvalues are negative. Both cases turn out to give the same result, so let us assume $v_yu_x - u_yv_x>0$ here. The saddle integral in \eqref{eq:superconvergence_integrandx0z_expr2} is then given by
\be
&\int d\tl x_0 d^{d-2}\vec y_z \p{\cdots}e^{J_F\x h(\tl x_0,\vec y_z,\O_{\vec n_0})}\nn \\
\approx & \p{\cdots}|_{\tl x_0 \to 0,\vec y_z \to 0}\x\p{\frac{v_x}{v_y}}^{J_F} e^{\frac{i\pi}{2}(d-1)}J_F^{-\frac{d-1}{2}}\sqrt{\frac{(2\pi)^{d-1}}{\mathrm{Det}M_{\vec y_0}}} + \ldots,
\ee
where $\p{\cdots}$ are the other factors in the integrand of \eqref{eq:superconvergence_integrandx0z_expr2}, and $\ldots$ are subleading terms at large $J_F$. Evaluating the remaining factors on the saddle locus $\tl x_0=0,\vec y_z=0$, we obtain
\be\label{eq:superconvergence_integrandx0z_expr3}
&(4u_y v_y)^{(d-\De_\f)}(\tfrac{4}{u_x v_x})^{(d-\De_{\f})}\int d\O_{\vec n_0}\frac{(u_y v_y)^{\frac{d-4}{2}}}{8}\p{\frac{v_x}{v_y}}^{J_F} e^{\frac{i\pi}{2}(d-1)}J_F^{-\frac{d-1}{2}}\sqrt{\frac{(2\pi)^{d-1}}{\mathrm{Det}M_{\vec y_0}}}\x \nn \\
&\left.\frac{2^{-J_F}}{(-x_{2^{+}4^{-}}^2)^{\frac{2\tl{\De}_\f-d+2-J_F}{2}}(x_{4^{-}0}^2)^{\frac{d-2}{2}}(x_{2^{+}0}^2)^{\frac{d-2}{2}}}\frac{1}{(-2z\.x_{1^{+}0})^{\frac{d-2}{2}}(-2z\.x_{30})^{\frac{d-2}{2}}(-2x_{1^{+}0}\.x_{30})^{\frac{2\tl{\de}_\f-d+2-J_F}{2}}}\right|_{\tl x_0 \to 0,\vec y_z \to 0} \nn \\
&=2^{-\frac{9}{2}+2d+J_F}\pi^{\frac{d-1}{2}}\vol(S^{d-3})e^{\frac{i\pi}{2}(d-1)}J_F^{-\frac{d-1}{2}}\x \nn \\
&\qquad \frac{(v_yu_x)^{\frac{d-2-J_F}{2}}(u_yv_x)^{\frac{J_F}{2}+d-2}(v_yu_x-u_yv_x)^{\frac{2-d}{2}}(1-u_yv_yu_xv_x)^{3-d}}{\sqrt{(1-u_y^2v_x^2)(1-v_y^2u_x^2)}}.
\ee 

We can write the result in terms of the conformally-invariant cross-ratios $r$ and $\eta$ defined as
\be
u'=\frac{(1-2r\eta+r^2)^2}{16r^2},\quad v'=\frac{(1+2r\eta+r^2)^2}{16r^2},
\ee
where $u',v'$ are defined in \eqref{eq:cross_ratios}. In terms of $r,\eta$, the $u$-channel Regge limit corresponds to $r\to 0$ with fixed $\eta$. In our gauge fixing, they are given by
\be
r=|x||y|=(u_yv_yu_xv_x)^{\frac{1}{2}},\qquad \eta = -\frac{x\.y}{|x||y|} = \frac{v_yu_x + u_yv_x}{2(u_yv_yu_xv_x)^{\frac{1}{2}}}.
\ee

Then, in terms of $r,\eta$, \eqref{eq:superconvergence_integrandx0z_expr3} becomes
\be\label{eq:superconvergence_integrandx0z_expr4}
2^{-\frac{9}{2}+2d+J_F}\pi^{\frac{d-1}{2}}\vol(S^{d-3})e^{\frac{i\pi}{4}(d-1)}\nu^{-\frac{d-1}{2}}\frac{(1-r^2)^{3-d}r^{d-2}w^{-i\nu}(w-\tfrac{1}{w})^{\frac{2-d}{2}}}{((1+r^2)^2-4r^2\eta^2)^{\frac 12}},
\ee
where we have used $J_F=\frac{2-d}{2}+i\nu$ and $\eta=\frac{w+1/w}{2}$. One can then notice that the leading large $\nu$ behavior is similar to
\be
\cP_{\frac{2-d}{2}+i\nu}\p{\tfrac{w+1/w}{2}} \sim \frac{\G(d-2)}{\G(\tfrac{d-2}{2})}\frac{\nu^{\frac{2-d}{2}}}{(w-1/w)^{\frac{d-2}{2}}}\p{e^{-\frac{i\pi}{4}(d-2)}w^{i\nu} + e^{\frac{i\pi}{4}(d-2)}w^{-i\nu} },\quad \nu \to \oo, |w|>1.
\ee
In particular, the second term agrees with \eqref{eq:superconvergence_integrandx0z_expr4}, and the first term comes from the other saddle locus. Therefore, after including both saddle loci, the integral \eqref{eq:superconvergence_integrandx0z_expr2} becomes
\be\label{eq:superconvergence_integrandx0z_expr5}
2^{-\frac{9}{2}+2d+J_F}\pi^{\frac{d-1}{2}}e^{\frac{i\pi}{4}}\vol(S^{d-3})\nu^{-\frac 12}\frac{\G(\tfrac{d-2}{2})}{\G(d-2)}\frac{(1-r^2)^{3-d}r^{d-2}}{((1+r^2)^2-4r^2\eta^2)^{\frac 12}}\cP_{\frac{2-d}{2}+i\nu}(\eta).
\ee

The expression we start with, \eqref{eq:superconvergence_integrandx0z}, is essentially the shadow representation of the conformal block $G^{\tl{\De}_\f}_{\frac{d}{2}+i\nu,J=-1}$, where one of the three-point structures becomes a celestial structure and delta functions due to setting $J=-1$ \cite{Chang:2020qpj}. What we have shown here is that its large $\nu$ limit is given by \eqref{eq:superconvergence_integrandx0z_expr5}. Although the same result can be obtained in the cross-ratio space using the Casimir equation \cite{Kravchuk:2018htv}, we believe that the above calculation from saddle point can be more straightforwardly generalized to spinning operators. Since $G^{\tl{\De}_\f}_{\frac{d}{2}+i\nu,J=-1}$ is exactly the kernel appearing in the Lorentzian inversion formula for $C(\De=\tfrac{d}{2}+i\nu,J=-1)$, the calculation here could also be helpful for understanding the OPE data at large $\De$ \cite{Mukhametzhanov:2018zja}. 

Finally, to reproduce the heavy action formula, we can take the conformally-invariant expression \eqref{eq:superconvergence_integrandx0z_expr5} and plug it back in \eqref{eq:scalarfunctional_plus_final_app}. Similar to the main text, we can then study its action on conformal block and choose the gauge fixing \eqref{eq:gaugefixing}, which introduces the Faddeev-Popov factor \eqref{eq:FP_Det}. This gives \eqref{eq:scalarfunctional_plus_gaugefixed} with the kernel part replaced with \eqref{eq:superconvergence_integrandx0z_expr5}. Combining with the subtraction factors \eqref{eq:subtracction_scalars}, we find that we get an integral of the form
\be
&\int d^dx d^dy \frac{r^{k+d-1}(1-r^4)\eta}{((1+r^2)^2-4r^2\eta^2)^{\frac{k+1}{2}}}\x \nn \\
&\quad \frac{1}{(-x^2)^{2d}}\frac{1}{(-x_{12}^2)^{\tl{\De}_\f}(-x_{34}^2)^{\tl{\De}_\f}}\left.|\<\f_{1^{+}}\f_{2^{+}}\cO^{\mu_1\cdots\mu_J}(x_5)\>| |\<\tl{\cO}^{\dag}_{\mu_1\cdots\mu_J}(x_5)\f_3\f_4\>|\right|_{\mathrm{gauge-fixed}}.
\ee
The only difference between this integral and the one considered in \cite{Caron-Huot:2021enk} (see (3.25) and (3.37)) is a factor of $\eta$, which becomes $1$ in the bulk-point limit. One can further check that the overall factors also agree. Hence, the calculation in \cite{Caron-Huot:2021enk} implies that we recover the same heavy action formula \eqref{eq:scalar_sumrule_CFTtoflatspace}.

\section{Details on matching partial waves}\label{app:partial_waves_more}
In this appendix, we give some more details on the partial wave discussion in section \ref{sec:partial_waves}.
\subsection{Partial waves in monomial basis}
In the main text, we define the vertices $v(n,e_1,e_2)$ directly from evaluating the CFT three-point structures at the saddle and applying the polarization map. From the flat space perspective, it is more convenient to define the vertices as monomials, which can be easily expressed using Young tableaux, as given in \cite{Caron-Huot:2022jli}. Furthermore, the bootstrap calculation done in \cite{Caron-Huot:2022jli} also uses the monomial basis (up to a Gram-Schmidt procedure to convert it into a orthonormal basis). The dictionary found in the main text tells us that we can take the flat space functional found in \cite{Caron-Huot:2022jli} and construct a positive CFT functional with positive action on heavy blocks. If we want to further study the OPE coefficients in the bootstrap calculation, we will need to know the relation between the two different bases.

One can find the relation between the two bases by explicitly evaluating the CFT three-point structures at the saddle for all the exchanged representations $\r$ in the photon and graviton case (see \cite{Caron-Huot:2022jli} for a complete list). It turns out that there is a clear map between the building blocks of CFT three-point tensor structures and the building blocks of the monomial basis, given by columns of the Young tableaux. We find that the map between the structures of $\<\cO_1\cO_2\cO_5\>$ and $v(n,e_1,e_2)$ is given by
\be\label{eq:Young_vs_CFT_map}
\young(n) &\leftrightarrow [X_5,Z_5]\.[X_1,X_2],\qquad\qquad e_1\.e_2 \leftrightarrow [X_1,Z_1]\.[X_2,Z_2], \nn \\
\young(\eone,n) &\leftrightarrow [X_5,Z_5,W_5]\.[X_1,Z_1,X_2], \qquad \young(\etwo,n) \leftrightarrow [X_5,Z_5,W_5]\.[X_2,Z_2,X_1], \nn \\
\young(\eone,\etwo,n) &\leftrightarrow [X_5,Z_5,W_5,\tl{W}_5]\.[X_1,Z_1,X_2,Z_2],
\ee
where we use the notation \eqref{eq:commutator_contraction_definition} for the CFT structures. The Young tableaux columns represent the antisymmetrization of the vectors in the boxes (see \cite{Caron-Huot:2022jli} for the precise definition). For example,
\be
\young(\eone,n) =e_1\.w_1n\.w_2-n\.w_1e_1\.w_2,
\ee
where $w_1,w_2$ are the polarizations of the vertex. Both sides of the map agree up to an overall factor after one evaluates the CFT structure at the scattering-crystal configuration \eqref{eq:scattering_crystal} and applies the polarization map \eqref{eq:polarvec_125_CFTtoflat}.

The map given in \eqref{eq:Young_vs_CFT_map} allows us to unambiguously relate the structure labels of the two bases, and we find that the partial waves computed in the two bases are related by
\be\label{eq:partial_waves_CFT_monomial_relation}
\pi^{\mathrm{CFT}}_{\r,(a'b')}= 2^{2J-2j-n^{a'}_{15}-n^{a'}_{25}-n^{b'}_{36}-n^{b'}_{46}}(-1)^{n^{a'}_{12}+n^{a'}_{25}+n^{b'}_{34}+n^{b'}_{46}}\pi^{\mathrm{tableaux}}_{\r,(a'b')},
\ee
where $n^{a'}_{15}$ counts the number of $H_{15}$ of the $(a')$ tensor structure, and other $n^{a'}_{ij},n^{b'}_{ij}$ are defined similarly (see \eqref{eq:3pt_convention_spinning}). Note that the above relation is true for each $a',b'$. Alternatively, one can think of the factor relating the two partial waves as two diagonal matrices that rescale the vertices.

We also see that this change of basis preserves the positivity of the partial wave actions, so we can reuse the positive functional in the monomial basis. In the forward limit, where $n=n',e_3=e_1^{*},e_4=e_2^{*}$, the partial wave in the monomial basis is positive (or positive semi-definite for multiple tensor structures) by construction. In the CFT basis, \eqref{eq:partial_waves_CFT_monomial_relation} implies that the partial wave is also positive. (Note that if we instead define $\bar v(n',e_3,e_4)$ in \eqref{eq:CFT_vertices_definition_bar}, then the partial wave wouldn't be positive in the forward limit.)

\subsection{Polarization map and CRT symmetry}
We now explain why the two definitions of vertices \eqref{eq:CFT_vertices_definition}, \eqref{eq:CFT_vertices_definition_bar} are consistent with each other. In particular, we will show that by using the definition of $v(n,e_1,e_2)$ and the CRT symmetry, we can recover the definition of $\bar v(-n',e_3,e_4)$.

For any CFT, there is an anti-unitary CRT symmetry $J$ satisfying $J^2=1$, and it acts on local operators as (assuming the operator is bosonic)
\be
J\cO(x,z)J^{-1} = \cO^{\dag}(\bar x,\bar z^{*}),
\ee
where $\bar x = (-x^0,-x^1,x^2,\ldots)$ (and similarly for $\bar z^{*}$) is a Rindler reflection. Since $J$ is anti-unitary, any CFT correlation function should satisfy
\be
\<\cO_1(x_1,z_1)\cdots\cO_n(x_n,z_n)\> = \<\cO_1^{\dag}(\bar x_1,\bar z_1^{*})\cdots\cO^{\dag}_n(\bar x_n,\bar z_n^{*})\>^{*},
\ee
where we have applied a CRT to each operator.

More generally, as explained in \cite{Kologlu:2019bco}, for any spacelike points $A$ and $B$, one can define two Rindler wedges $B>x>A^{-}$ and $A>x>B^{-}$. Then, there exists a Rindler conjugation $J_{AB}$ that exchanges the two wedges. Explicitly,
\be
J_{AB}(X) = X- 2\frac{X\.X_A}{X_A\.X_B}X_B- 2\frac{X\.X_B}{X_A\.X_B}X_A.
\ee
It turns out that if we choose $A$ to be the future infinity and $B^{-}$ to be the origin, under the corresponding Rindler conjugation $J_{AB}$ we have $3 \to 1^{+}, 4\to 2^{+}, 6^{+}\to 5$ in the scattering crystal configuration \eqref{eq:scattering_crystal}. Therefore, using this $J_{AB}$, one might expect
\be
\left.\<0|\cO^{\dag}(x_{6^{+}},z_6)\cO_4(x_4,z_4)\cO_3(x_3,z_3)|0\>\right|_{\mathrm{saddle}} \stackrel{?}{=} \left.\<0|\cO(x_5,\bar z_6^{*})\cO^\dag_4(x_{2^+},\bar z_4^{*})\cO^\dag_3(x_{1^{+}},\bar z_3^{*})|0\>^{*}\right|_{\mathrm{saddle}}.
\ee
This is however too fast. A funny feature of our saddle configuration is that although the points $x_{1,2,3,4}$ are timelike, they all have purely imaginary spatial components. The correct way to think about this should be we first consider e.g.\ $x_3=(u_y,v_y,0)$ for real $u_y,v_y$, and at the end analytically continue to $u_y \to -i\frac{m-\sqrt{m^2-\nu^2}}{\nu}, v_y \to i\frac{m-\sqrt{m^2-\nu^2}}{\nu}$. To take into account the fact that there are imaginary spatial components, we should apply an additional reflection to them, which can then be shifted to act on the polarizations. In summary, the correct CRT relation becomes
\be\label{eq:CRT_forsaddle}
&\left.\<0|\cO^{\dag}(x_{6^{+}},z_6)\cO_4(x_4,z_4)\cO_3(x_3,z_3)|0\>\right|_{\mathrm{saddle}} \nn \\
&= \left.\<0|\cO(x_5,I_e\.z_6^{*})\cO^\dag_4(x_{2^+},I_e\.z_4^{*})\cO^\dag_3(x_{1^{+}},I_e\.z_3^{*})|0\>^{*}\right|_{\mathrm{saddle}},
\ee
where the polarizations are reflected only in the time direction. Since \eqref{eq:CRT_forsaddle} is simply a statement about the CFT structures at the saddle configuration, we can explicitly verify that it is true for different representations $\r$.

Using the CRT relation \eqref{eq:CRT_forsaddle}, we can write
\be\label{eq:CRT_vertices_expr0}
&(-x_{34}^2)^{\frac{\De_3+\De_4}{2}}\p{\frac{x_{36}^2}{x_{46}^2}}^{\frac{\De_3-\De_4}{2}}\mathcal{I}_e^{\r} \left.\frac{\<0|\cO^{\dag}(x_{6^+},z_6)[\cO_4(x_{4},z_4),\cO_3(x_3,z_3)]|0\>_{(b)}}{2i(\sin(\pi\tfrac{\tl \tau_\r-\De_3-J_3-\De_4-J_4}{2}))}\right|_{\mathrm{saddle}} \nn \\
&=(-x_{12}^2)^{\frac{\De_3+\De_4}{2}}\p{\frac{x_{15}^2}{x_{25}^2}}^{\frac{\De_3-\De_4}{2}}\left.\frac{\<0|[\cO_3(x_{1^+},z_3^{*}),\cO_4(x_{2^+},z_4^{*})]\cO(x_{5},z_6^{*})|0\>^{*}_{(b)}}{-2i(\sin(\pi\tfrac{\tl \tau_\r-\De_3-J_3-\De_4-J_4}{2}))}\right|_{\mathrm{saddle}}.
\ee
We assume there is only one polarization $z_6$ for simplicity, but the argument for more polarizations $\vec w_6$ is the same. We have also imposed conservation to set the time components of $z_3,z_4$ to zero. So, $I_e\.z_3=z_3,I_e\.z_4=z_4$, and the reflection tensor $\mathcal{I}_e^{\r}$ cancels with the $I_e$ acting on $z_6$. The additional minus sign in the denominator comes from changing the operator order in the commutator.

Since the positions are now at $x_{1^+},x_{2^+},x_5$, we can use the definition \eqref{eq:CFT_vertices_definition} for $v(n,e_1,e_2)$ (although now it is a vertex for external particles with spin $J_3,J_4$). We can then write \eqref{eq:CRT_vertices_expr0} as
\be
&(-x_{34}^2)^{\frac{\De_3+\De_4}{2}}\p{\frac{x_{36}^2}{x_{46}^2}}^{\frac{\De_3-\De_4}{2}}\mathcal{I}_e^{\r} \left.\frac{\<0|\cO^{\dag}(x_{6^+},z_6)[\cO_4(x_{4},z_4),\cO_3(x_3,z_3)]|0\>_{(b)}}{2i(\sin(\pi\tfrac{\tl \tau_\r-\De_3-J_3-\De_4-J_4}{2}))}\right|_{\mathrm{saddle}} \nn \\
&=v(n,z_{3}^{x*}n_{\perp}+\vec z_{3\perp}^*,-z_{4}^{x*}n_{\perp}+\vec z_{4\perp}^*,w_1=(0,iz_6^{t*},z_6^{x*},\vec z_{6\perp}^*))^*.
\ee
When contracting the indices of the vertex $\bar v$, we should contract the polarizations after taking the complex conjugate. So, using $(v\.z)^*=v^*\.z^*$, we can write the above equation as a Schwarz reflection,
\be
&(-x_{34}^2)^{\frac{\De_3+\De_4}{2}}\p{\frac{x_{36}^2}{x_{46}^2}}^{\frac{\De_3-\De_4}{2}}\mathcal{I}_e^{\r} \left.\frac{\<0|\cO^{\dag}(x_{6^+},z_6)[\cO_4(x_{4},z_4),\cO_3(x_3,z_3)]|0\>_{(b)}}{2i(\sin(\pi\tfrac{\tl \tau_\r-\De_3-J_3-\De_4-J_4}{2}))}\right|_{\mathrm{saddle}} \nn \\
&=\bar v(n,z_{3}^{x}n_{\perp}+\vec z_{3\perp},-z_{4}^{x}n_{\perp}+\vec z_{4\perp},w_1=(0,-iz_6^{t},z_6^{x},\vec z_{6\perp})).
\ee
Rotational invariance demands that the vertex should just contain dot products of the polarizations and $n^{\mu}$. Thus we are free to apply a reflection to all the vectors since it leaves the dot products invariant. In the bulk Minkowski coordinates $(t^{\mathrm{bulk}},x_1^{\mathrm{bulk}},x_2^{\mathrm{bulk}},\ldots)$, we will apply a reflection in the $x_1^{\mathrm{bulk}}$ direction. This will send $n^{\mu} \to -n^{\prime\mu}, n_{\perp}^{\mu} \to n_{\perp}^{\prime\mu}$. (See section \ref{sec:partial_waves} for their expressions in Minkowski coordinates.) After this reflection, we finally arrive at
\be
&(-x_{34}^2)^{\frac{\De_3+\De_4}{2}}\p{\frac{x_{36}^2}{x_{46}^2}}^{\frac{\De_3-\De_4}{2}}\mathcal{I}_e^{\r} \left.\frac{\<0|\cO^{\dag}(x_{6^+},z_6)[\cO_4(x_{4},z_4),\cO_3(x_3,z_3)]|0\>_{(b)}}{2i(\sin(\pi\tfrac{\tl \tau_\r-\De_3-J_3-\De_4-J_4}{2}))}\right|_{\mathrm{saddle}} \nn \\
&=\bar v(-n',z_{3}^{x}n'_{\perp}+\vec z_{3\perp},-z_{4}^{x}n'_{\perp}+\vec z_{4\perp},w_1=(0,iz_6^{t},z_6^{x},\vec z_{6\perp})),
\ee
which agrees with the $\bar v(-n',e_3,e_4)$ definition \eqref{eq:CFT_vertices_definition_bar} and the polarization map \eqref{eq:polarvec_346_CFTtoflat}.

\subsection{Computing $R_{\r}$}
In the main text, we claim that
\be\label{eq:Rrho_derivation_eq0}
\left.\frac{2^d \mathrm{dim}(\r)}{\vol(\SO(d))}\frac{\<\tl{\cO}^{\dag}_a(x_5)\tl{\cO}^{\bar b}(x_6^{+})\>}{\p{\<\tl{\cO}^{\dag}\tl\cO\>,\<\cO^\dag \cO\>}}\right|_{\mathrm{saddle}} = 2^{-J+2\De}(-1)^{J-j+\tl{j}}R_\r \p{\mathcal{I}_e^{\r}}_a{}^{\bar b},
\ee
where the $R_{\r}$ coefficient is given by \eqref{eq:Rrhocoeff_definition}. Here, we give a derivation for \eqref{eq:Rrhocoeff_definition} in the case where $\r=(J,j)$. 

By the two-point pairing definition \eqref{eq:Euclidean_2pt_pairing}, we can rewrite the left-hand side of the above equation as
\be\label{eq:Rrho_derivation_eq1}
\mathrm{dim}(\r)\frac{\<\tl{\cO}^{\dag}_a(e)\tl{\cO}^{\bar b}((-e)^{+})\>}{\<\tl{\cO}^{\dag}_{a'}(e)\tl{\cO}^{\bar b'}((-e)^{+})\>\<\cO^{a'}(e)\cO^{\dag}_{\bar b'}((-e)^{+})\>}.
\ee
It is not hard to see that both the shadow two-point $\<\tl{\cO}^\dag\tl{\cO}\>$ and the two-point structure $\<\cO^\dag \cO\>$ are proportional to the reflection tensor $\mathcal{I}^\r_e$. However, the coefficient $R_\r$ should only depend on the convention of the two-point structure $\<\cO^\dag \cO\>$. From \eqref{eq:2pt_withtransverserep_convention}, we get
\be\label{eq:Rrho_derivation_eq2}
&\<\cO^\dag(e,z_5,w_5,\tl{w}_5)\cO((-e)^{+},z_6,w_6,\tl{w}_6)\> \nn \\
&= 2^{J-2\De}(-1)^{J-j+\tl{j}}\hat I^\r_{e}(z_5,w_5,\tl{w}_5;z_6,w_6,\tl{w}_6),
\ee
where
\be
&\hat I^\r_{e}(z_5,w_5,\tl{w}_5;z_6,w_6,\tl{w}_6)\nn \\
&=(z_6\.I_e\.z_5)^{J-j}\p{\sum_{\s\in S_2}z_{6\a_1}w_{6\a_2}I_e^{\a_1}{}_{\b_1}I_e^{\a_2}{}_{\b_2}z_5^{\b_{\s(1)}}w_5^{\b_{\s(2)}}}^{j-\tl{j}}\nn \\
&\x\p{\sum_{\s\in S_3}z_{6\a_1}w_{6\a_2}\tl{w}_{6\a_3}I_e^{\a_1}{}_{\b_1}I_e^{\a_2}{}_{\b_2}I_e^{\a_3}{}_{\b_3}z_5^{\b_{\s(1)}}w_5^{\b_{\s(2)}}\tl{w}_5^{\b_{\s(3)}}}^{\tl{j}},
\ee
where $I_e^{\mu}{}_{\nu} = \de^{\mu}{}_{\nu}+2e^{\mu}e_{\nu}$, and we have introduced polarization vectors $z,w,\tl{w}$ for the three rows of the Young diagram of $\r$. By comparing \eqref{eq:Rrho_derivation_eq0}, \eqref{eq:Rrho_derivation_eq1}, \eqref{eq:Rrho_derivation_eq2}, and using the fact that $\p{\mathcal{I}_e^{\r}}_a{}^{\bar b}\p{\mathcal{I}_e^{\r}}^a{}_{\bar b} = \dim \r$,  we obtain
\be\label{eq:Rrho_derivation_eq3}
&\mathcal{I}_e^{\r} = R_\r \hat I^\r_{e}.
\ee
This implies
\be\label{eq:Rrho_derivation_eq4}
(\hat I^\r_{e})_a{}^{\bar b}(\hat I^\r_{e})^a{}_{\bar b} = R_{\r}^{-2}\dim\r.
\ee

Now, let us specialize to $\r=(J,j)$ and compute $(\hat I^\r_{e})_a{}^{\bar b}(\hat I^\r_{e})^a{}_{\bar b}$. The main idea is that we can use weight-shifting operators to derive a recursion relation \cite{Karateev:2017jgd, Karateev:2018oml}. In particular, we will use
\be
\left.\cD_{z,w}^{0+\mu}\right|_{J,j} &= (j-J)w^{\mu} + z^{\mu}z\.\frac{\ptl}{\ptl w}, \nn \\
\left.\cD_{z,w}^{0-\mu}\right|_{J,j} &= \p{(-J-d+4-j)\de^{\mu}{}_{\nu} + z^\mu \frac{\ptl}{\ptl z^\nu}}\p{(d-6+2j)\frac{\ptl}{\ptl w_\nu}-w^\nu\ptl^2_w}, 
\ee
where $\cD^{0+}_{z,w}$ increases the transverse spin $j$ by $1$ and $\cD^{0-}_{z,w}$ decreases $j$ by $1$. These operators are the weight-shifting operators of the $\SO(d-1,1)$ group in the vector representation \cite{Karateev:2017jgd}. Using them, we can build a ``bubble diagram,"
\be
\cD^{0+}_{z_5,w_5}\.\cD^{0-}_{z_5,w_5} \hat I^{\r}_e(z_5,w_5;z_6,w_6) = j(d-6+2j)(J+j+d-4)(J-j+2) \hat I^{\r}_e(z_5,w_5;z_6,w_6).
\ee
Furthermore, one can perform crossing on a weight-shift operator and move it to the other leg of $\hat I_e$,
\be
\cD^{0-\mu}_{z_5,w_5} \hat I^{\r}_e(z_5,w_5;z_6,w_6) = \frac{j(d-6+2j)(J+j+d-4)}{J-j+1}\cD^{0+\mu}_{z_6,w_6}\hat I^{\r'}_e(z_5,w_5;z_6,w_6),
\ee
where $\r'=(J,j-1)$. Combining the above two equations, we get
\be\label{eq:WS_relation_forIe}
\hat I^{\r}_e(z_5,w_5;z_6,w_6) = \frac{1}{(J-j+1)(J-j+2)} \cD^{0+}_{z_5,w_5}\.\cD^{0+}_{z_6,w_6}\hat I^{\r'}_e(z_5,w_5;z_6,w_6).
\ee

Our goal is to compute the index contraction of two $\hat I^{\r}_e$ tensors. Using \eqref{eq:WS_relation_forIe} we can rewrite it as the index contraction of $\cD^{0+}_{z_5,w_5}\.\cD^{0+}_{z_6,w_6}\hat I^{\r'}_e$ and $\hat I^{\r}_e$. Then, we can ``integrate by parts" to move the weight-shifting operators to act on $\hat I^{\r}_e$. More precisely, let us denote the index contraction by a pairing $\p{\cdots,\cdots}$. Then the integration by parts relation is
\be
\p{\cD^{0+\mu}_{z,w} \hat I^{\r'}_e, \hat I^{\r}_e} = \frac{J-j+2}{j(d-6+2j)(J+j+d-4)}\p{I^{\r'}_e,\cD^{0-\mu}_{z,w}  I^{\r}_e}.
\ee
Note that this relation is different from the one given in \cite{Karateev:2018oml}. This is simply because they are adjoint relations with respect to different pairings. To obtain the adjoint with respect to index contraction, a simple way is to use the identity of the $\cD^{(h)}$ operator given in \cite{Caron-Huot:2022jli}. (Note that the $\cD^{0-}$ operator defined here is just a special case of $\cD^{(h)}$ with $h=2$.)

The adjoint relation enables us to turn the computation into the index contraction of $\hat I^{\r'}_e$ and $\cD^{0-}_{z_5,w_5}\.\cD^{0-}_{z_6,w_6}\hat I^{\r}_e$, which then becomes the contraction of two $\hat I^{\r'}_e$'s. Together with \eqref{eq:Rrho_derivation_eq4}, this gives the recursion relation
\be
R_{\r}^{-2}\dim\r= \frac{(j+d-5)(2j+d-4)(J-j+2)(J+j+d-3)}{j(2j+d-6)(J-j+1)(J+j+d-4)} R_{\r'}^{-2}\dim\r'.
\ee
The relation between $\dim\r$ and $\dim\r'$ can be obtained from standard formula \cite{Kravchuk:2017dzd, Karateev:2018oml}, or from the recursion relation for the Plancherel measure of the $\SO(d-1,1)$ Lorentz group \cite{Karateev:2018oml}. Eventually, we get
\be
\frac{R_{\r'=(J,j-1)}^2}{R_{\r=(J,j)}^2} = \frac{(J-j+2)^2}{(J-j+1)^2}.
\ee
With the initial condition $R_{\r=(J,0)} =1$, this leads to \eqref{eq:Rrhocoeff_definition} in the $\tl{j}=0$ case.

\section{Dual structures at large $\nu$}\label{app:dual_large_nu}
In this appendix, we discuss how to compute the dual structure $\p{\<0|\cO_4\wL[\cO]\cO_2|0\>^{(a)}}^{-1}$ that appears in the kernel of the functional. In particular, we are interested in the large $\nu$ limit, where $\nu$ parametrizes the scaling dimension of $\cO$ as $\De=\frac{d}{2}+i\nu$.

\subsection{Light transform at large dimension}

To understand the dual structure, we should first study the light transform of the three-point function $\<0|\cO_4\cO\cO_2|0\>$, and then consider its Lorentzian three-point pairing. Hence, let us first consider the light transform $\wL[\cO]$ in the limit where $\cO$ has large scaling dimension. Recall that the light transformed three-point function is given by
\be
\<0|\cO_4\wL[\cO_{\De,J,\l'}](X_0,Z_0)\cO_2|0\> = \int d\a\ \<0|\cO_4\cO_{\De,J,\l'}(Z_0-\a X_0,-X_0)\cO_2|0\>
\ee
We will focus on the case $4>0>2^{-}$ and consider the limit $\De=\frac{d}{2}+i\nu$, $\nu \to \oo$. In this limit, we see that the quickly varying part of the integrand is
\be\label{eq:L_alphafunc_full}
 e^{-\frac{i\nu}{2}(\log(-2(Z_0-\a X_0)\.X_2) +\log(-2 (Z_0-\a X_0)\.X_4))},
\ee
which implies that the integral has a saddle point at 
\be
\a_* =\frac{2X_0\.X_2 Z_0\.X_4 + 2X_0\.X_4 Z_0\.X_2}{X_{02}X_{04}}.
\ee
As one can check, this point is spacelike from both 2 and 4. Therefore, the light transform gets localized to the point in the middle of the region that is spacelike from 2 and 4.

\begin{figure}[t!]
\centering
\includegraphics[width=9.5cm]{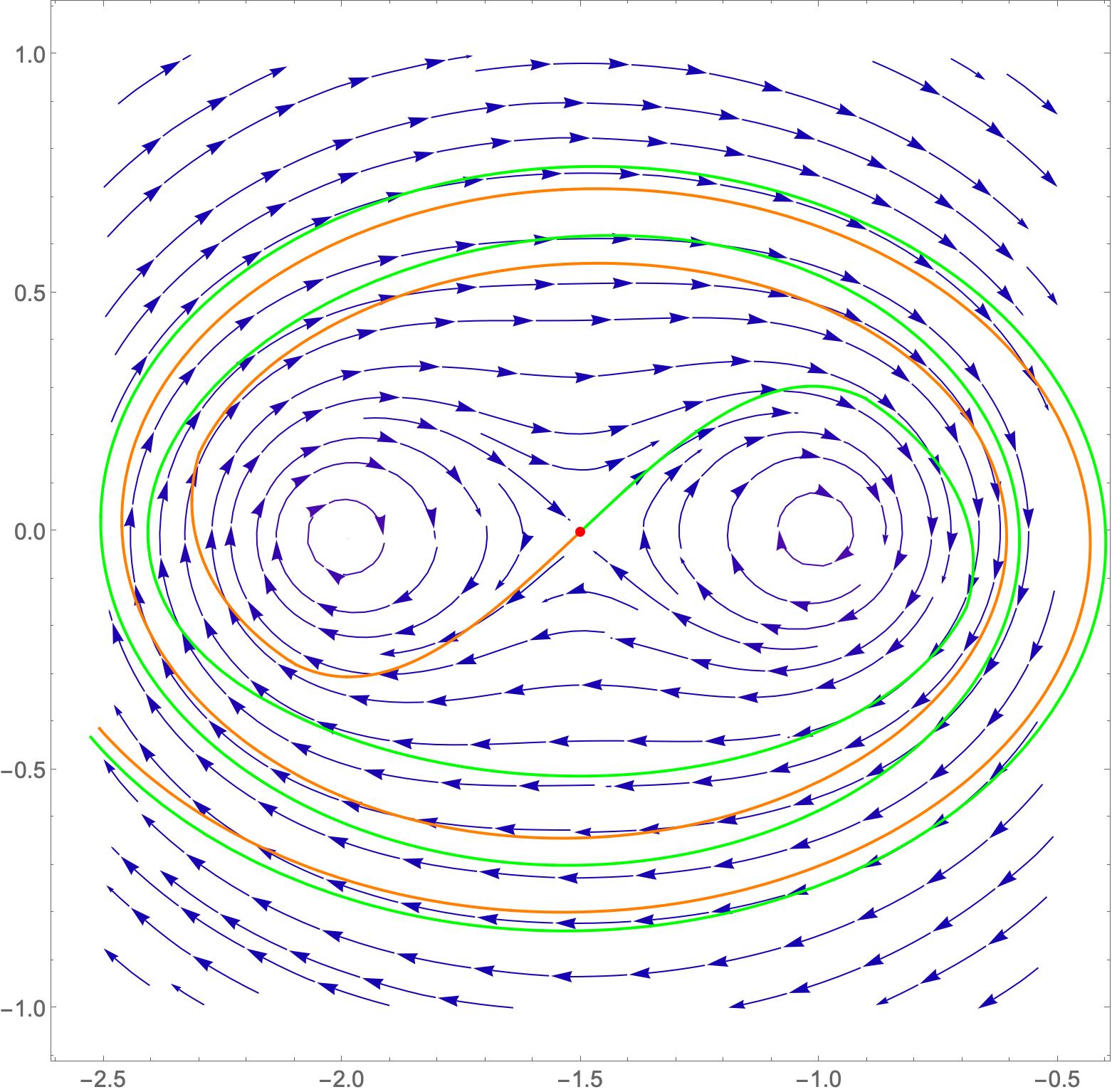}
\caption{The steepest descent flow of the saddle integral \eqref{eq:lighttransform_saddleintegral}. The red point is the saddle point. We show the deformed contour passing through the saddle in green and orange. The green part starts at infinity, goes in the opposite direction of the flow, and gradually spirals in toward the saddle. Along the green contour, the integrand keeps increasing and reaches maximum at the saddle point. It passes through the saddle point along a direction that is rotated by $e^{-\frac{3\pi i}{4}}$ relative to the real axis. After passing through the saddle point, the contour (in orange) goes in the same direction as the flow and gradually spirals out to infinity. The integrand at infinity goes as $\a^{-i\nu}$, so by making the contour spiral many times clockwise at infinity, we ensure that the contribution $\a^{-i\nu} \propto e^{2\pi \nu \,\mathrm{arg}(\a)}$ can be made arbitrarily small.}
\label{fig:contour_lighttransform}
\end{figure}

Expanding the quickly varying part around $\a_*$, we find that the saddle integral we have to do is
\be\label{eq:lighttransform_saddleintegral}
\int d\a\ e^{\frac{i\nu}{2}\p{\frac{X_{02}X_{04}}{V_{0,24} X_{24}}}^2(\a-\a_*)^2} =  e^{-\frac{3i\pi}{4}}\sqrt{\frac{2\pi}{\nu}} \frac{(-V_{0,42})X_{24}}{X_{02}(-X_{04})}.
\ee
The phase factor $e^{-\frac{3i\pi}{4}}$ comes from deforming the contour to a steepest descent contour. We show the steepest descent flow and the deformed contour in figure \ref{fig:contour_lighttransform}. The contour passes through the saddle point with angle $-\frac{3\pi}{4}$, and hence we have the $e^{-\frac{3i\pi}{4}}$ factor in \eqref{eq:lighttransform_saddleintegral}. Also, note that when taking $\p{\frac{X_{02}X_{04}}{V_{0,42} X_{24}}}^2$ out of the square root, we have to make sure all factors are positive. In summary, the above analysis implies that the large $\De$ limit of $\<0|\cO_4\wL[\cO_{\De,J}](X_0,Z_0)\cO_2|0\>$ should be given by
\be
\lim_{\substack{\De=\frac{d}{2}+i\nu \\ \nu\gg1}} \<0|\cO_4\wL[\cO_{\De,J,\l'}](X_0,Z_0)\cO_2|0\> = e^{-\frac{3i\pi}{4}}\sqrt{\frac{2\pi}{\nu}} \frac{(-V_{0,42})X_{24}}{X_{02}(-X_{04})}\<0|\cO_4\cO_{\De,J,\l'}(Z_0-\a_{*} X_0,-X_0)\cO_2|0\>.
\ee

Furthermore, we can study how each building block of the three-point structure transforms under $X_0 \to Z_0-\a_{*} X_0,Z_0\to -X_0$. We will focus on the structures that appear when $\l'$ has a single row of length $j'$. We find
\be\label{eq:light_transform_replacementrule}
&X_{02} \to \frac{(-V_{0,42})X_{24}}{-X_{04}},\quad  V_{2,04} \to -\p{V_{2,04} + \frac{H_{20}}{V_{0,42}}},\quad U_{0,24}\to \frac{(-V_{0,42})X_{24}}{(-X_{04})X_{02}}U_{0,24},\nn \\
&X_{04} \to \frac{(-V_{0,42})X_{24}}{X_{02}}, \quad  V_{4,20} \to -\p{V_{4,20} + \frac{H_{40}}{V_{0,42}}}, \quad U_{0,42}\to \frac{(-V_{0,42})X_{24}}{X_{04}X_{02}}U_{0,42},
\ee
and all other structures are invariant (assuming $\cO_2$ and $\cO_4$ have no transverse spin). Therefore, starting with the three-point structure \eqref{eq:3pt_convention_spinning}, its light transform at large $\De$ is given by
\be\label{eq:light_transform_largeDelta_spinning}
&\lim_{\substack{\De=\frac{d}{2}+i\nu \\ \nu\gg1}}\<0|\cO_4\wL[\cO_{\De,J,j'}](X_0,Z_0,W_0)\cO_2|0\>=2^{\De+J-1}e^{-\frac{3i\pi}{4}}\sqrt{\frac{2\pi}{\nu}}  \nn \\
&\x \frac{(-2V_{0,42})^{1-\De-J+m_0}\p{-V_{2,04}-\frac{H_{20}}{V_{0,42}}}^{m_2}\p{-V_{4,20}-\frac{H_{40}}{V_{0,42}}}^{m_4}H_{24}^{n_{24}}H_{02}^{n_{02}}H_{04}^{n_{04}}(-2U_{0,24})^{k_{02}}(2U_{0,42})^{k_{04}}}{X_{24}^{\frac{\De_2+J_2+\De_4+J_4-2+\De+J-j'}{2}}\p{-X_{04}}^{\frac{\De_4+J_4+2-\De-J+j'-\De_2-J_2}{2}}X_{02}^{\frac{\De_2+J_2+2-\De-J+j'+\De_2+J_2}{2}}}.
\ee
The result can also be written as
\be
\left.2^{\De+J-1}e^{-\frac{3i\pi}{4}}\sqrt{\frac{2\pi}{\nu}} \<0|\cO_4\cO^L(X_0,Z_0,W_0)\cO_2|0\>\right|_{V_{2,04} \to -\p{V_{2,04} + \frac{H_{20}}{V_{0,42}}},V_{4,20} \to -\p{V_{4,20} + \frac{H_{40}}{V_{0,42}}}, U_{0,42} \to -U_{0,42}},
\ee
where the operator $\cO^L$ has quantum numbers $(1-J,1-\De,j')$.

Computation of light transform of general spinning three-point functions has been discussed in \cite{Kologlu:2019bco} for general $\De$ (see eq. (5.69)). One can check that the above result agrees with the large $\De_1$ limit of the result in \cite{Kologlu:2019bco}, which contains an Appell $F_2$. The Appell $F_2$ function in this limit will simplify and lead to the replacement rule $V_{2,04} \to -\p{V_{2,04} + \frac{H_{20}}{V_{0,42}}},V_{4,20} \to -\p{V_{4,20} + \frac{H_{40}}{V_{0,42}}}$ given above.

We also need to understand how to impose conservation condition. In section \ref{sec:conservation_largeDelta}, we study the conservation condition when the exchanged operator $\cO$ has large dimension. Here, after the light transform, the dimension and spin are swapped, and therefore we have to instead consider conservation at large spin. However, the strategy in both cases are the same. We should identify the quickly-varying part of the three-point structure and take its derivative. By taking a $x_4$-derivative of the $\De$-dependent factor of \eqref{eq:light_transform_largeDelta_spinning}, we find that the conservation condition $\ptl_{x_4}\<\cO_4\cO\cO_2\>=0$ after the light transform becomes equivalent to
\be\label{eq:light_transform_conservation_largenu}
\p{V_{4,20} + \frac{H_{40}}{V_{0,42}}}_{A} D^{A}_{Z_4} \<0|\cO_4\wL[\cO]\cO_2|0\> = 0,
\ee
where $A$ is an embedding space index, and its meaning as a subscript is that we should take the structures $V_{4,20},H_{40}$ and strip off the $Z_4$. $D^A_{Z_4}$ is the Todorov/Thomas operator \cite{Costa:2011mg}. This condition is perfectly consistent with the original large $\De$ conservation \eqref{eq:conservation_largeDelta} and the replacement rule from the light transform given by \eqref{eq:light_transform_replacementrule}.

\subsection{Dual structures}

To get the dual structures, we can start with a standard basis of continuous-spin three-point structures $\<0|\tl{\cO}_4^{\dag}\cO^F\tl{\cO}_2^{\dag}|0\>^{(a)}$ by adding spinning structures to the scalar convention \eqref{eq:continuous_spin_convention} following \eqref{eq:3pt_convention_spinning}. The quantum numbers of $\tl{\cO}_2^{\dag},\tl{\cO}_4^{\dag}$ are $(d-\De_2,J_2)$ and $(d-\De_4,J_4)$. For $\cO^F$, we should have $(\De_F,J_F) = (J+d-1,\De-d+1)$, where $\De=\frac{d}{2}+i\nu$ and $J=J_1+J_3-1$ depends on the spin of the external operators. The actual dual structures should be given by
\be
\p{\<0|\cO_4\wL[\cO]\cO_2|0\>^{(a)}}^{-1} = \a^{(a)}{}_{(b)}\<0|\tl{\cO}_4^{\dag}\cO^F\tl{\cO}_2^{\dag}|0\>^{(b)},
\ee
and our goal is to find the coefficients $\a^{(a)}{}_{(b)}$.

Following the definition of dual structures \eqref{eq:dual_struct_definition_spinning}, we have
\be
\a^{(a)}{}_{(b)}\p{\<0|\tl{\cO}_4^{\dag}\cO^F\tl{\cO}_2^{\dag}|0\>^{(b)},\<0|\cO_4\wL[\cO]\cO_2|0\>_{(c)}}_L = \de^{a}_{c},
\ee
which gives
\be\label{eq:dual_struct_coeff_eq}
\a^{(a)}{}_{(b)} =& \p{\<0|\tl{\cO}_4^{\dag}\cO^F\tl{\cO}_2^{\dag}|0\>^{(b)},\<0|\cO_4\wL[\cO]\cO_2|0\>_{(a)}}^{-1}_L \nn \\
=&2^{2d-2}\vol(\SO(d-2))\p{\<0|\tl{\cO}_4^{\dag}(0^{+})\cO^F(\oo,z_0^{*})\tl{\cO}_2^{\dag}(e)|0\>^{(b)}\<0|\cO_4(0^{+})\wL[\cO](\oo,z_0^{*})\cO_2(e)|0\>_{(a)}}^{-1},
\ee
where the in the second line we have used the conformal group to fixed all the points to the configuration $x_4=0^{+}, x_2=e, x_0=\oo$ and $z_0=(1,1,\vec 0)$ (in lightcone coordinates). The prefactor $2^{2d-2}\vol(\SO(d-2))$ comes from the Faddeev-Popov determinant and volume of the stabilizer group of the gauge-fixed configuration \cite{Kravchuk:2018htv}.

Then, the calculation of dual structures is now reduced to evaluating the continuous-spin basis and the light transformed structures in this standard configuration. Since the continuous-spin basis can be obtained from \eqref{eq:continuous_spin_convention}, \eqref{eq:3pt_convention_spinning}, and the light transformed structures at large $\nu$ are given by \eqref{eq:light_transform_largeDelta_spinning}, the dual structures at large $\nu$ can be computed straightforwardly using \eqref{eq:dual_struct_coeff_eq}.

Lastly, when we impose conservation on the external operators $\cO_2,\cO_4$, the dual structures get a gauge redundancy due to the conservation equations. At large $\nu$, the statement becomes
\be
\<0|\cO_4\wL[\cO]\cO_2|0\>^{-1} \sim \<0|\cO_4\wL[\cO]\cO_2|0\>^{-1} + \p{V_{4,20} + \frac{H_{40}}{V_{0,42}}}(\ldots) + \p{V_{2,04} + \frac{H_{20}}{V_{0,42}}}(\ldots).
\ee
The idea is that whenever we have a $\p{V_{4,20} + \frac{H_{40}}{V_{0,42}}}$ factor, we can always integrate it by parts in the pairing \cite{Karateev:2018oml} and get the large-$\nu$ conservation equation \eqref{eq:light_transform_conservation_largenu}. The argument for the other factor is similar. Fortunately, thanks to the identity \eqref{eq:kernel_identity_saddle}, when we evaluate the structures at the saddle and set the time component of the external polarizations to zero, the result is independent of this gauge redundancy.

In summary, to compute the dual structures for conserved external operators, we can use the gauge redundancy to remove all the $V_i$ structures for the external operators and simply consider a basis with just the $H_{i0},H_{ij}$ structures, as stated in section \ref{sec:kernel_shock}. We can then use \eqref{eq:dual_struct_coeff_eq} to compute the dual structure coefficients, where the label $(b)$ only includes continuous-spin structures without the $V_i$'s, and $(a)$ are the conserved structures. These two structures both have the same counting given by \eqref{eq:spinning_dual_counting}.

\section{CFT four-point structures}\label{app:4pt_structures}
In this appendix, we give the expressions of the CFT four-point structures that we use in the main text. In particular, they should satisfy \eqref{eq:CFT_4pt_toflatspace}. From the group-theoretic counting argument \cite{Kravchuk:2016qvl}, we already know that the number of CFT four-point structures should agree with the number of flat space polarization structures. To write down the CFT structures, let us first define
\be
H^{CFT}_{ijk} &= [X_i,Z_i]^{A_1A_2}[X_j,Z_j]^{A_2A_3}[X_k,Z_k]^{A_3A_1}, \nn \\
H^{CFT}_{ijkl} &= [X_i,Z_i]^{A_1A_2}[X_j,Z_j]^{A_2A_3}[X_k,Z_k]^{A_3A_4}[X_l,Z_l]^{A_4A_1}, \nn \\
X^{CFT}_{ijkl} &=H^{CFT}_{ijkl}-\frac{1}{4}\left(H^{CFT}_{14}H^{CFT}_{23} + H^{CFT}_{13}H^{CFT}_{24} + H^{CFT}_{12}H^{CFT}_{34} \right), \nn \\
S^{CFT} &=V_{1,24}^{CFT}H^{CFT}_{234} + V^{CFT}_{2,31}H^{CFT}_{341} + V^{CFT}_{3,42}H^{CFT}_{412} + V^{CFT}_{4,13}H^{CFT}_{123},
\ee
where $A_i$ are embedding space indices, and $[X,Z]^{AB}=X^AZ^B-X^BZ^A$. The superscript $CFT$ is to distinguish the CFT structure and the flat space structure, and $V^{CFT}, H_{ij}^{CFT}$ are given in \eqref{eq:CFT_building_blocks_VH}. Then, we find
\be
&\left.\{H^{CFT}_{14}H^{CFT}_{23},H^{CFT}_{13}H^{CFT}_{24},H^{CFT}_{12}H^{CFT}_{34},X^{CFT}_{1243}, X^{CFT}_{1234}, X^{CFT}_{1324}, S^{CFT}\}\right|_{\mathrm{saddle},\eqref{eq:polarvec_125_CFTtoflat},\eqref{eq:polarvec_346_CFTtoflat}} \nn \\
=&\frac{16}{\nu^4}\left\{H_{14}H_{23},H_{13}H_{24},H_{12}H_{34},X_{1243},X_{1234},X_{1324},\frac{2}{\nu^2}S\right\}.
\ee
Thus, we choose the left hand side of the above equation to be our four-point structure basis $Q^I$ in the main text. 

The CFT four-point structures we choose all have homogeneity $1$ for all $X_{1,2,3,4},Z_{1,2,3,4}$. Under the map of polarizations, they reproduce the expected amplitude structures, with the correct Regge behavior. The additional factors of $\nu$ should not be a big issue since we can absorb them into the prefactors. The only different case is that we have to introduce a $\frac{2}{\nu^2}$ factor for the $S$ structure. This is reasonable from dimensional analysis since the $S$ structure has an additional $p_i\.p_j$. It also seems to suggest that from the CFT point of view, $\frac{2}{\nu^2}S$ is a more natural choice for the structure.

For the four-graviton case, the construction given above can be easily generalized. The only four-graviton flat space structure that can't be written in terms of $H_{ij},X_{ijkl},S$ is the Gram determinant of all dot products between $(p_1,p_2,p_3,e_1,e_2,e_3,e_4)$, which we write as $\cG_{p_1,p_2,p_3,e_1,e_2,e_3,e_4}$. Even though this structure does not appear in the graviton sum rule dictionary given in section \ref{sec:graviton_sum_rules} as it grows more slowly in the Regge limit, let us still give the corresponding CFT structure for completeness. The corresponding CFT structure can be written as
\be\label{eq:graviton_4pt_forGstruct}
X_{24}\p{\frac{X_{14}X_{34}}{X_{12}X_{23}}}^{\frac{1}{2}}\cG_{X_1,X_2,X_3,Z_1,Z_2,Z_3,Z_4} - (\ldots) \cG_{X_1,X_2,X_3,Z_1,Z_2,Z_3,X_4}.
\ee
This structure is manifestly invariant under $Z_i\to Z_i + \#X_i$ thanks to properties of determinant. The only exception is the $Z_4\to Z_4+ \# X_4$ gauge redundancy. This is why we introduce the second term, and $(\ldots)$ is a factor that fixes homogeneity. The structure is constructed such that its homogeneity is the same as the other structures (has homogeneity 1 for all $X_i,Z_i$). When evaluating the above expression at the saddle and applying the polarization map, we find that the second term vanishes, and we have
\be
\left. X_{24}\p{\frac{X_{14}X_{34}}{X_{12}X_{23}}}^{\frac{1}{2}}\cG_{X_1,X_2,X_3,Z_1,Z_2,Z_3,Z_4}\right|_{\mathrm{saddle},\eqref{eq:polarvec_125_CFTtoflat},\eqref{eq:polarvec_346_CFTtoflat}} = \frac{256}{\nu^6}\cG_{p_1,p_2,p_3,e_1,e_2,e_3,e_4}.
\ee
So, \eqref{eq:graviton_4pt_forGstruct} should be the correct CFT four-point structure that gives $\cG_{p_1,p_2,p_3,e_1,e_2,e_3,e_4}$.

\bibliographystyle{JHEP}
\bibliography{refs}

\end{document}